\documentclass[
    aps,
    prb,
    reprint,
    superscriptaddress
    ]{revtex4-2}

\usepackage{amsmath}
\usepackage{amssymb}
\usepackage{physics}

\usepackage{graphicx}

\usepackage{dcolumn}

\usepackage{bm}

\usepackage{mathtools}

\usepackage{xr-hyper}

\usepackage{hyperref}

\usepackage[mathlines]{lineno}

\usepackage[dvipsnames]{xcolor}

\makeatletter
\newcommand*{\addFileDependency}[1]{
  \typeout{(#1)}
  \@addtofilelist{#1}
  \IfFileExists{#1}{}{\typeout{No file #1.}}
}
\makeatother

\newcommand*{\myexternaldocument}[1]{
    \externaldocument{#1}
    \addFileDependency{#1.tex}
    \addFileDependency{#1.auxx}
}

\myexternaldocument{si}

\setcitestyle{sort&compress,numbers,super} 

\begin{document}


\title{Near-Field Enhancement of Optical Second Harmonic Generation in Hybrid Gold-Lithium Niobate Nanostructures
}

\author{Rana Faryad Ali}
\thanks{Authors contributed equally}
\affiliation{Department of Chemistry and 4D LABS, Simon Fraser University, Burnaby, British Columbia V5A 1S6, Canada.}

\author{Jacob A. Busche}
\thanks{Authors contributed equally}
\affiliation{Department of Chemistry, University of Washington, Seattle, Washington 98195, USA.}

\author{Saeid Kamal}
\affiliation{Department of Chemistry and 4D LABS, Simon Fraser University, Burnaby, British Columbia V5A 1S6, Canada.}

\author{David J. Masiello}
\affiliation{Department of Chemistry, University of Washington, Seattle, Washington 98195, USA.}

\author{Byron D. Gates}
\email{bgates@sfu.ca}
\affiliation{Department of Chemistry and 4D LABS, Simon Fraser University, Burnaby, British Columbia V5A 1S6, Canada.}

\begin{abstract}
Nanophotonics research has focused recently on the ability of non-linear optical processes to mediate and transform optical signals in a myriad of novel devices, including optical modulators, transducers, color filters, photodetectors, photon sources, and ultrafast optical switches. The inherent weakness of optical nonlinearities at smaller scales has, however, hindered the realization of efficient miniaturized devices, and strategies for enhancing both device efficiencies and synthesis throughput via nanoengineering remain limited. Here, we demonstrate a novel mechanism by which second harmonic generation, a prototypical non-linear optical phenomenon, from individual lithium niobate particles can be significantly enhanced through nonradiative coupling to the localized surface plasmon resonances of embedded gold nanoparticles. A joint experimental and theoretical investigation of single mesoporous lithium niobate particles coated with a dispersed layer of $\sim$10-nm diameter gold nanoparticles shows that a $\sim$32-fold enhancement of second harmonic generation can be achieved without introducing finely tailored radiative nanoantennas to mediate photon transfer to or from the non-linear material. This work highlights the limitations of current strategies for enhancing non-linear optical phenomena and proposes a route through which a new class of subwavelength nonlinear optical platforms can be designed to maximize non-linear efficiencies through near-field energy exchange.
\end{abstract}

\maketitle



\section{Introduction}

Ultrafast optical frequency conversion using nonlinear optical (NLO) harmonic generation in micro- and nanoscale materials is emerging as an important phenomenon in the basic and applied study of photonics,\cite{firstenberg2013attractive} as well as for the development of novel sensing and imaging techniques in materials science, chemistry, and biology.\cite{garmire2013nonlinear} Perhaps the simplest of the ultrafast NLO processes relevant to each of these fields is second harmonic generation (SHG), in which two input ﬁelds oscillating at a fundamental frequency are coherently combined into an output ﬁeld oscillating at twice the fundamental.\cite{ali2019one-pot,ali2022lithium} In general, SHG is prized for its narrowband operation, relatively high efficiency among NLO processes, and natural occurrence in a wide array of inorganic materials.\cite{ali2019one-pot,ali2022lithium} 

However, even in ``good'' NLO materials, the nonlinear wave motion underpinning SHG is generally either weak\cite{walther2018giant} or slower than the $\sim$1-ps timescales desired in ultrafast applications.\cite{gao2022single-layer} To overcome this difficulty, SHG photonics designs have until recently centered around the use of large NLO crystals within which the fundamental wavelength (FW) and SH light waves can be constructively interfered to overcome intrinsically low SHG efficiencies.\cite{kauranen2013freeing,zhang2017phase-matching} It has become clear, however, that much smaller micro- and nanoscopic SHG devices are more desirable for cutting-edge applications like few-molecule sensing,\cite{butet2012sensing} bioimaging,\cite{malkinson2020fast} and phase-resolved light emission\cite{tanaka2021unidirectional} due to the unique degree of control over the spectrum and spatial distribution of the scattered fields that miniaturized optical systems provide.\cite{fratalocchi2015nano} 

Unfortunately, due to the complicated field profiles of nanostructured particles, interference and phase-matching techniques are rendered ineffective. As a result, the absolute efficiency of an ultrafast NLO process at the nanoscale tends to depend on the microscopic volume of NLO material used.\cite{savo2020broadband} Researchers continue to pursue fundamentally new NLO enhancement techniques and novel nanofabrication solutions to optimize or supersede this limit, exploring the NLO properties of light confined within resonant dielectric,\cite{wang2021high} metallic,\cite{li2021light-induced} and hybrid\cite{pu2010nonlinear} micro- and nanostructures, as well as within nanopatterned waveguides.\cite{suchowski2013phase} 


Our investigation addresses both the scientific and engineering challenges posed in NLO nanophotonics by revealing a new and practical strategy for enhancing SHG radiation. Specifically, we use the surface-localized near-fields of disordered arrays of Au nanoparticles (NPs) to create hitherto unexplored energy-transfer pathways for nanolocalized upconverted light. These disordered nanostructures, comprised of $\sim$1-$\mu$m diameter mesoporous lithium niobate (LiNbO$_3$) microspheres coated in a dispersed layer of 10-nm diameter Au NPs, allow for a straightforward and high-throughput synthetic process devoid of the difficulties of precision nanofabrication while simultaneously producing an NLO system with a large number of regions of enhanced field intensity (so-called ``hot spots'') between the LiNbO$_3$ and Au surfaces.

In stark contrast, many past investigations\cite{lehr2015enhancing,timpu2017enhanced,yi2019doubly,gurdal2020enhancement,chauvet2020hybrid,bonacina2020harmonic} have achieved enhancement factors between 5 and 20 by employing carefully-tailored hybrid nanostructures. The manufacture of such structures is technically challenging, causing difficulties with the reproduction of large enhancement ratios between samples.\cite{chauvet2020hybrid} Other studies have demonstrated enhancement factors of $\sim$100--1000 more reliably using simple core-shell SHG systems.\cite{richter2014coreshell,pu2010nonlinear} However, the field profiles and resonance structures of shells, which are often assembled as collections of large NPs, can be complex and difficult to separate from the influence of imperfections,\cite{weber2015far-} and can be susceptible to degradation under exposure to high-intensity lasers.\cite{cavicchi2013single} As such, we have chosen to use a more rarefied ensemble of smaller and more easily modeled NPs to isolate the NLO enhancing abilities and governing physical principles of an optically robust and reproducible proof-of-concept nanostructure design.





As will be shown below, our multiple-hot-spot geometry enhances SHG by a factor of 32. Further, this enhancement is clearly attributed through a complete make-measure-model synthesis and characterization approach to the near-field localization and resonant behaviors of metallic nanoantennas that are generally considered unsuitable for the task of enhancing far-field (i.e., radiative) emission. Although a few very recent studies\cite{taitt2021gold-seeded,gao2022single-layer} have measured moderate (e.g., $\sim$7$\times$) SHG enhancements using similarly disordered NP ensembles, each was either unable to isolate single nanostructures (Ref. \citenum{taitt2021gold-seeded}) or to thoroughly explore the governing mechanism behind the observed SHG enhancements (Ref. \citenum{gao2022single-layer}). Our results, therefore, represent both the first demonstration of NLO enhancement values of $>$10 from single disordered nanostructures, as well as the discovery of as-yet unknown mechanisms by which strongly-subwavelength NPs can enhance SHG.

\section{Results}

\subsection{Preparation of hybrid SHG microspheres}

\begin{figure*}[!ht] 
\centering
\includegraphics[width = 0.9\textwidth]{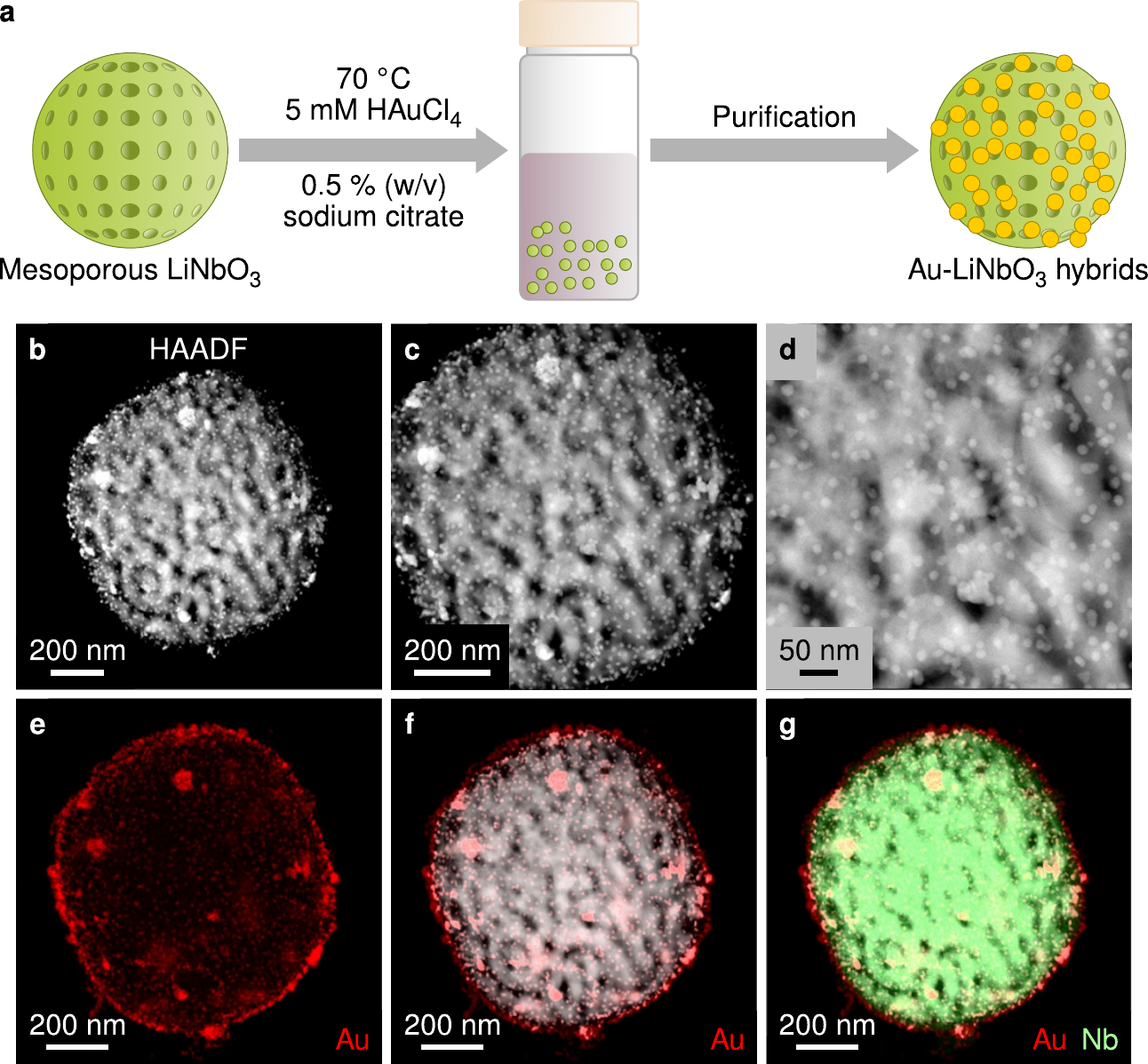}
\caption{
\textbf{Synthesis of hyrbid Au-LiNbO$_{\bm{3}}$ nanostructures.} \textbf{(a)} Representative schematic to prepare the hybrids of Au NPs with LiNbO$_3$ as achieved using an \textit{in situ} synthesis method. Assemblies of Au-LiNbO$_3$ hybrid particles as characterized by scanning transmission electron microscopy (STEM) operating in: \textbf{(b, c, d)} a high-angle annular dark-field (HAADF) mode. Energy dispersive X-ray spectroscopy (EDS) analysis of the Au-LiNbO$_3$ hybrids obtained by STEM techniques was used to create EDS maps of: \textbf{(e)} Au NPs; \textbf{(f)} Au NPs overlaid on the HAADF image of the  assemblies; and \textbf{(g)} overlaid Au-Nb signals within these assemblies.
\label{fig:fig1}
}
\end{figure*}

Selecting an NLO material with a large NLO susceptibility coefficient is critical to improve the efficiency of the SHG conversion process at smaller scales. Relative to many NLO materials, LiNbO$_3$ is a unique photonic material, often referred to as the “silicon of photonics” due to its relatively large second-order susceptibilities (e.g., 41.7 pm/V), higher optical damage resistance (e.g., 20 W/cm$^{2}$ for congruent LiNbO$_3$ crystals), and a relatively wide window of optical transparency (e.g., 400 to 5 000 nm).\cite{ali2018synthesis,ali2019one-pot} In the first step, monodisperse mesoporous LiNbO$_3$ particles were prepared by modifying a previously reported method (details in Materials and Methods Section \ref{sec:methodsSupp}).\cite{ali2022lithium} Scanning transmission electron microscopy (STEM) images confirm that diameters of the LiNbO$_3$ particles were $\sim$1 000 nm (Figure \ref{fig:fig1}b--d and Figure \ref{fig:figS1}) and that each particle has a mesoporous structure. These mesoporous particles contained a randomly disordered network of distinct grains in its bulk and a textured surface that contains randomly distributed $\sim$20-nm diameter pores. In addition, X-ray diffraction (XRD) patterns of the LiNbO$_3$ particles in a powder form were acquired and correlate well with a reported LiNbO$_3$ reference (Figure \ref{fig:figS2}), displaying a distinct rhombohedral phase (space group \textit{R3c}) that accompanies strong SHG behavior.\cite{ali2018synthesis,ali2019one-pot} Further evidence of the rhombohedral phase of the LiNbO$_3$ particles was provided by Raman spectroscopy of a powdered sample (Figure \ref{fig:figS3}), in which the observed bands are characteristic of rhombohedral LiNbO$_3$ and agree well with a commercially available LiNbO$_3$ standard.\cite{ali2018synthesis}

In the second step to prepare the hybrid Au-LiNbO$_3$ particles, we utilized an \textit{in situ} seed-mediated growth to load the surfaces of the mesoporous LiNbO$_3$ with Au NPs (Figure \ref{fig:fig1}a). This approach used a one-pot synthetic technique that eliminated the need for precision nanoengineering techniques to position the plasmonic materials upon their NLO support.\cite{sun2021progress} Briefly, an aqueous suspension of bare, mesoporous LiNbO$_3$ particles in the presence of Au chloride as a precursor was heated at 70 $^\circ$C for 3 h. The \textit{in situ} reduction of the Au precursor using sodium citrate as a reducing agent led to the formation of the hybrid structures, wherein the Au NPs were observed both through electron microscopy imaging (Figure \ref{fig:fig1}b--d) and energy dispersive X-ray spectroscopy (EDS) based mapping techniques (Figure \ref{fig:fig1}e--g and Figure \ref{fig:figS4}) to be uniformly distributed over the surfaces of the LiNbO$_3$ particles. Moreover, the Au particles were limited to diameters of $\sim$10 nm (Figure \ref{fig:fig1}b--d and Figure \ref{fig:figS5}). These Au nanoparticles were deposited on the outer surfaces, as well as within the pores of the mesoporous LiNbO$_3$. Synthesis of hybrid particles prepared from porous LiNbO$_3$ supports revealed that the number of Au NPs loaded onto each LiNbO$_3$ particle was $\sim$900--1 000.

\subsection{Linear and nonlinear optical spectroscopy}

\begin{figure}[ht!] 
\centering
\includegraphics[width = 0.4\textwidth]{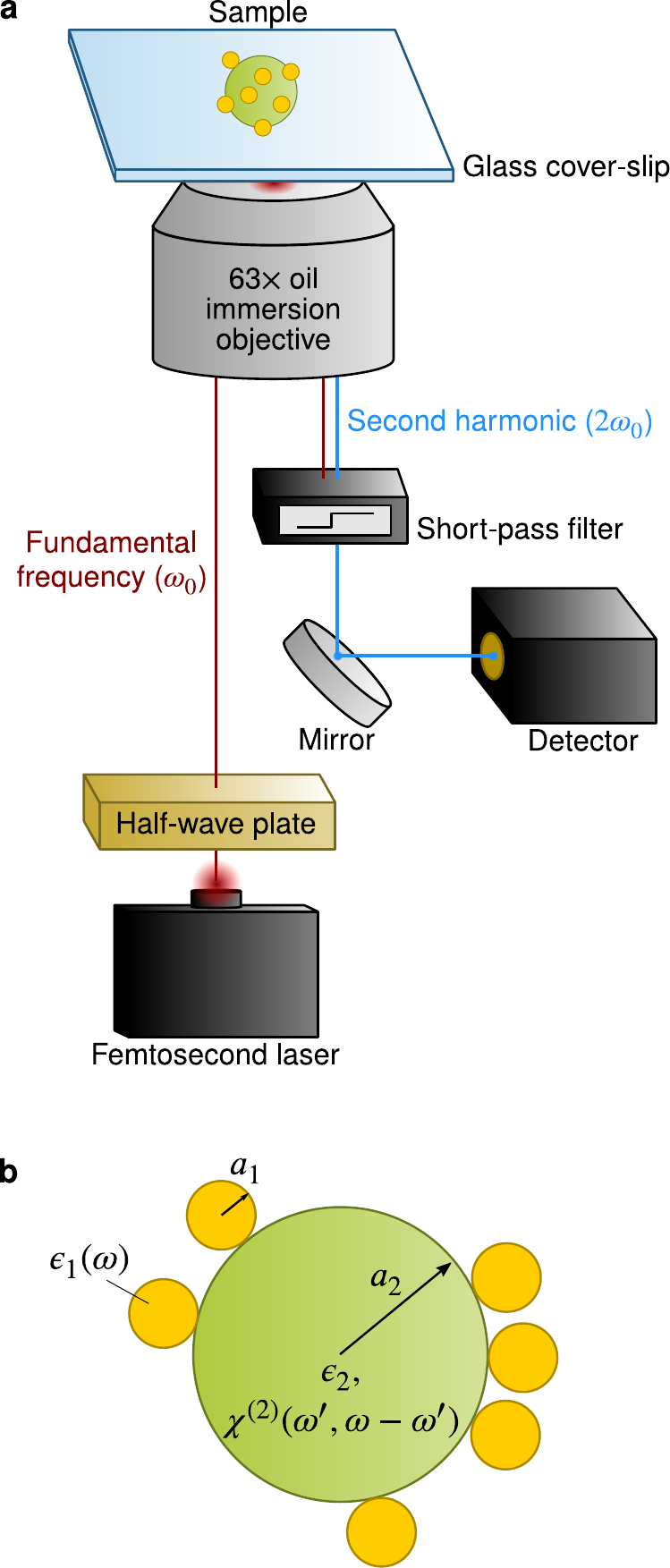}
\caption{
\textbf{Experimental geometry.} \textbf{(a)} Schematic of the microscope configuration used to characterize the second harmonic generation (SHG) response of single particles. This microscope was operated in a reflection mode setup and was equipped with a femtosecond laser, half-wave plate, and objective lenses of different magnifications including a 63$\times$ oil immersion lens. \textbf{(b)} Sketch of the sample geometry of the hybrid Au-LiNbO$_3$ particles.
\label{fig:fig2}
}
\end{figure}

Analysis of the SHG emission from the hybrid nanostructures was performed using a Leica SP5 commercial two-photon microscope. The incident light was generated using a pulsed, mode-locked Ti:sapphire laser with a pulse width of $\sim$140 fs. The resonances of the NLO particles have periods between 0.1 and 0.3 fs such that their driven oscillatory behavior and associated SHG scattering are independent of the pulse envelope. Additionally, with a repetition rate of 80 MHz, the spacing between pulses (12.5 ns) is much longer than the lifetimes of the target resonances (2--30 fs) such that scattering data from each pulse is considered to be uncorrelated to the data obtained from prior pulses. Moreover, the use of short, well-separated pulses also enabled the laser to deliver the high-intensity field required to generate the observable SHG signal. The scattered light and SHG emission from single particles were each analyzed by collecting the light with a microscope objective lens and analyzing the signal using a spectrometer that enabled separation of the input signal (e.g., fundamental wavelength) from the output signal (e.g., SH wavelength). Finally, The laser frequency was tuned from 680 nm to 1 080 nm to analyze the wide spectral response of these NLO materials. Additional details of the experimental setup are shown in Figure \ref{fig:fig2}.

\begin{figure}[!ht] 
\centering
\includegraphics[width = 0.4\textwidth]{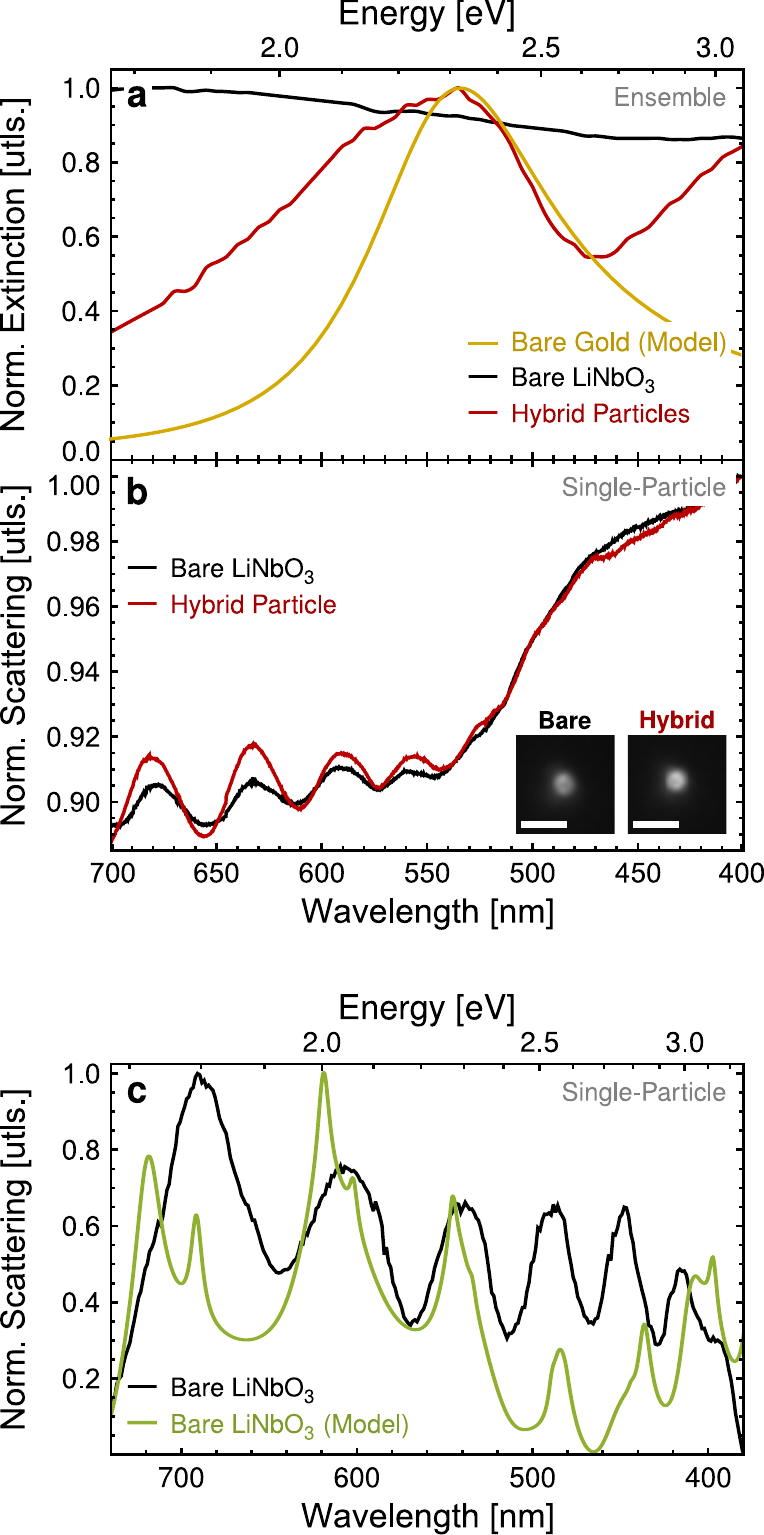}
\caption{
\textbf{Experimental and theoretical optical spectra.} \textbf{(a)} Measured extinction spectra for ensembles of bare LiNbO$_3$ particles (black) and of hybrid Au-LiNbO$_3$ particles (red) compared to the modeled dipole plasmon extinction (gold). These spectra show a plasmonic peak at $\sim$530 nm for the hybrid Au-LiNbO$_3$ particles, while the bare LiNbO$_3$ particles were optically transparent. \textbf{(b)} Light scattering spectra for individual, bare LiNbO$_3$ particles and individual Au-LiNbO$_3$ hybrid particles measured in a dark field (DF) configuration with white light illumination. The insets in (b) portray grey color images (scale bars = 2 $\mu$m) for (black) a single LiNbO$_3$ particle and (red) a single Au-LiNbO$_3$ hybrid particle supported on glass coverslips. The data were normalized to the maximum value of the scattering intensity. \textbf{(c)} Theoretical scattering cross sections (green) from a model microsphere with dielectric constant $6.3 + 0.05i$ and radius 700 nm in comparison with experimentally measured scattering data from a single, bare LiNbO$_3$ particle (black), normalized to range from 0 to 1 within the selected frequency window.
\label{fig:fig3}
}
\end{figure}

First, we acquired and compared the extinction spectra of the hybrid Au-LiNbO$_3$ particles, pristine LiNbO$_3$ particles, and Au NPs. Aqueous suspensions of each sample were prepared and characterized using a UV-visible spectrometer. No extinction peak was observed for the pristine LiNbO$_3$ particles indicating their optical transparency in the region from 400 to 800 nm (Figure \ref{fig:fig3}a). We prepared $\sim$10-nm diameter Au NPs and acquired and characterized their extinction spectra to obtain a detailed comparison to the hybrid Au-LiNbO$_3$ particles. This suspension of Au NPs had an extinction peak at $\sim$520 nm due to their plasmonic response (Figure \ref{fig:figS6}). The hybrid particles of Au-LiNbO$_3$, however, exhibited a broad extinction band from the visible to the near-IR with a maximum centered at 530 nm (Figure \ref{fig:fig3}a). The presence of this peak in the hybrid Au-LiNbO$_3$ particles was due to the contributions of the Au NPs. The broadness and red shift in the resonance peak of the Au-LiNbO$_3$ particles could be attributed to the formation of Au NP aggregates on the surfaces of the LiNbO$_3$ particles.\cite{madzharova2019gold}

For the linear scattering and nonlinear optical analyses of individual particles, we prepared separate, dilute aqueous suspensions (0.1 mg/mL) of both the LiNbO$_3$ and Au-LiNbO$_3$ particles via sonication. The resultant suspensions were drop cast onto glass coverslips and dried under a vacuum to evaporate the solvent. Microscopy analyses of the resulting substrates indicated the presence of well-dispersed, individual particles. The linear optical spectra of individual Au-LiNbO$_3$ and pristine LiNbO$_3$ particles were obtained to evaluate their scattering profiles (Figure \ref{fig:fig3}b). A dark-field optical spectroscopy setup was used for these scattering measurements where individual particles were imaged using a 50$\times$ dark-field objective. The individual particles were illuminated by white light generated from a halogen lamp. The scattered light was collected by the same objective lens and detected with an imaging spectrometer through a pinhole that restricted the collection of signal from the area around the measured nanoparticle. We determined the position and the number of the scattering peaks in the linear optical spectra of each type of particle. Four distinct scattering peaks were observed for both the Au-LiNbO$_3$ hybrids and pristine LiNbO$_3$ particles. As the diameters of the particles were comparable to the wavelength of the incident light, the linear scattering response of these materials does not resemble a Rayleigh scattering profile. Due to a negligible change in the size of Au-LiNbO$_3$ hybrids relative to the pristine LiNbO$_3$, the electromagnetic field inside each type of particle was consistent, resulting in indistinguishable Mie scattered modes between pristine and hybrid particles.\cite{bohren2004absorption} These scattering results are comparable to the Mie scattering spectra previously reported for LiNbO$_3$ based materials having comparable sizes.\cite{timpu2019lithium} The presence of distinct resonances in the scattering spectra indicate that both the Au-LiNbO$_3$ and LiNbO$_3$ particles act as optical Mie resonators. By considering these optical measurements of individual Au-LiNbO$_3$ and LiNbO$_3$ particles, it can be concluded that the Au NPs play no significant role in the Mie resonances in these materials. The features of Mie scattering were not measurable below 400 nm and above 700 nm due to the loss of sensitivity of the measurement setup in these regions. The linear optical spectra of individual Au NPs with diameters of $\sim$10 nm could not be detected due to their smaller size and limitations of the objective lens to locate individual Au particles at the nanoscale. A further improvement of the dark-field spectroscopy setup in the UV range is challenging because of the lack of objective lenses that have both a high transmission in the UV range and no chromatic aberrations across both the UV and visible range. 

We acquired input wavelength-dependent (Figure \ref{fig:fig4}a) and power-dependent (Figure \ref{fig:fig4}b) NLO spectra for the prepared mesoporous Au-LiNbO$_3$ hybrids to ensure the materials are SHG active and to verify the second-order nature of the detected signals. The power dependent SHG signals were collected by varying the power of the laser at the FW of 800 nm while imaging individual Au-LiNbO$_3$ particles on the substrate. The image emission matrix was averaged to yield a mean value in regions of interest selected to exclude the majority of the dark pixels. The selected regions of interest were maintained unchanged throughout the series of acquired images. The logarithms of the mean values for the SHG signals were plotted against the logarithm of the laser power. The slopes of the linear fits reflect the power dependence of the SHG on the pumping intensity. A slope of $\sim$2.0 for the Au-LiNbO$_3$ particles confirm that the signal is generated from a second-order NLO process (i.e., SHG). The SHG for individual Au-LiNbO$_3$ particles was also assessed for a series of discrete fundamental wavelengths. The resultant SH responses were each normalized by dividing their intensities by the maximum scattering intensity within each SH band when assessing the frequency doubling behaviour of the SHG process. A tunable SH response was observed at 400, 420, 440, 460 and 480 nm when the product was excited with FWs of 800, 840, 880, 920, and 960 nm, respectively (Figure \ref{fig:fig4}a). These results correlated well to the anticipated frequency doubling of the fundamentals.

The SHG spectra were acquired for a cluster of Au NPs using the same experimental setup to verify whether the strong SHG emission originated from the surface of the gold rather than from the NLO LiNbO$_3$ core. We used a cluster of solid Au NPs since individual 10-nm diameter gold NPs were difficult to drop-cast and locate using the 63$\times$ objective lens. We scanned the FWs over the range from 850 nm to 1 060 nm and collected the SHG response for the clusters of Au NPs (Figure \ref{fig:figS7}). No SHG response was generated from 400 to 530 nm for the Au NPs. The SHG emitted by the Au-LiNbO$_3$ hybrids is, therefore, exclusively generated by the LiNbO$_3$ particles. For NPs made out of centrosymmetric materials such as gold, SHG is mainly emitted from an Angstrom-thin layer near the particle surface since SHG emission is mainly induced by the surface normal component of the second-order nonlinear susceptibility tensor.\cite{sipe1980analysis, capretti2014size-dependent} The Au NPs used in this work are, however, spherical and the SHG generated from the bulk of the gold nanoparticle and through the surface-tangential component of the second-order nonlinear susceptibility is expected to be very low and are ignored in this study (Figure \ref{fig:figS7}).\cite{capretti2014size-dependent}

We measured the SHG spectra of individual pristine LiNbO$_3$ and hybrid Au-LiNbO$_3$ particles by sweeping the incident laser over the wavelength range from 850 to 1 070 nm in steps of 5 nm to investigate whether Au NPs play a role in the SHG enhancement of the hybrid structures (Figures \ref{fig:fig4}c and \ref{fig:figS14}a). The corresponding SHG response was recorded from 425 to 530 nm. A comparison was made between the SHG intensity from individual Au-LiNbO$_3$ particles with individual pristine LiNbO$_3$ particles examined under the same experimental parameters to quantitatively determine the NLO enhancement. This method to calculate the SHG enhancement has been used in the reported literature, such as where the enhancement of the NLO signals from metasurfaces is calculated by comparing the patterned metasurface to the unpatterned original surface.\cite{liu2016resonantly,yang2015nonlinear} After normalization of the SHG output of the Au-LiNbO$_3$ hybrids to the SHG output of the bare LiNbO$_3$, we were able to calculate the enhancement factors. The obtained enhancement factor was plotted against the FWs of the measurement. The highest enhancement value was obtained at a FW of 1 030 nm, reaching an enhancement of 32 times (Figures \ref{fig:fig4}c and \ref{fig:figS14}a). This evaluation demonstrates a corresponding enhancement in the SHG response at 515 nm by a factor of 32 for the mesoporous LiNbO$_3$ particles loaded with Au NPs. We prepared Au-LiNbO$_3$ hybrids with a lower loading of Au NPs to understand what role the Au NPs play in enhancing the SHG response (Figures \ref{fig:figS8}, \ref{fig:figS9}). The average number of Au NPs on each LiNbO$_3$ particle was $\sim$6. We acquired the SHG response of these Au-LiNbO$_3$ hybrids with their lower loading of Au NPs and compared this response to the NLO response of the bare LiNbO$_3$ particles. Interestingly, no enhancement in the SHG response was observed for these hybrids of Au-LiNbO$_3$ with only a few Au NPs (Figure \ref{fig:figS10}). These observations indicated the need to have a higher loading of Au NPs on the surfaces of the LiNbO$_3$ to enhance the SHG response.

\subsection{Interpretation of scattering signals}

With plasmon-induced enhancements greater than 30, the hybrid particles in this investigation show commensurate performance with other NLO nanostructures of a similar size without the requirements of being prepared using precision nanoengineering techniques. For example, an SHG emission enhancement factor of $\lesssim$ 30 is reported in Ref. \citenum{gili2018metal-dielectric} and a factor of $\sim$22 is reported in Ref. \citenum{renaut2019reshaping}. However, the plasmonic Au NPs used in this study are not properly tuned to act as efficient nanoantennas, obscuring the phenomena underpinning their ability to magnify SHG.

More precisely, hybrid SHG nanostructures generally employ metal NPs to perform one of two functions relevant to upconversion. The first function, in which plasmons enhance the in-coupling of light to the SHG material, employs the near-fields of highly polarizable plasmons that are tuned to the FW to augment the pump field of an NLO process.\cite{pu2010nonlinear,gili2018metal-dielectric} In the second function, radiative plasmons tuned to the up- or down-converted frequency enhance the out-coupling of light by extracting energy from the NLO material through near-field interactions and then quickly scattering it to the far-field, i.e., via the Purcell effect.\cite{renaut2019reshaping,timpu2017enhanced} Carefully designed nanostructures can make use of both strategies,\cite{chauvet2020hybrid} and antenna effects can be significant whether or not the NLO material itself is optically resonant.\cite{gurdal2020enhancement}

Our nanoantennas are far too small to act as radiative out-coupling nanoantennas yet have a plasmon resonance that is aligned with the SH output maximum. The most straightforward explanations for the surprising efficacy of these ``bad antenna'' NPs are that the NPs' enhanced SHG emission is a result of either of the following hypotheses: (i) coupling effects, in which optical energy is routed through the system via more complicated interactions than can be captured by a simple antenna picture; or (ii) superradiant plasmon scattering generated by their collective interaction with the radiation field. 

No existing model can fully address either of these questions. Recent theoretical studies have investigated several related phenomena, including radiation from dipole-driven Mie scatterers, \cite{hall2017unified} NLO scattering from bare dielectric spheres,\cite{smirnova2018multipolar} NLO emission from quantum emitters in idealized Fabry-P\'{e}rot cavities,\cite{chang2016deterministic,welakuh2021down-conversion} and light emission from ensembles of oscillating dipoles.\cite{choudhary2019weak,masson2022universality} However, a novel synthesis of the ideas from each study is required to uncover the operative SHG enhancement mechanism(s) in our nanostructures.

Beginning with hypothesis (i), we first simplify the particle geometries highlighted in Figure \ref{fig:fig1} in a manner consistent with prior investigations of complicated multicrystalline scatterers.\cite{savo2020broadband} In detail, we model each microsphere as a smooth, isotropic spherical particle of radius $a_2 = 500$ nm. We also simplify the form of the incoming light, assuming the laser field $\mathbf{E}_0(\mathbf{r},\omega)$ to be a monochromatic plane wave of characteristic field strength $E_0$ and frequency $\omega_0$ that travels in the $z$-direction and is linearly polarized along $x$. That is to say, $\mathbf{E}_0(\mathbf{r},\omega) = E_0[\pi\delta(\omega - \omega_0)\exp(\mathrm{i}\omega_0z/c) + \pi\delta(\omega + \omega_0)\exp(-\mathrm{i}\omega_0z/c)]\hat{\mathbf{x}}$.

The combination of LiNbO$_3$'s nearly constant dielectric function $\epsilon_2 = 5.5 + 0.035i$ (see Ref. \citenum{palik1998handbook} and also the Supplementary Information [SI], Section \ref{sec:materialModels}) and relatively weak second-order nonlinear susceptibility $|\bm{\chi}^{(2)}(\omega_0,\omega - \omega_0)|\sim10^{-6}$ cm/statV (SI, Section \ref{sec:materialModels} and Figure \ref{fig:figS11}) for observation frequencies $\omega$ and laser frequencies in the optical region of interest allows for simplification of the LiNbO$_3$'s NLO behaviors. Specifically, the nonlinear wave motion of an excited microsphere can be perturbatively expanded as the superposition of Mie resonances at the laser FW, SH, third harmonic, and so on (SI, Section \ref{sec:nearField}).

The FW scattered electric field $\mathbf{E}_\mathrm{sca}(\mathbf{r},\omega)$ is simply the component of the total electric field for which the output frequency matches the laser frequency. The FW resonances in general have both appreciable magnetic and electric fields, have nondipolar symmetry, and produce a field $\mathbf{E}_\mathrm{sca}(\mathbf{r},\omega) = \sum_{\bm{\alpha}}a_2^{-\ell - 2}[ \rho_{\bm{\alpha}}(\omega) \mathbf{X}_{\bm{\alpha}}(\mathbf{r},k) + \rho_{\bm{\alpha}}^*(-\omega) \mathbf{X}^*_{\bm{\alpha}}(\mathbf{r},k)]$ outside the microsphere's surface. Here, $k = \omega/c$ is the wavenumber and $\bm{\alpha} = \{T,p,\ell,m\}$ is the collective index of each Mie mode near the FW, with $T = E,M$ signifying the type (electric or magnetic) of the mode, $p$ (even or odd) its $x$-axis reflection symmetry, $\ell = 1,2,\ldots$ its spherical harmonic order, and $m = 0,\ldots,\ell$ its spherical harmonic degree. The vector spherical harmonic mode functions $\mathbf{X}_{\bm{\alpha}}(\mathbf{r},k)$ and multipole moments $\rho_{\bm{\alpha}}(\omega)$ encode the spatial lobe structure and spectrum, respectively, of each resonance.

From here, in accordance with previous work,\cite{smirnova2018multipolar} we assume that upconversion in the LiNbO$_3$ spheres during the experiment is coherent and occurs when a small portion the energy bound in the laser and FW scattered fields is converted into a second material polarization oscillating at the SH ($2\omega_0$). This so-called nonlinear polarization field is always proportional to $|\bm{\chi}_2^{(2)}(\omega_0,\omega_0)|E_0^2$ (see Eq. [\ref{eq:currents}]) and generates two SH electric fields of its own. 

The first SH field, $\bm{\mathcal{E}}_0(\mathbf{r},\omega)$, describes direct radiation by the nonlinear polarization. It does not interact with the LiNbO$_3$ sphere directly but does excite the Au NP ensemble. The second, $\bm{\mathcal{E}}_\mathrm{sca}(\mathbf{r},\omega) = \sum_{\bm{\beta}}a_2^{-\ell - 2}[ \rho_{\bm{\beta}}(\omega) \mathbf{X}_{\bm{\beta}}(\mathbf{r},k) + \rho_{\bm{\beta}}^*(-\omega) \mathbf{X}^*_{\bm{\beta}}(\mathbf{r},k)]$, is a scattered field similar to the scattered FW electric field $\mathbf{E}_\mathrm{sca}(\mathbf{r},\omega)$ but is driven by the nonlinear polarization rather than directly by the laser light. In fact, $\bm{\mathcal{E}}_\mathrm{sca}(\mathbf{r},\omega)$ can be defined identically to its FW counterpart but with the sum over $\bm{\alpha}$, i.e. the sum over the set of Mie resonances near the FW, replaced by a sum over the set of Mie resonances $\bm{\beta} = \{T',p',\ell',m'\}$ near the SH. The SH scattered fields can also interact with the NP ensemble which, in this case, acts as a set of symmetry-breaking resonant cavities that allow mixing between the otherwise orthogonal SH Mie resonances of the LiNbO$_3$ sphere.

\begin{figure*}[ht!] 
\centering
\includegraphics[width = 0.9\textwidth]{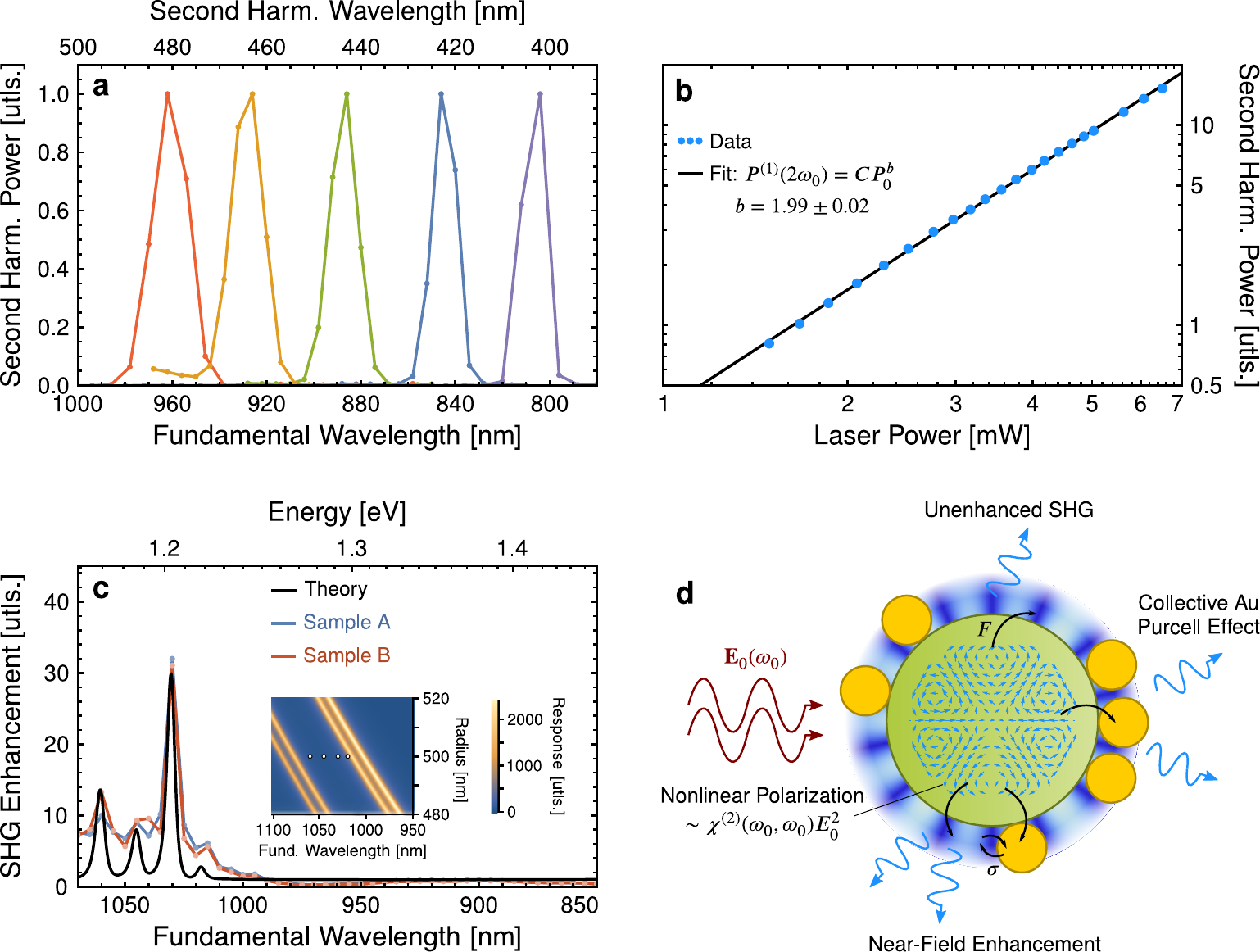}
\caption{\small
\textbf{Second harmonic generation enhancement.} \textbf{(a)} SHG spectra of the Au-LiNbO$_3$ hybrid particles as the FW is tuned from 800 nm (violet) to 960 nm (red). \textbf{(b)} Power dependence of SHG at 800 nm. The power scattered at the SH (blue dots) is proportional to the input laser power at the FW to the power $b = 1.99$ (95\% confidence interval of $\pm0.02$, see fit [black line]) times a normalization constant $C$. \textbf{(c)} Enhancement of the scattered SHG power as a function of the fundamental wavelength for an individual hybrid Au-LiNbO$_3$ nanostructure recorded by sweeping the excitation laser wavelength from 850 to 1 070 nm with a step size of 5 nm. The experimental results (blue, red) are overlaid by the modeled near-field coupling enhancements (black). The plotted curves are shifted phenomenologically by 10--25 nm from the solutions of Eq. \eqref{eq:EOMmain} to align with the experiment, as is highlighted in the Discussion. Inset: response functions (SI, Secion \ref{sec:oscModelConstr}) of the narrow Mie resonances of a bare LiNbO$_3$ sphere as a function of the sphere radius and photon energy. The response function maxima (gold/white) lie at the same resonance energies $\omega_{\bm{\beta}}$ as the SHG enhancement peaks, implying a red (blue) shift in the narrow SHG enhancement features with increasing (decreasing) $a_2$. The white dots indicate the shifted peak positions shown in the main panel. Further details are available in the SI. \textbf{(d)} A diagram of the pathways for energy transfer of light within the system. Clockwise from left: incident light (dark red), SHG (blue arrows) through energy transfer from nonlinear polarization (blue vector field) to Mie resonances (halo), superradiant Prucell-enhanced SHG, and near-field enhanced SHG.
\label{fig:fig4}
}
\end{figure*}

Whether excited by $\bm{\mathcal{E}}_0$ or $\bm{\mathcal{E}}_\mathrm{sca}$, the excited NPs can transfer energy back to the LiNbO$_3$ sphere through their near-fields. Thus, the two SH field-NP interactions provide additional pathways for energy to be transferred from the abundant incident power\cite{bohren2004absorption} $P_0 = ca_2^3E_0^2/8$ of the laser to the LiNbO$_3$ microsphere, which then scatters more SH light as a result. Figure \ref{fig:fig4}d provides a sketch of these processes. 

Quantification of the resulting SHG enhancements is dramatically simplified by the fact that each Mie resonance $\bm{\beta}$ is well-approximated as a discrete oscillator with resonance frequency $\omega_{\bm{\beta}}$ and damping rate $\gamma_{\bm{\beta}}$ (see SI, Eq. [\ref{eq:Xharmonics}] and Section \ref{sec:paramInf}). The plasmons in each NP, here assumed to have radii $a_1 = 5$ nm, have similarly simple oscillator properties (SI, Section \ref{sec:paramInf}), with resonance frequencies $\omega_1$ and damping rates $\gamma_1$. 

However, even with these conveniences, explicitly describing the numerous energy-transfer pathways between the $N$ NPs of the ensemble and the LiNbO$_3$ is difficult. Therefore, we first model the near-field energy exchange between the LiNbO$_3$ and Au by considering a reduced system with only a single Au NP coupled to the LiNbO$_3$ modes. Letting the NP dipole be oriented along the $\nu$ Cartesian axis with a moment $d_\nu(\omega)$, its motion along with the motion of each SH Mie moment $\rho_{\bm{\beta}}(\omega)$ can be quantified from coupled equations of motion (SI, Section \ref{sec:nearField}), here given in the Fourier domain:
\begin{widetext}
\begin{equation}\label{eq:EOMmain}
\begin{split}
d_\nu(\omega)\left(\Omega_1^2 - \omega^2 - i\omega\gamma_1\right) &= \frac{f_\nu a_1^3}{2\omega_1e^2}\eta_1(\omega)\left(eF_{1\nu}(\mathbf{r}_0,\omega) + \sum_{\bm{\beta}}\frac{1}{a_2^{\ell - 1}}\left[\sigma_{\bm{\beta}\nu}(\mathbf{r}_0)\rho_{\bm{\beta}}(\omega) + \sigma_{\bm{\beta}\nu}^*(\mathbf{r}_0)\rho_{\bm{\beta}}^*(-\omega)\right]\right),\\
\rho_{\bm{\beta}}(\omega)\left(\Omega_{\bm{\beta}}^2 - \omega^2 - i\omega\gamma_{\bm{\beta}}\right) &= \frac{f_{\bm{\beta}}a_2^{\ell + 2}}{2\omega_{\bm{\beta}}e^2}\eta_{\bm{\beta}}(\omega)\left(eF_{2\bm{\beta}}(\omega) + \sum_{\nu}\sigma_{\bm{\beta}\nu}(\mathbf{r}_0)d_\nu(\omega)\right).
\end{split}
\end{equation}
\end{widetext}

In addition to the previously defined constants, each of the moments has a natural frequency $\Omega_{1,\bm{\beta}} = \sqrt{\omega_{1,\bm{\beta}}^2 + \gamma_{1,\bm{\beta}}^2/4}$, an oscillator strength $f_{\nu,\bm{\beta}}$, and a charge $e$. On the right hand side of Eq. \eqref{eq:EOMmain}, the forces $F_{1\nu}(\mathbf{r}_0,\omega)$ and $F_{2\bm{\beta}}(\omega)$ describe the driving of the Au NP dipoles and Mie modes, respectively, by the $\bm{\chi}_2^{(2)}$-upconverted incident light with the NP centered at $\mathbf{r}_0$. The analytical form of these forces is complicated and is left for the SI (see Section \ref{sec:nearField}). The phases of the moments' responses to these forces are encoded by their characteristic phase delays $\eta_1(\omega) = 2\omega_1\cos\psi_1 + \gamma_1\sin\psi_1 + 2i\omega\sin\psi_1$ and $\eta_{\bm{\beta}}(\omega) = (\Omega_{\bm{\beta}}^* + \omega)\exp(-i\psi_{\bm{\beta}})$, respectively, wherein $\psi_{1,\bm{\beta}}$ are the characteristic phase lag parameters of the NP plasmons and Mie resonances, respectively. Numerical values for these oscillator parameters are given in Tables \ref{tab:tabS1} and \ref{tab:tabS2} and the particles' individual resonance profiles are plotted in Figures \ref{fig:figS12} and \ref{fig:figS13}.

In addition to the upconversion forces, the moments of the system are coupled with strengths $\sigma_{\bm{\beta}\nu}(\mathbf{r}_0) = (e^2/a_2^3)\hat{\mathbf{e}}_\nu\cdot\mathbf{X}_{\bm{\beta}}(\mathbf{r}_0,k_{\bm{\beta}})$, wherein $\hat{\mathbf{e}}_\nu$ is the $\nu$-oriented unit vector, $\mathbf{X}_{\bm{\beta}}(\mathbf{r},k_{\bm{\beta}})$ is the vector spherical harmonic that describes the spatial variations of the fields of the $\bm{\beta}^\mathrm{th}$ mode, and $k_{\bm{\beta}} = \omega_{\bm{\beta}}/c$. In the case where the couplings strengths are set to zero, the LiNbO$_3$ and Au NP radiate independently and no energy passes between them. As the magnitude of $\sigma_{\bm{\beta}\nu}(\mathbf{r}_0)$ increases, the likelihood that a photon will be exchanged between the LiNbO$_3$ and Au multiple times before being radiated away also increases. 

One can see from the scattering spectra of Figure \ref{fig:fig3}b that no mode splitting is generated by the addition of the Au NPs to the LiNbO$_3$ sphere in our system. Splitting is readily observed in strongly-coupled NLO polariton systems,\cite{chervy2016high-efficiency} such that weak interactions between the Au NPs and microsphere are expected. The theory is in good agreement with this observation, as the weak coupling condition\cite{kockum2019ultrastrong,liberko2021probing} $2a_2^3|\sigma_{\bm{\beta}\nu}(\mathbf{r}_0)|\sqrt{f_{\bm{\beta}}f_\nu}/e^2(\gamma_1 + \gamma_{\bm{\beta}})\sqrt{\omega_{\bm{\beta}}\omega_1} < 0.1$ is satisfied for all $\mathbf{r}_0$. The equations of motion of the LiNbO$_3$ Mie modes can, thus, be solved perturbatively to capture the effects of these exchanges on the power radiated from the microsphere. 

Explicitly, letting $\rho_{\bm{\beta}}(\omega) = \sum_{n=1}^\infty\rho_{\bm{\beta}}^{(n)}(\omega)$ where each order in the expansion includes the effects of $n$ exchanges of energy, the power scattered from the $\bm{\beta}^\mathrm{th}$ mode is $P_{\bm{\beta}}(2\omega_0) = (c^3/2\pi^2\omega_0^2\omega_{\bm{\beta}}a_2^{2\ell + 4}) \int_{-\pi/2\omega_0}^{\pi/2\omega_0} \mathrm{Re}\{\dot{\rho}_{\bm{\beta}}(t)\}^2\;\mathrm{d}t$ (SI, Section \ref{sec:scattering}). The power scattered by $\bm{\beta}$ in the absence of the NP dipoles, $P_{\bm{\beta}}^{(1)}(2\omega_0)$, can be calculated simply by replacing $\rho_{\bm{\beta}}(t)$ with $\rho_{\bm{\beta}}^{(1)}(t)$. 

Due to a convenient orthogonality condition (Eq. [\ref{eq:XcurlXint}]), the total power scattered by the microsphere at the SH frequency, $P(2\omega_0)$, is then simply a sum over the power scattered from each mode, such that $P(2\omega_0) = \sum_{\bm{\beta}}P_{\bm{\beta}}(2\omega_0)$, and the scattering enhancement can be simply defined as $P(2\omega_0)/P^{(1)}(2\omega_0)$, wherein $P^{(1)}(2\omega_0) = \sum_{\bm{\beta}}P_{\bm{\beta}}^{(1)}(2\omega_0)$ is the total SH power scattered from a bare LiNbO$_3$ sphere. In good agreement with the experiment (see Figure \ref{fig:figS10}), this ratio is very nearly $1$ for a model with a single NP due to the weak coupling between the microsphere and the dipole of the plasmons. However, when generalizing to a model with $N = 1 000$ NPs in the ensemble (SI, Section \ref{sec:methodsSupp}), the enhancement ratio reaches a peak value of $\sim$30 in the analyzed SH frequency range and reflects the underlying resonant structure of the Mie modes.

Figure \ref{fig:fig4}c shows these results in detail, demonstrating the excellent quantitative agreement between the SHG enhancement theory and the experimental data. Importantly, the model provides not only a numerical reproduction of the enhancement peaks but straightforwardly explains their origin. Beginning with the lowest-order nontrivial terms in the perturbation series of the Mie resonance $\bm{\beta}$, one can see that the second-order term $\rho_{\bm{\beta}}^{(2)}$ describes the transfer of energy from the upconversion-induced dipoles of the Au NP to a Mie resonance of the LiNbO$_3$ sphere, a process that is impossible without the presence of the NP ensemble and, as described above, represents a newly discovered pathway through which energy can be transferred from the energy bath of the laser to the LiNbO$_3$ particle and enhance the overall SHG signal. Accordingly, the third-order term describes the process in which energy is transferred from one Mie mode to another through two separate energy exchanges with the NP ensemble, a process which tends to shuttle energy from broad, more strongly-pumped Mie modes to narrower resonances more weakly excited by the upconversion process.

Higher-order processes can be derived from the analytical solution to $\rho_{\bm{\beta}}^{(n)}(\omega)$ but contribute increasingly small corrections to the total LiNbO$_3$ polarization, as is shown in the SI, Section \ref{sec:methodsSupp}. Further, after comparing the magnitudes of the forces of the system, one can conclude that the Au NP ensemble contributes to the enhancement of SHG primarily by mediating energy transfer from the broader, lower-order Mie resonances to their narrower counterparts. The second-order terms do add a small boost to SHG through the augmentations $\sigma_{\bm{\beta}\nu}(\mathbf{r}_0)F_{1\nu}(\mathbf{r}_0,\omega)$ to the driving forces of each mode (SI, Section \ref{sec:methodsSupp}), but while their total contribution is important for relatively broad modes $\bm{\beta}$, it is minor for narrow Mie resonances. Figure \ref{fig:figS14} shows this effect explicitly, with the total enhancement signal well-approximated by a second-order expansion in regions of small enhancement but requiring third-order corrections near the high-order mode resonance positions.

Our second (ii) hypothesis, in contrast to the coupling model developed above, is concerned with the radiation from the Au NPs themselves. Generally, Au NPs less than $\sim$30 nm in diameter can be assumed to be weak scatterers as their characteristic length scale lies below one tenth of their surface plasmon resonance wavelength, implying that our choice to this point to neglect light emission from the NP ensemble is appropriate.  However, small dipole oscillators such as our NPs are known to exhibit superradiance, an effect which is observed when coupling between resonant emitters causes each member of a large ensemble to emit more light than it would on its own. This phenomenon has been observed recently in collections of coupled dipole emitters in many prior investigations\cite{choudhary2019weak,masson2022universality} and thus is of interest here. 

To evaluate whether the radiation from each NP in the experimental ensemble is boosted sufficiently by inter-NP interactions to generate overall SHG enhancement from the collections of NPs in our experiments, we combine per-particle optical cross-sections collected from multiple-Mie scattering simulations\cite{pellegrini2007interacting} with analytical models to calculate the time-averaged power radiated at the SH from the ``typical'' NP in square ensembles of sizes $N = 1$ through 225. We label this power as $P_\mathrm{pl}(2\omega_0)$, such that the power radiated from the ensemble is $NP_\mathrm{pl}(2\omega_0)$. With the details left for the SI, Section \ref{sec:scattering}, the ratio of the powers scattered from the ensemble and the bare LiNbO$_3$ particle is then given by:
\begin{equation}
\frac{NP_\mathrm{pl}(2\omega_0)}{P^{(1)}(2\omega_0)} = \frac{2\pi\omega_0NA(N)\gamma_1^\mathrm{rad}f_1f_2a_1^3}{c^3\gamma_1^2\gamma_2}\kappa(2\omega_0).
\end{equation}
Here, $\kappa(2\omega_0)$ is a spectral profile (Eq. [\ref{eq:kappa}]) that reflects the resonance structure of the LiNbO$_3$ modes and has a small maximum value of $\sim$$10^{-3}$. Further, $A(N)$ is a phenomenological scattering enhancement factor of the Au NPs extracted from the data in Figure \ref{fig:figS15}a that builds in superradiance effects and corresponds to a value $\sim$300 for $N \sim1 000$ that is not greatly influenced by ensemble disorder (see Figure \ref{fig:figS15}b). 

Combined with the radiative contribution to the damping rate of the Au NPs $\gamma_1^\mathrm{rad}\sim10^{-4}\gamma_1$, the mean Mie oscillator strength $f_2$ and damping rate $\gamma_2$, and the rest of the previously defined constants, the superradiant enhancement factor does not overcome the small value of $a_1^3\kappa(2\omega_0)$ to provide an overall scattering boost from the NP ensemble. Explicitly, $NP_\mathrm{pl}(2\omega_0)/P^{(1)}(2\omega_0)\ll1$ for all relevant wavelengths, varying between $\sim$$2\times10^{-2}$ and $3\times10^{-2}$ for $2\omega_0$ in the visible spectrum. Superradiance, therefore, cannot account for the observed enhancements to SHG.

\section{Discussion}

The invention of efficient miniaturized NLO devices remains an open and important challenge in the field of nanophotonics. Moreover, with current strategies for the maximization of NLO signal strength at the nanoscale limited by the difficulties and expense of precisely nanoengineering single NLO nano- and microstructures, multiply-resonant nanoantenna systems, and/or individual electric field hotspots, new hybrid nanomaterials that can provide meaningful NLO enhancements without unduly complicated fabrication processes are highly desirable. To this end, we have designed a new class of disordered Au-LiNbO$_3$ NLO nanostructures that are straightforward to synthesize and demonstrate state-of-the-art enhancements. These hybrid structures were prepared through a high-throughput solution-phase self-assembly process through which the mesoporous structures of micrometer-scale LiNbO$_3$ spheres were coated with a disperse layer of $\sim$10-nm diameter Au NPs with a highly tunable density ($\sim$6--1 000 Au NPs/LiNbO$_3$ sphere) and an even distribution.

Single nanostructures were then isolated, characterized, and modeled in order to interrogate the mechanisms governing the enhancement of their SHG. Broadband linear and nonlinear optical scattering and extinction measurements of these hybrid structures demonstrate Mie-like resonant behaviors within the LiNbO$_3$ and minimal rearrangement of the LiNbO$_3$ resonance structure by the Au NPs. The Mie-like behaviors are consistent with prior investigations of multicrystalline NLO microspheres.\cite{savo2020broadband} Further, analytical and numerical models predict that the coupling of each deeply subwavelength, nonradiative NP with its neighboring NPs and the microsphere is weak, such that modifications of the microsphere's spectrum should, in agreement with observations, be minimal. These models also reveal that SHG enhancement occurs within the hyrbid structures through a previously unexplored near-field coupling mechanism, available only at the nanoscale, that enhances the conversion of energy from incoming light into NLO material resonances and then confines it within the nanostructures' narrow resonances. 

As a result, the hybrid materials described herein broaden the optical nanomaterial design space by highlighting the NLO enhancement capabilities of subwavelength or otherwise nonradiative nanoantenna systems. For example, the observable SH resonances of the LiNbO$_3$ are long-lived with $\gamma_{\bm{\beta}}/\omega_{\bm{\beta}}\sim10^{-3}$--$10^{-2}$ such that they are candidates for use in sensing and light emission applications where measurable impurity-induced peak shifts and/or tunable monochromatic emission at the nanoscale are desired. We have shown that the SH excitation of such narrow Mie resonances is preferentially and significantly enhanced by weak coupling to a relatively low and disordered volume of a second resonant material (Au) that is resonant at the output SH frequency of the LiNbO$_3$ sphere. Therefore, new design strategies for sensors and emitters can specifically incorporate imprecise ensembles of very small nanoantennas as needed to optimize the NLO signal intensity.

In addition, the sensitivity of the SH Mie resonances' output frequencies to intrinsic and environmental defects is not degraded by the presence of the Au NPs. As evidence, the strong dependence of the microsphere's Mie resonance spectral positions, widths, and heights on its geometry and dielectric function is preserved in hybrid structures, as is shown in Figure \ref{fig:fig3}b. This preservation is supported by the excellent agreement with the model, which specifies that the weak coupling between the Mie resonances and plasmon dipoles precludes rearrangement or obfuscation of the LiNbO$_3$ Mie spectrum.

Determination of the exact value of the dielectric function $\epsilon_2$ of the structures analyzed in the experiment is difficult, as the mesoporous material is novel, single-particle ellipsometry through e.g. electron beam spectroscopy\cite{olafsson2020electron} is in its infancy, and substrate effects can cause nontrivial reorganization of the sphere's Mie spectrum (SI, Section \ref{sec:paramInf}). However, as is detailed in the SI, Section \ref{sec:materialModels}, the estimate $\epsilon_2 = 5.5 + 0.035i$ used alongside the chosen value $a_2 = 500$ nm as well as the HAADF images and linear and nonlinear scattering spectra of Figures \ref{fig:fig1}, \ref{fig:fig3}, and \ref{fig:fig4} produces uniquely precise agreement between theory, experiment, and prior art\cite{palik1998handbook} in the SH spectral region. Variation of either $a_2$ or $\epsilon_2$ by $\sim$5\% can generate noticeable $\sim$15--35 nm shifts in the Mie resonance positions of both bare LiNbO$_3$ and hybrid structures (see Figure \ref{fig:fig4}d and \ref{fig:figS14}c), such that we take $\pm5\%$ to be a rough estimate of the uncertainty in the microsphere parameters. The small ($\sim$25 nm) phenomenological shifts made to align the theory with the experimental data of Figure \ref{fig:fig4}c fall well within these bounds.

By contrast, the Mie resonances are \textit{not} particularly sensitive to the parameters of the NP plasmon resonances. The NPs' extinction profiles simply provide a spectral envelope inside which SHG enhancements are possible. Finally, the microsphere's surface coverage by the NP ensemble is limited, allowing for ready interaction between external NPs, cavities, waveguides, molecules, etc. and the LiNbO$_3$.

Future advancements in the fabrication and characterization processes of our (or similar) nanostructures that provide greater control over simple parameters like the microsphere's shape and internal nanocrystal density will allow researchers to more precisely tune these systems. Further, the simplicity of the coupling phenomena governing the SHG output intensity enhancements suggests that nanostructures displaying NLO behaviors beyond SHG should be able to take advantage of the same effects. In combination with the flexible perturbative model developed here, our results suggest that nanoscale NLO enhancements can be straightforwardly modeled, designed, and realized in a wide array of near-field coupled nanoscopic systems displaying all manner of frequency conversion behaviors without the need for expensive nanoengineering.

\section{Materials and methods}

Further details regarding the synthesis, characterization, and modeling of the nanostructures used in this work can be found in the SI.\\
\\
\noindent\textbf{Data and materials availability:} All data needed to evaluate the conclusions in the paper are present in the paper and/or the Supplementary Information.

\begin{acknowledgments}
Work at Simon Fraser University was supported by the Natural Sciences and Engineering Research Council of Canada (Discovery Grant No. RGPIN-2020-06522) and CMC Microsystems (MNT Grant No. 6345) (R.F.A., S.K., B.D.G.). This work made use of 4D LABS, LASIR, and the Center for Soft Materials shared facilities supported by the Canada Foundation for Innovation (CFI), British Columbia Knowledge Development Fund (BCKDF), Western Economic Diversification Canada, and Simon Fraser University. Work at the University of Washington was supported by the U.S. National Science Foundation under Award No. CHE-1954393 (J.A.B., D.J.M.).\\
\\
\noindent\textbf{Author contributions:} R.F.A. and B.D.G. designed the experimental study; R.F.A. prepared and characterized the nanostructures; R.F.A., S.K. and B.D.G. contributed to the SHG and scattering measurements; J.A.B. and D.J.M. designed the theoretical study; J.A.B. conducted the theoretical calculations; All authors analyzed the experimental data and results of the theoretical study, and contributed to the preparation of the manuscript.
\end{acknowledgments}

\noindent\textbf{Conflict of interest:} The authors declare no conflict of interest.



%



\begin{thebibliography}{50}%
\makeatletter
\providecommand \@ifxundefined [1]{%
 \@ifx{#1\undefined}
}%
\providecommand \@ifnum [1]{%
 \ifnum #1\expandafter \@firstoftwo
 \else \expandafter \@secondoftwo
 \fi
}%
\providecommand \@ifx [1]{%
 \ifx #1\expandafter \@firstoftwo
 \else \expandafter \@secondoftwo
 \fi
}%
\providecommand \natexlab [1]{#1}%
\providecommand \enquote  [1]{``#1''}%
\providecommand \bibnamefont  [1]{#1}%
\providecommand \bibfnamefont [1]{#1}%
\providecommand \citenamefont [1]{#1}%
\providecommand \href@noop [0]{\@secondoftwo}%
\providecommand \href [0]{\begingroup \@sanitize@url \@href}%
\providecommand \@href[1]{\@@startlink{#1}\@@href}%
\providecommand \@@href[1]{\endgroup#1\@@endlink}%
\providecommand \@sanitize@url [0]{\catcode `\\12\catcode `\$12\catcode
  `\&12\catcode `\#12\catcode `\^12\catcode `\_12\catcode `\%12\relax}%
\providecommand \@@startlink[1]{}%
\providecommand \@@endlink[0]{}%
\providecommand \url  [0]{\begingroup\@sanitize@url \@url }%
\providecommand \@url [1]{\endgroup\@href {#1}{\urlprefix }}%
\providecommand \urlprefix  [0]{URL }%
\providecommand \Eprint [0]{\href }%
\providecommand \doibase [0]{https://doi.org/}%
\providecommand \selectlanguage [0]{\@gobble}%
\providecommand \bibinfo  [0]{\@secondoftwo}%
\providecommand \bibfield  [0]{\@secondoftwo}%
\providecommand \translation [1]{[#1]}%
\providecommand \BibitemOpen [0]{}%
\providecommand \bibitemStop [0]{}%
\providecommand \bibitemNoStop [0]{.\EOS\space}%
\providecommand \EOS [0]{\spacefactor3000\relax}%
\providecommand \BibitemShut  [1]{\csname bibitem#1\endcsname}%
\let\auto@bib@innerbib\@empty
\bibitem [{\citenamefont {Firstenberg}\ \emph {et~al.}(2013)\citenamefont
  {Firstenberg}, \citenamefont {Peyronel}, \citenamefont {Liang}, \citenamefont
  {Gorshkov}, \citenamefont {Lukin},\ and\ \citenamefont
  {Vuletić}}]{firstenberg2013attractive}%
  \BibitemOpen
  \bibfield  {author} {\bibinfo {author} {\bibfnamefont {O.}~\bibnamefont
  {Firstenberg}}, \bibinfo {author} {\bibfnamefont {T.}~\bibnamefont
  {Peyronel}}, \bibinfo {author} {\bibfnamefont {Q.-Y.}\ \bibnamefont {Liang}},
  \bibinfo {author} {\bibfnamefont {A.~V.}\ \bibnamefont {Gorshkov}}, \bibinfo
  {author} {\bibfnamefont {M.~D.}\ \bibnamefont {Lukin}},\ and\ \bibinfo
  {author} {\bibfnamefont {V.}~\bibnamefont {Vuletić}},\ }\bibfield  {title}
  {\bibinfo {title} {Attractive photons in a quantum nonlinear medium},\
  }\href@noop {} {\bibfield  {journal} {\bibinfo  {journal} {Nature}\ }\textbf
  {\bibinfo {volume} {502}},\ \bibinfo {pages} {71} (\bibinfo {year}
  {2013})}\BibitemShut {NoStop}%
\bibitem [{\citenamefont {Garmire}(2013)}]{garmire2013nonlinear}%
  \BibitemOpen
  \bibfield  {author} {\bibinfo {author} {\bibfnamefont {E.}~\bibnamefont
  {Garmire}},\ }\bibfield  {title} {\bibinfo {title} {Nonlinear optics in daily
  life},\ }\href@noop {} {\bibfield  {journal} {\bibinfo  {journal} {Optics
  Express}\ }\textbf {\bibinfo {volume} {21}},\ \bibinfo {pages} {30532}
  (\bibinfo {year} {2013})}\BibitemShut {NoStop}%
\bibitem [{\citenamefont {Ali}\ \emph {et~al.}(2019)\citenamefont {Ali},
  \citenamefont {Bilton},\ and\ \citenamefont {Gates}}]{ali2019one-pot}%
  \BibitemOpen
  \bibfield  {author} {\bibinfo {author} {\bibfnamefont {R.~F.}\ \bibnamefont
  {Ali}}, \bibinfo {author} {\bibfnamefont {M.}~\bibnamefont {Bilton}},\ and\
  \bibinfo {author} {\bibfnamefont {B.~D.}\ \bibnamefont {Gates}},\ }\bibfield
  {title} {\bibinfo {title} {One-pot synthesis of sub-10 nm {LiNbO$_3$}
  nanocrystals exhibiting a tunable optical second harmonic response},\
  }\href@noop {} {\bibfield  {journal} {\bibinfo  {journal} {Nanoscale Adv.}\
  }\textbf {\bibinfo {volume} {1}},\ \bibinfo {pages} {2268} (\bibinfo {year}
  {2019})}\BibitemShut {NoStop}%
\bibitem [{\citenamefont {Ali}\ and\ \citenamefont
  {Gates}(2022)}]{ali2022lithium}%
  \BibitemOpen
  \bibfield  {author} {\bibinfo {author} {\bibfnamefont {R.~F.}\ \bibnamefont
  {Ali}}\ and\ \bibinfo {author} {\bibfnamefont {B.~D.}\ \bibnamefont
  {Gates}},\ }\bibfield  {title} {\bibinfo {title} {Lithium niobate particles
  with a tunable diameter and porosity for optical second harmonic
  generation},\ }\href@noop {} {\bibfield  {journal} {\bibinfo  {journal} {RSC
  Adv.}\ }\textbf {\bibinfo {volume} {12}},\ \bibinfo {pages} {822} (\bibinfo
  {year} {2022})}\BibitemShut {NoStop}%
\bibitem [{\citenamefont {Walther}\ \emph {et~al.}(2018)\citenamefont
  {Walther}, \citenamefont {Johne},\ and\ \citenamefont
  {Pohl}}]{walther2018giant}%
  \BibitemOpen
  \bibfield  {author} {\bibinfo {author} {\bibfnamefont {V.}~\bibnamefont
  {Walther}}, \bibinfo {author} {\bibfnamefont {R.}~\bibnamefont {Johne}},\
  and\ \bibinfo {author} {\bibfnamefont {T.}~\bibnamefont {Pohl}},\ }\bibfield
  {title} {\bibinfo {title} {Giant optical nonlinearities from {Rydberg}
  excitons in semiconductor microcavities},\ }\href@noop {} {\bibfield
  {journal} {\bibinfo  {journal} {Nat. Commun.}\ }\textbf {\bibinfo {volume}
  {9}},\ \bibinfo {pages} {1309} (\bibinfo {year} {2018})}\BibitemShut
  {NoStop}%
\bibitem [{\citenamefont {Gao}\ \emph {et~al.}(2022)\citenamefont {Gao},
  \citenamefont {Han}, \citenamefont {Wang}, \citenamefont {Sun}, \citenamefont
  {Lu}, \citenamefont {Han}, \citenamefont {Yan}, \citenamefont {Liu},\ and\
  \citenamefont {Dong}}]{gao2022single-layer}%
  \BibitemOpen
  \bibfield  {author} {\bibinfo {author} {\bibfnamefont {W.}~\bibnamefont
  {Gao}}, \bibinfo {author} {\bibfnamefont {S.}~\bibnamefont {Han}}, \bibinfo
  {author} {\bibfnamefont {B.}~\bibnamefont {Wang}}, \bibinfo {author}
  {\bibfnamefont {Z.}~\bibnamefont {Sun}}, \bibinfo {author} {\bibfnamefont
  {Y.}~\bibnamefont {Lu}}, \bibinfo {author} {\bibfnamefont {Q.}~\bibnamefont
  {Han}}, \bibinfo {author} {\bibfnamefont {X.}~\bibnamefont {Yan}}, \bibinfo
  {author} {\bibfnamefont {J.}~\bibnamefont {Liu}},\ and\ \bibinfo {author}
  {\bibfnamefont {J.}~\bibnamefont {Dong}},\ }\bibfield  {title} {\bibinfo
  {title} {Single-layer gold nanoparticle film enhances the upconversion
  luminescence of a single {NaYbF$_4$}: 2\%{Er$^{3+}$} microdisk},\ }\href@noop
  {} {\bibfield  {journal} {\bibinfo  {journal} {J. Alloys Compd.}\ }\textbf
  {\bibinfo {volume} {900}},\ \bibinfo {pages} {163493} (\bibinfo {year}
  {2022})}\BibitemShut {NoStop}%
\bibitem [{\citenamefont {Kauranen}(2013)}]{kauranen2013freeing}%
  \BibitemOpen
  \bibfield  {author} {\bibinfo {author} {\bibfnamefont {M.}~\bibnamefont
  {Kauranen}},\ }\bibfield  {title} {\bibinfo {title} {Freeing nonlinear optics
  from phase matching},\ }\href@noop {} {\bibfield  {journal} {\bibinfo
  {journal} {Science}\ }\textbf {\bibinfo {volume} {342}},\ \bibinfo {pages}
  {1182} (\bibinfo {year} {2013})}\BibitemShut {NoStop}%
\bibitem [{\citenamefont {Zhang}\ \emph {et~al.}(2017)\citenamefont {Zhang},
  \citenamefont {Yu}, \citenamefont {Wu},\ and\ \citenamefont
  {Halasyamani}}]{zhang2017phase-matching}%
  \BibitemOpen
  \bibfield  {author} {\bibinfo {author} {\bibfnamefont {W.}~\bibnamefont
  {Zhang}}, \bibinfo {author} {\bibfnamefont {H.}~\bibnamefont {Yu}}, \bibinfo
  {author} {\bibfnamefont {H.}~\bibnamefont {Wu}},\ and\ \bibinfo {author}
  {\bibfnamefont {P.~S.}\ \bibnamefont {Halasyamani}},\ }\bibfield  {title}
  {\bibinfo {title} {Phase-matching in nonlinear optical compounds: A materials
  perspective},\ }\href@noop {} {\bibfield  {journal} {\bibinfo  {journal}
  {Chem. Mater.}\ }\textbf {\bibinfo {volume} {29}},\ \bibinfo {pages} {2655}
  (\bibinfo {year} {2017})}\BibitemShut {NoStop}%
\bibitem [{\citenamefont {Butet}\ \emph {et~al.}(2012)\citenamefont {Butet},
  \citenamefont {Russier-Antoine}, \citenamefont {Jonin}, \citenamefont
  {Lascoux}, \citenamefont {Benichou},\ and\ \citenamefont
  {Brevet}}]{butet2012sensing}%
  \BibitemOpen
  \bibfield  {author} {\bibinfo {author} {\bibfnamefont {J.}~\bibnamefont
  {Butet}}, \bibinfo {author} {\bibfnamefont {I.}~\bibnamefont
  {Russier-Antoine}}, \bibinfo {author} {\bibfnamefont {C.}~\bibnamefont
  {Jonin}}, \bibinfo {author} {\bibfnamefont {N.}~\bibnamefont {Lascoux}},
  \bibinfo {author} {\bibfnamefont {E.}~\bibnamefont {Benichou}},\ and\
  \bibinfo {author} {\bibfnamefont {P.-F.}\ \bibnamefont {Brevet}},\ }\bibfield
   {title} {\bibinfo {title} {Sensing with {Multipolar} {Second} {Harmonic}
  {Generation} from {Spherical} {Metallic} {Nanoparticles}},\ }\href@noop {}
  {\bibfield  {journal} {\bibinfo  {journal} {Nano Lett.}\ }\textbf {\bibinfo
  {volume} {12}},\ \bibinfo {pages} {1697} (\bibinfo {year}
  {2012})}\BibitemShut {NoStop}%
\bibitem [{\citenamefont {Malkinson}\ \emph {et~al.}(2020)\citenamefont
  {Malkinson}, \citenamefont {Mahou}, \citenamefont {Chaudan}, \citenamefont
  {Gacoin}, \citenamefont {Sonay}, \citenamefont {Pantazis}, \citenamefont
  {Beaurepaire},\ and\ \citenamefont {Supatto}}]{malkinson2020fast}%
  \BibitemOpen
  \bibfield  {author} {\bibinfo {author} {\bibfnamefont {G.}~\bibnamefont
  {Malkinson}}, \bibinfo {author} {\bibfnamefont {P.}~\bibnamefont {Mahou}},
  \bibinfo {author} {\bibfnamefont {E.}~\bibnamefont {Chaudan}}, \bibinfo
  {author} {\bibfnamefont {T.}~\bibnamefont {Gacoin}}, \bibinfo {author}
  {\bibfnamefont {A.~Y.}\ \bibnamefont {Sonay}}, \bibinfo {author}
  {\bibfnamefont {P.}~\bibnamefont {Pantazis}}, \bibinfo {author}
  {\bibfnamefont {E.}~\bibnamefont {Beaurepaire}},\ and\ \bibinfo {author}
  {\bibfnamefont {W.}~\bibnamefont {Supatto}},\ }\bibfield  {title} {\bibinfo
  {title} {Fast {\textit{in}} {\textit{vivo}} {Imaging} of {SHG} {Nanoprobes}
  with {Multiphoton} {Light}-{Sheet} {Microscopy}},\ }\href@noop {} {\bibfield
  {journal} {\bibinfo  {journal} {ACS Photonics}\ }\textbf {\bibinfo {volume}
  {7}},\ \bibinfo {pages} {1036} (\bibinfo {year} {2020})}\BibitemShut
  {NoStop}%
\bibitem [{\citenamefont {Tanaka}\ \emph {et~al.}(2021)\citenamefont {Tanaka},
  \citenamefont {Kimura},\ and\ \citenamefont
  {Shimura}}]{tanaka2021unidirectional}%
  \BibitemOpen
  \bibfield  {author} {\bibinfo {author} {\bibfnamefont {Y.~Y.}\ \bibnamefont
  {Tanaka}}, \bibinfo {author} {\bibfnamefont {T.}~\bibnamefont {Kimura}},\
  and\ \bibinfo {author} {\bibfnamefont {T.}~\bibnamefont {Shimura}},\
  }\bibfield  {title} {\bibinfo {title} {Unidirectional emission of
  phase-controlled second harmonic generation from a plasmonic nanoantenna},\
  }\href@noop {} {\bibfield  {journal} {\bibinfo  {journal} {Nanophotonics}\
  }\textbf {\bibinfo {volume} {10}},\ \bibinfo {pages} {4601} (\bibinfo {year}
  {2021})}\BibitemShut {NoStop}%
\bibitem [{\citenamefont {Fratalocchi}\ \emph {et~al.}(2015)\citenamefont
  {Fratalocchi}, \citenamefont {Dodson}, \citenamefont {Zia}, \citenamefont
  {Genevet}, \citenamefont {Verhagen}, \citenamefont {Altug},\ and\
  \citenamefont {Sorger}}]{fratalocchi2015nano}%
  \BibitemOpen
  \bibfield  {author} {\bibinfo {author} {\bibfnamefont {A.}~\bibnamefont
  {Fratalocchi}}, \bibinfo {author} {\bibfnamefont {C.~M.}\ \bibnamefont
  {Dodson}}, \bibinfo {author} {\bibfnamefont {R.}~\bibnamefont {Zia}},
  \bibinfo {author} {\bibfnamefont {P.}~\bibnamefont {Genevet}}, \bibinfo
  {author} {\bibfnamefont {E.}~\bibnamefont {Verhagen}}, \bibinfo {author}
  {\bibfnamefont {H.}~\bibnamefont {Altug}},\ and\ \bibinfo {author}
  {\bibfnamefont {V.~J.}\ \bibnamefont {Sorger}},\ }\bibfield  {title}
  {\bibinfo {title} {Nano-optics gets practical},\ }\href@noop {} {\bibfield
  {journal} {\bibinfo  {journal} {Nat. Nanotechnol.}\ }\textbf {\bibinfo
  {volume} {10}},\ \bibinfo {pages} {11} (\bibinfo {year} {2015})}\BibitemShut
  {NoStop}%
\bibitem [{\citenamefont {Savo}\ \emph {et~al.}(2020)\citenamefont {Savo},
  \citenamefont {Morandi}, \citenamefont {M\"{u}ller}, \citenamefont
  {Kaufmann}, \citenamefont {Timpu}, \citenamefont {Reig~Escal\'{e}},
  \citenamefont {Zanini}, \citenamefont {Isa},\ and\ \citenamefont
  {Grange}}]{savo2020broadband}%
  \BibitemOpen
  \bibfield  {author} {\bibinfo {author} {\bibfnamefont {R.}~\bibnamefont
  {Savo}}, \bibinfo {author} {\bibfnamefont {A.}~\bibnamefont {Morandi}},
  \bibinfo {author} {\bibfnamefont {J.~S.}\ \bibnamefont {M\"{u}ller}},
  \bibinfo {author} {\bibfnamefont {F.}~\bibnamefont {Kaufmann}}, \bibinfo
  {author} {\bibfnamefont {F.}~\bibnamefont {Timpu}}, \bibinfo {author}
  {\bibfnamefont {M.}~\bibnamefont {Reig~Escal\'{e}}}, \bibinfo {author}
  {\bibfnamefont {M.}~\bibnamefont {Zanini}}, \bibinfo {author} {\bibfnamefont
  {L.}~\bibnamefont {Isa}},\ and\ \bibinfo {author} {\bibfnamefont
  {R.}~\bibnamefont {Grange}},\ }\bibfield  {title} {\bibinfo {title}
  {Broadband {Mie} driven random quasi-phase-matching},\ }\href@noop {}
  {\bibfield  {journal} {\bibinfo  {journal} {Nat. Photonics}\ }\textbf
  {\bibinfo {volume} {14}},\ \bibinfo {pages} {740} (\bibinfo {year}
  {2020})}\BibitemShut {NoStop}%
\bibitem [{\citenamefont {Wang}\ \emph {et~al.}(2021)\citenamefont {Wang},
  \citenamefont {Fang}, \citenamefont {Yi}, \citenamefont {Yang}, \citenamefont
  {Wang}, \citenamefont {Zhou}, \citenamefont {Shen}, \citenamefont {Zhu},
  \citenamefont {Zhou}, \citenamefont {Bao}, \citenamefont {Li}, \citenamefont
  {Chen}, \citenamefont {Huang}, \citenamefont {Zhang}, \citenamefont {Cheng},\
  and\ \citenamefont {Ou}}]{wang2021high}%
  \BibitemOpen
  \bibfield  {author} {\bibinfo {author} {\bibfnamefont {C.}~\bibnamefont
  {Wang}}, \bibinfo {author} {\bibfnamefont {Z.}~\bibnamefont {Fang}}, \bibinfo
  {author} {\bibfnamefont {A.}~\bibnamefont {Yi}}, \bibinfo {author}
  {\bibfnamefont {B.}~\bibnamefont {Yang}}, \bibinfo {author} {\bibfnamefont
  {Z.}~\bibnamefont {Wang}}, \bibinfo {author} {\bibfnamefont {L.}~\bibnamefont
  {Zhou}}, \bibinfo {author} {\bibfnamefont {C.}~\bibnamefont {Shen}}, \bibinfo
  {author} {\bibfnamefont {Y.}~\bibnamefont {Zhu}}, \bibinfo {author}
  {\bibfnamefont {Y.}~\bibnamefont {Zhou}}, \bibinfo {author} {\bibfnamefont
  {R.}~\bibnamefont {Bao}}, \bibinfo {author} {\bibfnamefont {Z.}~\bibnamefont
  {Li}}, \bibinfo {author} {\bibfnamefont {Y.}~\bibnamefont {Chen}}, \bibinfo
  {author} {\bibfnamefont {K.}~\bibnamefont {Huang}}, \bibinfo {author}
  {\bibfnamefont {J.}~\bibnamefont {Zhang}}, \bibinfo {author} {\bibfnamefont
  {Y.}~\bibnamefont {Cheng}},\ and\ \bibinfo {author} {\bibfnamefont
  {X.}~\bibnamefont {Ou}},\ }\bibfield  {title} {\bibinfo {title} {High-q
  microresonators on 4h-silicon-carbide-on-insulator platform for nonlinear
  photonics},\ }\href@noop {} {\bibfield  {journal} {\bibinfo  {journal} {Light
  Sci. Appl.}\ }\textbf {\bibinfo {volume} {10}},\ \bibinfo {pages} {1}
  (\bibinfo {year} {2021})}\BibitemShut {NoStop}%
\bibitem [{\citenamefont {Li}\ \emph {et~al.}(2021)\citenamefont {Li},
  \citenamefont {Lei}, \citenamefont {Qiu}, \citenamefont {Jin}, \citenamefont
  {Lan},\ and\ \citenamefont {Zayats}}]{li2021light-induced}%
  \BibitemOpen
  \bibfield  {author} {\bibinfo {author} {\bibfnamefont {G.-C.}\ \bibnamefont
  {Li}}, \bibinfo {author} {\bibfnamefont {D.}~\bibnamefont {Lei}}, \bibinfo
  {author} {\bibfnamefont {M.}~\bibnamefont {Qiu}}, \bibinfo {author}
  {\bibfnamefont {W.}~\bibnamefont {Jin}}, \bibinfo {author} {\bibfnamefont
  {S.}~\bibnamefont {Lan}},\ and\ \bibinfo {author} {\bibfnamefont {A.~V.}\
  \bibnamefont {Zayats}},\ }\bibfield  {title} {\bibinfo {title} {Light-induced
  symmetry breaking for enhancing second-harmonic generation from an ultrathin
  plasmonic nanocavity},\ }\href@noop {} {\bibfield  {journal} {\bibinfo
  {journal} {Nat. Commun.}\ }\textbf {\bibinfo {volume} {12}},\ \bibinfo
  {pages} {4326} (\bibinfo {year} {2021})}\BibitemShut {NoStop}%
\bibitem [{\citenamefont {Pu}\ \emph {et~al.}(2010)\citenamefont {Pu},
  \citenamefont {Grange}, \citenamefont {Hsieh},\ and\ \citenamefont
  {Psaltis}}]{pu2010nonlinear}%
  \BibitemOpen
  \bibfield  {author} {\bibinfo {author} {\bibfnamefont {Y.}~\bibnamefont
  {Pu}}, \bibinfo {author} {\bibfnamefont {R.}~\bibnamefont {Grange}}, \bibinfo
  {author} {\bibfnamefont {C.-L.}\ \bibnamefont {Hsieh}},\ and\ \bibinfo
  {author} {\bibfnamefont {D.}~\bibnamefont {Psaltis}},\ }\bibfield  {title}
  {\bibinfo {title} {Nonlinear {Optical} {Properties} of {Core}-{Shell}
  {Nanocavities} for {Enhanced} {Second}-{Harmonic} {Generation}},\ }\href@noop
  {} {\bibfield  {journal} {\bibinfo  {journal} {Phys. Rev. Lett.}\ }\textbf
  {\bibinfo {volume} {104}},\ \bibinfo {pages} {207402} (\bibinfo {year}
  {2010})}\BibitemShut {NoStop}%
\bibitem [{\citenamefont {Suchowski}\ \emph {et~al.}(2013)\citenamefont
  {Suchowski}, \citenamefont {O’Brien}, \citenamefont {Wong}, \citenamefont
  {Salandrino}, \citenamefont {Yin},\ and\ \citenamefont
  {Zhang}}]{suchowski2013phase}%
  \BibitemOpen
  \bibfield  {author} {\bibinfo {author} {\bibfnamefont {H.}~\bibnamefont
  {Suchowski}}, \bibinfo {author} {\bibfnamefont {K.}~\bibnamefont
  {O’Brien}}, \bibinfo {author} {\bibfnamefont {Z.~J.}\ \bibnamefont {Wong}},
  \bibinfo {author} {\bibfnamefont {A.}~\bibnamefont {Salandrino}}, \bibinfo
  {author} {\bibfnamefont {X.}~\bibnamefont {Yin}},\ and\ \bibinfo {author}
  {\bibfnamefont {X.}~\bibnamefont {Zhang}},\ }\bibfield  {title} {\bibinfo
  {title} {Phase mismatch–free nonlinear propagation in optical zero-index
  materials},\ }\href@noop {} {\bibfield  {journal} {\bibinfo  {journal}
  {Science}\ }\textbf {\bibinfo {volume} {342}},\ \bibinfo {pages} {1223}
  (\bibinfo {year} {2013})}\BibitemShut {NoStop}%
\bibitem [{\citenamefont {Lehr}\ \emph {et~al.}(2015)\citenamefont {Lehr},
  \citenamefont {Reinhold}, \citenamefont {Thiele}, \citenamefont {Hartung},
  \citenamefont {Dietrich}, \citenamefont {Menzel}, \citenamefont {Pertsch},
  \citenamefont {Kley},\ and\ \citenamefont
  {T\"{u}nnermann}}]{lehr2015enhancing}%
  \BibitemOpen
  \bibfield  {author} {\bibinfo {author} {\bibfnamefont {D.}~\bibnamefont
  {Lehr}}, \bibinfo {author} {\bibfnamefont {J.}~\bibnamefont {Reinhold}},
  \bibinfo {author} {\bibfnamefont {I.}~\bibnamefont {Thiele}}, \bibinfo
  {author} {\bibfnamefont {H.}~\bibnamefont {Hartung}}, \bibinfo {author}
  {\bibfnamefont {K.}~\bibnamefont {Dietrich}}, \bibinfo {author}
  {\bibfnamefont {C.}~\bibnamefont {Menzel}}, \bibinfo {author} {\bibfnamefont
  {T.}~\bibnamefont {Pertsch}}, \bibinfo {author} {\bibfnamefont {E.-B.}\
  \bibnamefont {Kley}},\ and\ \bibinfo {author} {\bibfnamefont
  {A.}~\bibnamefont {T\"{u}nnermann}},\ }\bibfield  {title} {\bibinfo {title}
  {Enhancing second harmonic generation in gold nanoring resonators filled with
  lithium niobate},\ }\href@noop {} {\bibfield  {journal} {\bibinfo  {journal}
  {Nano Lett.}\ }\textbf {\bibinfo {volume} {15}},\ \bibinfo {pages} {1025}
  (\bibinfo {year} {2015})}\BibitemShut {NoStop}%
\bibitem [{\citenamefont {Timpu}\ \emph {et~al.}(2017)\citenamefont {Timpu},
  \citenamefont {Hendricks}, \citenamefont {Petrov}, \citenamefont {Ni},
  \citenamefont {Renaut}, \citenamefont {Wolf}, \citenamefont {Isa},
  \citenamefont {Kivshar},\ and\ \citenamefont {Grange}}]{timpu2017enhanced}%
  \BibitemOpen
  \bibfield  {author} {\bibinfo {author} {\bibfnamefont {F.}~\bibnamefont
  {Timpu}}, \bibinfo {author} {\bibfnamefont {N.~R.}\ \bibnamefont
  {Hendricks}}, \bibinfo {author} {\bibfnamefont {M.}~\bibnamefont {Petrov}},
  \bibinfo {author} {\bibfnamefont {S.}~\bibnamefont {Ni}}, \bibinfo {author}
  {\bibfnamefont {C.}~\bibnamefont {Renaut}}, \bibinfo {author} {\bibfnamefont
  {H.}~\bibnamefont {Wolf}}, \bibinfo {author} {\bibfnamefont {L.}~\bibnamefont
  {Isa}}, \bibinfo {author} {\bibfnamefont {Y.}~\bibnamefont {Kivshar}},\ and\
  \bibinfo {author} {\bibfnamefont {R.}~\bibnamefont {Grange}},\ }\bibfield
  {title} {\bibinfo {title} {Enhanced second-harmonic generation from
  sequential capillarity-assisted particle assembly of hybrid nanodimers},\
  }\href@noop {} {\bibfield  {journal} {\bibinfo  {journal} {Nano Lett.}\
  }\textbf {\bibinfo {volume} {17}},\ \bibinfo {pages} {5381} (\bibinfo {year}
  {2017})}\BibitemShut {NoStop}%
\bibitem [{\citenamefont {Yi}\ \emph {et~al.}(2019)\citenamefont {Yi},
  \citenamefont {Wang}, \citenamefont {Schwarz}, \citenamefont {Zhong},
  \citenamefont {Chimeh}, \citenamefont {Korte}, \citenamefont {Zhan},
  \citenamefont {Schaaf}, \citenamefont {Runge},\ and\ \citenamefont
  {Lienau}}]{yi2019doubly}%
  \BibitemOpen
  \bibfield  {author} {\bibinfo {author} {\bibfnamefont {J.-M.}\ \bibnamefont
  {Yi}}, \bibinfo {author} {\bibfnamefont {D.}~\bibnamefont {Wang}}, \bibinfo
  {author} {\bibfnamefont {F.}~\bibnamefont {Schwarz}}, \bibinfo {author}
  {\bibfnamefont {J.}~\bibnamefont {Zhong}}, \bibinfo {author} {\bibfnamefont
  {A.}~\bibnamefont {Chimeh}}, \bibinfo {author} {\bibfnamefont
  {A.}~\bibnamefont {Korte}}, \bibinfo {author} {\bibfnamefont
  {J.}~\bibnamefont {Zhan}}, \bibinfo {author} {\bibfnamefont {P.}~\bibnamefont
  {Schaaf}}, \bibinfo {author} {\bibfnamefont {E.}~\bibnamefont {Runge}},\ and\
  \bibinfo {author} {\bibfnamefont {C.}~\bibnamefont {Lienau}},\ }\bibfield
  {title} {\bibinfo {title} {Doubly resonant plasmonic hot spot–exciton
  coupling enhances second harmonic generation from {Au}/{ZnO} hybrid porous
  nanosponges},\ }\href@noop {} {\bibfield  {journal} {\bibinfo  {journal} {ACS
  Photonics}\ }\textbf {\bibinfo {volume} {6}},\ \bibinfo {pages} {2779}
  (\bibinfo {year} {2019})}\BibitemShut {NoStop}%
\bibitem [{\citenamefont {G\"{u}rdal}\ \emph {et~al.}(2020)\citenamefont
  {G\"{u}rdal}, \citenamefont {Horneber}, \citenamefont {Meixner},
  \citenamefont {Kern}, \citenamefont {Zhang},\ and\ \citenamefont
  {Fleischer}}]{gurdal2020enhancement}%
  \BibitemOpen
  \bibfield  {author} {\bibinfo {author} {\bibfnamefont {E.}~\bibnamefont
  {G\"{u}rdal}}, \bibinfo {author} {\bibfnamefont {A.}~\bibnamefont
  {Horneber}}, \bibinfo {author} {\bibfnamefont {A.~J.}\ \bibnamefont
  {Meixner}}, \bibinfo {author} {\bibfnamefont {D.~P.}\ \bibnamefont {Kern}},
  \bibinfo {author} {\bibfnamefont {D.}~\bibnamefont {Zhang}},\ and\ \bibinfo
  {author} {\bibfnamefont {M.}~\bibnamefont {Fleischer}},\ }\bibfield  {title}
  {\bibinfo {title} {Enhancement of the second harmonic signal of nonlinear
  crystals by a single metal nanoantenna},\ }\href@noop {} {\bibfield
  {journal} {\bibinfo  {journal} {Nanoscale}\ }\textbf {\bibinfo {volume}
  {12}},\ \bibinfo {pages} {23105} (\bibinfo {year} {2020})}\BibitemShut
  {NoStop}%
\bibitem [{\citenamefont {Chauvet}\ \emph {et~al.}(2020)\citenamefont
  {Chauvet}, \citenamefont {Ethis~de Corny}, \citenamefont {Jeannin},
  \citenamefont {Laurent}, \citenamefont {Huant}, \citenamefont {Gacoin},
  \citenamefont {Dantelle}, \citenamefont {Nogues},\ and\ \citenamefont
  {Bachelier}}]{chauvet2020hybrid}%
  \BibitemOpen
  \bibfield  {author} {\bibinfo {author} {\bibfnamefont {N.}~\bibnamefont
  {Chauvet}}, \bibinfo {author} {\bibfnamefont {M.}~\bibnamefont {Ethis~de
  Corny}}, \bibinfo {author} {\bibfnamefont {M.}~\bibnamefont {Jeannin}},
  \bibinfo {author} {\bibfnamefont {G.}~\bibnamefont {Laurent}}, \bibinfo
  {author} {\bibfnamefont {S.}~\bibnamefont {Huant}}, \bibinfo {author}
  {\bibfnamefont {T.}~\bibnamefont {Gacoin}}, \bibinfo {author} {\bibfnamefont
  {G.}~\bibnamefont {Dantelle}}, \bibinfo {author} {\bibfnamefont
  {G.}~\bibnamefont {Nogues}},\ and\ \bibinfo {author} {\bibfnamefont
  {G.}~\bibnamefont {Bachelier}},\ }\bibfield  {title} {\bibinfo {title}
  {Hybrid {KTP}–plasmonic nanostructures for enhanced nonlinear optics at the
  nanoscale},\ }\href@noop {} {\bibfield  {journal} {\bibinfo  {journal} {ACS
  Photonics}\ }\textbf {\bibinfo {volume} {7}},\ \bibinfo {pages} {665}
  (\bibinfo {year} {2020})}\BibitemShut {NoStop}%
\bibitem [{\citenamefont {Bonacina}\ \emph {et~al.}(2020)\citenamefont
  {Bonacina}, \citenamefont {Brevet}, \citenamefont {Finazzi},\ and\
  \citenamefont {Celebrano}}]{bonacina2020harmonic}%
  \BibitemOpen
  \bibfield  {author} {\bibinfo {author} {\bibfnamefont {L.}~\bibnamefont
  {Bonacina}}, \bibinfo {author} {\bibfnamefont {P.-F.}\ \bibnamefont
  {Brevet}}, \bibinfo {author} {\bibfnamefont {M.}~\bibnamefont {Finazzi}},\
  and\ \bibinfo {author} {\bibfnamefont {M.}~\bibnamefont {Celebrano}},\
  }\bibfield  {title} {\bibinfo {title} {Harmonic generation at the
  nanoscale},\ }\href@noop {} {\bibfield  {journal} {\bibinfo  {journal} {J. of
  Appl. Phys.}\ }\textbf {\bibinfo {volume} {127}},\ \bibinfo {pages} {230901}
  (\bibinfo {year} {2020})}\BibitemShut {NoStop}%
\bibitem [{\citenamefont {Richter}\ \emph {et~al.}(2014)\citenamefont
  {Richter}, \citenamefont {Steinbr\"{u}ck}, \citenamefont {Zilk},
  \citenamefont {Sergeyev}, \citenamefont {Pertsch}, \citenamefont
  {T\"{u}nnermann},\ and\ \citenamefont {Grange}}]{richter2014coreshell}%
  \BibitemOpen
  \bibfield  {author} {\bibinfo {author} {\bibfnamefont {J.}~\bibnamefont
  {Richter}}, \bibinfo {author} {\bibfnamefont {A.}~\bibnamefont
  {Steinbr\"{u}ck}}, \bibinfo {author} {\bibfnamefont {M.}~\bibnamefont
  {Zilk}}, \bibinfo {author} {\bibfnamefont {A.}~\bibnamefont {Sergeyev}},
  \bibinfo {author} {\bibfnamefont {T.}~\bibnamefont {Pertsch}}, \bibinfo
  {author} {\bibfnamefont {A.}~\bibnamefont {T\"{u}nnermann}},\ and\ \bibinfo
  {author} {\bibfnamefont {R.}~\bibnamefont {Grange}},\ }\bibfield  {title}
  {\bibinfo {title} {Core–shell potassium niobate nanowires for enhanced
  nonlinear optical effects},\ }\href@noop {} {\bibfield  {journal} {\bibinfo
  {journal} {Nanoscale}\ }\textbf {\bibinfo {volume} {6}},\ \bibinfo {pages}
  {5200} (\bibinfo {year} {2014})}\BibitemShut {NoStop}%
\bibitem [{\citenamefont {Weber}\ \emph {et~al.}(2015)\citenamefont {Weber},
  \citenamefont {Feis}, \citenamefont {Gellini}, \citenamefont {Pilot},
  \citenamefont {Salvi},\ and\ \citenamefont {Signorini}}]{weber2015far-}%
  \BibitemOpen
  \bibfield  {author} {\bibinfo {author} {\bibfnamefont {V.}~\bibnamefont
  {Weber}}, \bibinfo {author} {\bibfnamefont {A.}~\bibnamefont {Feis}},
  \bibinfo {author} {\bibfnamefont {C.}~\bibnamefont {Gellini}}, \bibinfo
  {author} {\bibfnamefont {R.}~\bibnamefont {Pilot}}, \bibinfo {author}
  {\bibfnamefont {P.~R.}\ \bibnamefont {Salvi}},\ and\ \bibinfo {author}
  {\bibfnamefont {R.}~\bibnamefont {Signorini}},\ }\bibfield  {title} {\bibinfo
  {title} {Far- and near-field properties of gold nanoshells studied by
  photoacoustic and surface-enhanced {Raman} spectroscopies},\ }\href@noop {}
  {\bibfield  {journal} {\bibinfo  {journal} {Phys. Chem. Chem. Phys.}\
  }\textbf {\bibinfo {volume} {17}},\ \bibinfo {pages} {21190} (\bibinfo {year}
  {2015})}\BibitemShut {NoStop}%
\bibitem [{\citenamefont {Cavicchi}\ \emph {et~al.}(2013)\citenamefont
  {Cavicchi}, \citenamefont {Meier}, \citenamefont {Presser}, \citenamefont
  {Prabhu},\ and\ \citenamefont {Guha}}]{cavicchi2013single}%
  \BibitemOpen
  \bibfield  {author} {\bibinfo {author} {\bibfnamefont {R.~E.}\ \bibnamefont
  {Cavicchi}}, \bibinfo {author} {\bibfnamefont {D.~C.}\ \bibnamefont {Meier}},
  \bibinfo {author} {\bibfnamefont {C.}~\bibnamefont {Presser}}, \bibinfo
  {author} {\bibfnamefont {V.~M.}\ \bibnamefont {Prabhu}},\ and\ \bibinfo
  {author} {\bibfnamefont {S.}~\bibnamefont {Guha}},\ }\bibfield  {title}
  {\bibinfo {title} {Single laser pulse effects on suspended-au-nanoparticle
  size distributions and morphology},\ }\href@noop {} {\bibfield  {journal}
  {\bibinfo  {journal} {The Journal of Physical Chemistry C}\ }\textbf
  {\bibinfo {volume} {117}},\ \bibinfo {pages} {10866} (\bibinfo {year}
  {2013})}\BibitemShut {NoStop}%
\bibitem [{\citenamefont {Taitt}\ \emph {et~al.}(2021)\citenamefont {Taitt},
  \citenamefont {Urbain}, \citenamefont {Behel}, \citenamefont
  {Pablo-Sainz-Ezquerra}, \citenamefont {Kandybka}, \citenamefont {Millet},
  \citenamefont {Martinez-Rodriguez}, \citenamefont {Yeromonahos},
  \citenamefont {Beauquis}, \citenamefont {Le~Dantec}, \citenamefont {Mugnier},
  \citenamefont {Brevet}, \citenamefont {Chevolot},\ and\ \citenamefont
  {Monnier}}]{taitt2021gold-seeded}%
  \BibitemOpen
  \bibfield  {author} {\bibinfo {author} {\bibfnamefont {R.}~\bibnamefont
  {Taitt}}, \bibinfo {author} {\bibfnamefont {M.}~\bibnamefont {Urbain}},
  \bibinfo {author} {\bibfnamefont {Z.}~\bibnamefont {Behel}}, \bibinfo
  {author} {\bibfnamefont {A.-M.}\ \bibnamefont {Pablo-Sainz-Ezquerra}},
  \bibinfo {author} {\bibfnamefont {I.}~\bibnamefont {Kandybka}}, \bibinfo
  {author} {\bibfnamefont {E.}~\bibnamefont {Millet}}, \bibinfo {author}
  {\bibfnamefont {N.}~\bibnamefont {Martinez-Rodriguez}}, \bibinfo {author}
  {\bibfnamefont {C.}~\bibnamefont {Yeromonahos}}, \bibinfo {author}
  {\bibfnamefont {S.}~\bibnamefont {Beauquis}}, \bibinfo {author}
  {\bibfnamefont {R.}~\bibnamefont {Le~Dantec}}, \bibinfo {author}
  {\bibfnamefont {Y.}~\bibnamefont {Mugnier}}, \bibinfo {author} {\bibfnamefont
  {P.-F.}\ \bibnamefont {Brevet}}, \bibinfo {author} {\bibfnamefont
  {Y.}~\bibnamefont {Chevolot}},\ and\ \bibinfo {author} {\bibfnamefont
  {V.}~\bibnamefont {Monnier}},\ }\bibfield  {title} {\bibinfo {title}
  {Gold-seeded lithium niobate nanoparticles: Influence of gold surface
  coverage on second harmonic properties},\ }\href@noop {} {\bibfield
  {journal} {\bibinfo  {journal} {Nanomaterials}\ }\textbf {\bibinfo {volume}
  {11}},\ \bibinfo {pages} {950} (\bibinfo {year} {2021})}\BibitemShut
  {NoStop}%
\bibitem [{\citenamefont {Ali}\ and\ \citenamefont
  {Gates}(2018)}]{ali2018synthesis}%
  \BibitemOpen
  \bibfield  {author} {\bibinfo {author} {\bibfnamefont {R.~F.}\ \bibnamefont
  {Ali}}\ and\ \bibinfo {author} {\bibfnamefont {B.~D.}\ \bibnamefont
  {Gates}},\ }\bibfield  {title} {\bibinfo {title} {Synthesis of lithium
  niobate nanocrystals with size focusing through an ostwald ripening
  process},\ }\href@noop {} {\bibfield  {journal} {\bibinfo  {journal} {Chem.
  Mater.}\ }\textbf {\bibinfo {volume} {30}},\ \bibinfo {pages} {2028}
  (\bibinfo {year} {2018})}\BibitemShut {NoStop}%
\bibitem [{\citenamefont {Sun}\ \emph {et~al.}(2021)\citenamefont {Sun},
  \citenamefont {Chen}, \citenamefont {Yin}, \citenamefont {Curnan},
  \citenamefont {Han}, \citenamefont {Chen},\ and\ \citenamefont
  {Ma}}]{sun2021progress}%
  \BibitemOpen
  \bibfield  {author} {\bibinfo {author} {\bibfnamefont {X.}~\bibnamefont
  {Sun}}, \bibinfo {author} {\bibfnamefont {H.}~\bibnamefont {Chen}}, \bibinfo
  {author} {\bibfnamefont {Y.}~\bibnamefont {Yin}}, \bibinfo {author}
  {\bibfnamefont {M.~T.}\ \bibnamefont {Curnan}}, \bibinfo {author}
  {\bibfnamefont {J.~W.}\ \bibnamefont {Han}}, \bibinfo {author} {\bibfnamefont
  {Y.}~\bibnamefont {Chen}},\ and\ \bibinfo {author} {\bibfnamefont
  {Z.}~\bibnamefont {Ma}},\ }\bibfield  {title} {\bibinfo {title} {Progress of
  exsolved metal nanoparticles on oxides as high performance (electro)
  catalysts for the conversion of small molecules},\ }\href@noop {} {\bibfield
  {journal} {\bibinfo  {journal} {Small}\ }\textbf {\bibinfo {volume} {17}},\
  \bibinfo {pages} {2005383} (\bibinfo {year} {2021})}\BibitemShut {NoStop}%
\bibitem [{\citenamefont {Madzharova}\ \emph {et~al.}(2019)\citenamefont
  {Madzharova}, \citenamefont {Nodar}, \citenamefont {\v{Z}ivanovi\'{c}},
  \citenamefont {Huang}, \citenamefont {Koch}, \citenamefont {Esteban},
  \citenamefont {Aizpurua},\ and\ \citenamefont {Kneipp}}]{madzharova2019gold}%
  \BibitemOpen
  \bibfield  {author} {\bibinfo {author} {\bibfnamefont {F.}~\bibnamefont
  {Madzharova}}, \bibinfo {author} {\bibfnamefont {A.}~\bibnamefont {Nodar}},
  \bibinfo {author} {\bibfnamefont {V.}~\bibnamefont {\v{Z}ivanovi\'{c}}},
  \bibinfo {author} {\bibfnamefont {M.~R.~S.}\ \bibnamefont {Huang}}, \bibinfo
  {author} {\bibfnamefont {C.~T.}\ \bibnamefont {Koch}}, \bibinfo {author}
  {\bibfnamefont {R.}~\bibnamefont {Esteban}}, \bibinfo {author} {\bibfnamefont
  {J.}~\bibnamefont {Aizpurua}},\ and\ \bibinfo {author} {\bibfnamefont
  {J.}~\bibnamefont {Kneipp}},\ }\bibfield  {title} {\bibinfo {title} {Gold-
  and silver-coated barium titanate nanocomposites as probes for two-photon
  multimodal microspectroscopy},\ }\href@noop {} {\bibfield  {journal}
  {\bibinfo  {journal} {Adv. Funct. Mater.}\ }\textbf {\bibinfo {volume}
  {29}},\ \bibinfo {pages} {1904289} (\bibinfo {year} {2019})}\BibitemShut
  {NoStop}%
\bibitem [{\citenamefont {Bohren}\ and\ \citenamefont
  {Huffman}(2004)}]{bohren2004absorption}%
  \BibitemOpen
  \bibfield  {author} {\bibinfo {author} {\bibfnamefont {C.~F.}\ \bibnamefont
  {Bohren}}\ and\ \bibinfo {author} {\bibfnamefont {D.~R.}\ \bibnamefont
  {Huffman}},\ }\href@noop {} {\emph {\bibinfo {title} {Absorption and
  Scattering of Light by Small Particles}}}\ (\bibinfo  {publisher}
  {Wiley-VCH},\ \bibinfo {address} {Weinheim},\ \bibinfo {year} {2004})\ pp.\
  \bibinfo {pages} {57--154}\BibitemShut {NoStop}%
\bibitem [{\citenamefont {Timpu}\ \emph {et~al.}(2019)\citenamefont {Timpu},
  \citenamefont {Sendra}, \citenamefont {Renaut}, \citenamefont {Lang},
  \citenamefont {Timofeeva}, \citenamefont {Buscaglia}, \citenamefont
  {Buscaglia},\ and\ \citenamefont {Grange}}]{timpu2019lithium}%
  \BibitemOpen
  \bibfield  {author} {\bibinfo {author} {\bibfnamefont {F.}~\bibnamefont
  {Timpu}}, \bibinfo {author} {\bibfnamefont {J.}~\bibnamefont {Sendra}},
  \bibinfo {author} {\bibfnamefont {C.}~\bibnamefont {Renaut}}, \bibinfo
  {author} {\bibfnamefont {L.}~\bibnamefont {Lang}}, \bibinfo {author}
  {\bibfnamefont {M.}~\bibnamefont {Timofeeva}}, \bibinfo {author}
  {\bibfnamefont {M.~T.}\ \bibnamefont {Buscaglia}}, \bibinfo {author}
  {\bibfnamefont {V.}~\bibnamefont {Buscaglia}},\ and\ \bibinfo {author}
  {\bibfnamefont {R.}~\bibnamefont {Grange}},\ }\bibfield  {title} {\bibinfo
  {title} {Lithium niobate nanocubes as linear and nonlinear ultraviolet mie
  resonators},\ }\href@noop {} {\bibfield  {journal} {\bibinfo  {journal} {ACS
  Photonics}\ }\textbf {\bibinfo {volume} {6}},\ \bibinfo {pages} {545}
  (\bibinfo {year} {2019})}\BibitemShut {NoStop}%
\bibitem [{\citenamefont {Sipe}\ \emph {et~al.}(1980)\citenamefont {Sipe},
  \citenamefont {So}, \citenamefont {Fukui},\ and\ \citenamefont
  {Stegeman}}]{sipe1980analysis}%
  \BibitemOpen
  \bibfield  {author} {\bibinfo {author} {\bibfnamefont {J.~E.}\ \bibnamefont
  {Sipe}}, \bibinfo {author} {\bibfnamefont {V.~C.~Y.}\ \bibnamefont {So}},
  \bibinfo {author} {\bibfnamefont {M.}~\bibnamefont {Fukui}},\ and\ \bibinfo
  {author} {\bibfnamefont {G.~I.}\ \bibnamefont {Stegeman}},\ }\bibfield
  {title} {\bibinfo {title} {Analysis of second-harmonic generation at metal
  surfaces},\ }\href@noop {} {\bibfield  {journal} {\bibinfo  {journal} {Phys.
  Rev. B}\ }\textbf {\bibinfo {volume} {21}},\ \bibinfo {pages} {4389}
  (\bibinfo {year} {1980})}\BibitemShut {NoStop}%
\bibitem [{\citenamefont {Capretti}\ \emph {et~al.}(2014)\citenamefont
  {Capretti}, \citenamefont {Pecora}, \citenamefont {Forestiere}, \citenamefont
  {Dal~Negro},\ and\ \citenamefont {Miano}}]{capretti2014size-dependent}%
  \BibitemOpen
  \bibfield  {author} {\bibinfo {author} {\bibfnamefont {A.}~\bibnamefont
  {Capretti}}, \bibinfo {author} {\bibfnamefont {E.~F.}\ \bibnamefont
  {Pecora}}, \bibinfo {author} {\bibfnamefont {C.}~\bibnamefont {Forestiere}},
  \bibinfo {author} {\bibfnamefont {L.}~\bibnamefont {Dal~Negro}},\ and\
  \bibinfo {author} {\bibfnamefont {G.}~\bibnamefont {Miano}},\ }\bibfield
  {title} {\bibinfo {title} {Size-dependent second-harmonic generation from
  gold nanoparticles},\ }\href@noop {} {\bibfield  {journal} {\bibinfo
  {journal} {Phys. Rev. B}\ }\textbf {\bibinfo {volume} {89}},\ \bibinfo
  {pages} {125414} (\bibinfo {year} {2014})}\BibitemShut {NoStop}%
\bibitem [{\citenamefont {Liu}\ \emph {et~al.}(2016)\citenamefont {Liu},
  \citenamefont {Sinclair}, \citenamefont {Saravi}, \citenamefont {Keeler},
  \citenamefont {Yang}, \citenamefont {Reno}, \citenamefont {Peake},
  \citenamefont {Setzpfandt}, \citenamefont {Staude}, \citenamefont {Pertsch},\
  and\ \citenamefont {Brener}}]{liu2016resonantly}%
  \BibitemOpen
  \bibfield  {author} {\bibinfo {author} {\bibfnamefont {S.}~\bibnamefont
  {Liu}}, \bibinfo {author} {\bibfnamefont {M.~B.}\ \bibnamefont {Sinclair}},
  \bibinfo {author} {\bibfnamefont {S.}~\bibnamefont {Saravi}}, \bibinfo
  {author} {\bibfnamefont {G.~A.}\ \bibnamefont {Keeler}}, \bibinfo {author}
  {\bibfnamefont {Y.}~\bibnamefont {Yang}}, \bibinfo {author} {\bibfnamefont
  {J.}~\bibnamefont {Reno}}, \bibinfo {author} {\bibfnamefont {G.~M.}\
  \bibnamefont {Peake}}, \bibinfo {author} {\bibfnamefont {F.}~\bibnamefont
  {Setzpfandt}}, \bibinfo {author} {\bibfnamefont {I.}~\bibnamefont {Staude}},
  \bibinfo {author} {\bibfnamefont {T.}~\bibnamefont {Pertsch}},\ and\ \bibinfo
  {author} {\bibfnamefont {I.}~\bibnamefont {Brener}},\ }\bibfield  {title}
  {\bibinfo {title} {Resonantly enhanced second-harmonic generation using
  {III}–{V} semiconductor all-dielectric metasurfaces},\ }\href@noop {}
  {\bibfield  {journal} {\bibinfo  {journal} {Nano Lett.}\ }\textbf {\bibinfo
  {volume} {16}},\ \bibinfo {pages} {5426} (\bibinfo {year}
  {2016})}\BibitemShut {NoStop}%
\bibitem [{\citenamefont {Yang}\ \emph {et~al.}(2015)\citenamefont {Yang},
  \citenamefont {Wang}, \citenamefont {Boulesbaa}, \citenamefont {Kravchenko},
  \citenamefont {Briggs}, \citenamefont {Puretzky}, \citenamefont {Geohegan},\
  and\ \citenamefont {Valentine}}]{yang2015nonlinear}%
  \BibitemOpen
  \bibfield  {author} {\bibinfo {author} {\bibfnamefont {Y.}~\bibnamefont
  {Yang}}, \bibinfo {author} {\bibfnamefont {W.}~\bibnamefont {Wang}}, \bibinfo
  {author} {\bibfnamefont {A.}~\bibnamefont {Boulesbaa}}, \bibinfo {author}
  {\bibfnamefont {I.~I.}\ \bibnamefont {Kravchenko}}, \bibinfo {author}
  {\bibfnamefont {D.~P.}\ \bibnamefont {Briggs}}, \bibinfo {author}
  {\bibfnamefont {A.}~\bibnamefont {Puretzky}}, \bibinfo {author}
  {\bibfnamefont {D.}~\bibnamefont {Geohegan}},\ and\ \bibinfo {author}
  {\bibfnamefont {J.}~\bibnamefont {Valentine}},\ }\bibfield  {title} {\bibinfo
  {title} {Nonlinear fano-resonant dielectric metasurfaces},\ }\href@noop {}
  {\bibfield  {journal} {\bibinfo  {journal} {Nano Lett.}\ }\textbf {\bibinfo
  {volume} {15}},\ \bibinfo {pages} {7388} (\bibinfo {year}
  {2015})}\BibitemShut {NoStop}%
\bibitem [{\citenamefont {Gili}\ \emph {et~al.}(2018)\citenamefont {Gili},
  \citenamefont {Ghirardini}, \citenamefont {Rocco}, \citenamefont {Marino},
  \citenamefont {Favero}, \citenamefont {Roland}, \citenamefont {Pellegrini},
  \citenamefont {Duò}, \citenamefont {Finazzi}, \citenamefont {Carletti},
  \citenamefont {Locatelli}, \citenamefont {Lemaître}, \citenamefont {Neshev},
  \citenamefont {De~Angelis}, \citenamefont {Leo},\ and\ \citenamefont
  {Celebrano}}]{gili2018metal-dielectric}%
  \BibitemOpen
  \bibfield  {author} {\bibinfo {author} {\bibfnamefont {V.~F.}\ \bibnamefont
  {Gili}}, \bibinfo {author} {\bibfnamefont {L.}~\bibnamefont {Ghirardini}},
  \bibinfo {author} {\bibfnamefont {D.}~\bibnamefont {Rocco}}, \bibinfo
  {author} {\bibfnamefont {G.}~\bibnamefont {Marino}}, \bibinfo {author}
  {\bibfnamefont {I.}~\bibnamefont {Favero}}, \bibinfo {author} {\bibfnamefont
  {I.}~\bibnamefont {Roland}}, \bibinfo {author} {\bibfnamefont
  {G.}~\bibnamefont {Pellegrini}}, \bibinfo {author} {\bibfnamefont
  {L.}~\bibnamefont {Duò}}, \bibinfo {author} {\bibfnamefont {M.}~\bibnamefont
  {Finazzi}}, \bibinfo {author} {\bibfnamefont {L.}~\bibnamefont {Carletti}},
  \bibinfo {author} {\bibfnamefont {A.}~\bibnamefont {Locatelli}}, \bibinfo
  {author} {\bibfnamefont {A.}~\bibnamefont {Lemaître}}, \bibinfo {author}
  {\bibfnamefont {D.}~\bibnamefont {Neshev}}, \bibinfo {author} {\bibfnamefont
  {C.}~\bibnamefont {De~Angelis}}, \bibinfo {author} {\bibfnamefont
  {G.}~\bibnamefont {Leo}},\ and\ \bibinfo {author} {\bibfnamefont
  {M.}~\bibnamefont {Celebrano}},\ }\bibfield  {title} {\bibinfo {title}
  {Metal–dielectric hybrid nanoantennas for efficient frequency conversion at
  the anapole mode},\ }\href@noop {} {\bibfield  {journal} {\bibinfo  {journal}
  {Beilstein J. Nanotechnol.}\ }\textbf {\bibinfo {volume} {9}},\ \bibinfo
  {pages} {2306} (\bibinfo {year} {2018})}\BibitemShut {NoStop}%
\bibitem [{\citenamefont {Renaut}\ \emph {et~al.}(2019)\citenamefont {Renaut},
  \citenamefont {Lang}, \citenamefont {Frizyuk}, \citenamefont {Timofeeva},
  \citenamefont {Komissarenko}, \citenamefont {Mukhin}, \citenamefont
  {Smirnova}, \citenamefont {Timpu}, \citenamefont {Petrov}, \citenamefont
  {Kivshar},\ and\ \citenamefont {Grange}}]{renaut2019reshaping}%
  \BibitemOpen
  \bibfield  {author} {\bibinfo {author} {\bibfnamefont {C.}~\bibnamefont
  {Renaut}}, \bibinfo {author} {\bibfnamefont {L.}~\bibnamefont {Lang}},
  \bibinfo {author} {\bibfnamefont {K.}~\bibnamefont {Frizyuk}}, \bibinfo
  {author} {\bibfnamefont {M.}~\bibnamefont {Timofeeva}}, \bibinfo {author}
  {\bibfnamefont {F.~E.}\ \bibnamefont {Komissarenko}}, \bibinfo {author}
  {\bibfnamefont {I.~S.}\ \bibnamefont {Mukhin}}, \bibinfo {author}
  {\bibfnamefont {D.}~\bibnamefont {Smirnova}}, \bibinfo {author}
  {\bibfnamefont {F.}~\bibnamefont {Timpu}}, \bibinfo {author} {\bibfnamefont
  {M.}~\bibnamefont {Petrov}}, \bibinfo {author} {\bibfnamefont
  {Y.}~\bibnamefont {Kivshar}},\ and\ \bibinfo {author} {\bibfnamefont
  {R.}~\bibnamefont {Grange}},\ }\bibfield  {title} {\bibinfo {title}
  {Reshaping the second-order polar response of hybrid metal–dielectric
  nanodimers},\ }\href@noop {} {\bibfield  {journal} {\bibinfo  {journal} {Nano
  Lett.}\ }\textbf {\bibinfo {volume} {19}},\ \bibinfo {pages} {877} (\bibinfo
  {year} {2019})}\BibitemShut {NoStop}%
\bibitem [{\citenamefont {Hall}\ \emph {et~al.}(2017)\citenamefont {Hall},
  \citenamefont {Reynolds}, \citenamefont {Henderson}, \citenamefont {Riesen},
  \citenamefont {Monro},\ and\ \citenamefont {Afshar}}]{hall2017unified}%
  \BibitemOpen
  \bibfield  {author} {\bibinfo {author} {\bibfnamefont {J.~M.~M.}\
  \bibnamefont {Hall}}, \bibinfo {author} {\bibfnamefont {T.}~\bibnamefont
  {Reynolds}}, \bibinfo {author} {\bibfnamefont {M.~R.}\ \bibnamefont
  {Henderson}}, \bibinfo {author} {\bibfnamefont {N.}~\bibnamefont {Riesen}},
  \bibinfo {author} {\bibfnamefont {T.~M.}\ \bibnamefont {Monro}},\ and\
  \bibinfo {author} {\bibfnamefont {S.}~\bibnamefont {Afshar}},\ }\bibfield
  {title} {\bibinfo {title} {Unified theory of whispering gallery multilayer
  microspheres with single dipole or active layer sources},\ }\href@noop {}
  {\bibfield  {journal} {\bibinfo  {journal} {Opt. Express}\ }\textbf {\bibinfo
  {volume} {25}},\ \bibinfo {pages} {6192} (\bibinfo {year}
  {2017})}\BibitemShut {NoStop}%
\bibitem [{\citenamefont {Smirnova}\ \emph {et~al.}(2018)\citenamefont
  {Smirnova}, \citenamefont {Smirnov},\ and\ \citenamefont
  {Kivshar}}]{smirnova2018multipolar}%
  \BibitemOpen
  \bibfield  {author} {\bibinfo {author} {\bibfnamefont {D.}~\bibnamefont
  {Smirnova}}, \bibinfo {author} {\bibfnamefont {A.~I.}\ \bibnamefont
  {Smirnov}},\ and\ \bibinfo {author} {\bibfnamefont {Y.~S.}\ \bibnamefont
  {Kivshar}},\ }\bibfield  {title} {\bibinfo {title} {Multipolar
  second-harmonic generation by {Mie}-resonant dielectric nanoparticles},\
  }\href@noop {} {\bibfield  {journal} {\bibinfo  {journal} {Phys. Rev. A}\
  }\textbf {\bibinfo {volume} {97}},\ \bibinfo {pages} {013807} (\bibinfo
  {year} {2018})}\BibitemShut {NoStop}%
\bibitem [{\citenamefont {Chang}\ \emph {et~al.}(2016)\citenamefont {Chang},
  \citenamefont {Gonz\'{a}lez-Tudela}, \citenamefont {S\'{a}nchez~Muñoz},
  \citenamefont {Navarrete-Benlloch},\ and\ \citenamefont
  {Shi}}]{chang2016deterministic}%
  \BibitemOpen
  \bibfield  {author} {\bibinfo {author} {\bibfnamefont {Y.}~\bibnamefont
  {Chang}}, \bibinfo {author} {\bibfnamefont {A.}~\bibnamefont
  {Gonz\'{a}lez-Tudela}}, \bibinfo {author} {\bibfnamefont {C.}~\bibnamefont
  {S\'{a}nchez~Muñoz}}, \bibinfo {author} {\bibfnamefont {C.}~\bibnamefont
  {Navarrete-Benlloch}},\ and\ \bibinfo {author} {\bibfnamefont
  {T.}~\bibnamefont {Shi}},\ }\bibfield  {title} {\bibinfo {title}
  {Deterministic down-converter and continuous photon-pair source within the
  bad-cavity limit},\ }\href@noop {} {\bibfield  {journal} {\bibinfo  {journal}
  {Phys. Rev. Lett.}\ }\textbf {\bibinfo {volume} {117}},\ \bibinfo {pages}
  {203602} (\bibinfo {year} {2016})}\BibitemShut {NoStop}%
\bibitem [{\citenamefont {Welakuh}\ \emph {et~al.}(2021)\citenamefont
  {Welakuh}, \citenamefont {Ruggenthaler}, \citenamefont {Tchenkoue},
  \citenamefont {Appel},\ and\ \citenamefont
  {Rubio}}]{welakuh2021down-conversion}%
  \BibitemOpen
  \bibfield  {author} {\bibinfo {author} {\bibfnamefont {D.~M.}\ \bibnamefont
  {Welakuh}}, \bibinfo {author} {\bibfnamefont {M.}~\bibnamefont
  {Ruggenthaler}}, \bibinfo {author} {\bibfnamefont {M.-L.~M.}\ \bibnamefont
  {Tchenkoue}}, \bibinfo {author} {\bibfnamefont {H.}~\bibnamefont {Appel}},\
  and\ \bibinfo {author} {\bibfnamefont {A.}~\bibnamefont {Rubio}},\ }\bibfield
   {title} {\bibinfo {title} {Down-conversion processes in ab initio
  nonrelativistic quantum electrodynamics},\ }\href@noop {} {\bibfield
  {journal} {\bibinfo  {journal} {Phys. Rev. Res.}\ }\textbf {\bibinfo {volume}
  {3}},\ \bibinfo {pages} {033067} (\bibinfo {year} {2021})}\BibitemShut
  {NoStop}%
\bibitem [{\citenamefont {Choudhary}\ \emph {et~al.}(2019)\citenamefont
  {Choudhary}, \citenamefont {De~Leon}, \citenamefont {Swiecicki},
  \citenamefont {Awan}, \citenamefont {Schulz}, \citenamefont {Upham},
  \citenamefont {Alam}, \citenamefont {Sipe},\ and\ \citenamefont
  {Boyd}}]{choudhary2019weak}%
  \BibitemOpen
  \bibfield  {author} {\bibinfo {author} {\bibfnamefont {S.}~\bibnamefont
  {Choudhary}}, \bibinfo {author} {\bibfnamefont {I.}~\bibnamefont {De~Leon}},
  \bibinfo {author} {\bibfnamefont {S.}~\bibnamefont {Swiecicki}}, \bibinfo
  {author} {\bibfnamefont {K.~M.}\ \bibnamefont {Awan}}, \bibinfo {author}
  {\bibfnamefont {S.~A.}\ \bibnamefont {Schulz}}, \bibinfo {author}
  {\bibfnamefont {J.}~\bibnamefont {Upham}}, \bibinfo {author} {\bibfnamefont
  {M.~Z.}\ \bibnamefont {Alam}}, \bibinfo {author} {\bibfnamefont {J.~E.}\
  \bibnamefont {Sipe}},\ and\ \bibinfo {author} {\bibfnamefont {R.~W.}\
  \bibnamefont {Boyd}},\ }\bibfield  {title} {\bibinfo {title} {Weak
  superradiance in arrays of plasmonic nanoantennas},\ }\href@noop {}
  {\bibfield  {journal} {\bibinfo  {journal} {Phys. Rev. A}\ }\textbf {\bibinfo
  {volume} {100}},\ \bibinfo {pages} {043814} (\bibinfo {year}
  {2019})}\BibitemShut {NoStop}%
\bibitem [{\citenamefont {Masson}\ and\ \citenamefont
  {Asenjo-Garcia}(2022)}]{masson2022universality}%
  \BibitemOpen
  \bibfield  {author} {\bibinfo {author} {\bibfnamefont {S.~J.}\ \bibnamefont
  {Masson}}\ and\ \bibinfo {author} {\bibfnamefont {A.}~\bibnamefont
  {Asenjo-Garcia}},\ }\bibfield  {title} {\bibinfo {title} {Universality of
  {Dicke} superradiance in arrays of quantum emitters},\ }\href@noop {}
  {\bibfield  {journal} {\bibinfo  {journal} {Nat. Commun.}\ }\textbf {\bibinfo
  {volume} {13}},\ \bibinfo {pages} {2285} (\bibinfo {year}
  {2022})}\BibitemShut {NoStop}%
\bibitem [{\citenamefont {Palik}\ and\ \citenamefont
  {Ghosh}(1998)}]{palik1998handbook}%
  \BibitemOpen
  \bibinfo {editor} {\bibfnamefont {E.~D.}\ \bibnamefont {Palik}}\ and\
  \bibinfo {editor} {\bibfnamefont {G.}~\bibnamefont {Ghosh}},\ eds.,\
  \href@noop {} {\emph {\bibinfo {title} {Handbook of Optical Constants of
  Solids}}}\ (\bibinfo  {publisher} {Academic Press},\ \bibinfo {address} {San
  Diego},\ \bibinfo {year} {1998})\ pp.\ \bibinfo {pages}
  {695--702}\BibitemShut {NoStop}%
\bibitem [{\citenamefont {Chervy}\ \emph {et~al.}(2016)\citenamefont {Chervy},
  \citenamefont {Xu}, \citenamefont {Duan}, \citenamefont {Wang}, \citenamefont
  {Mager}, \citenamefont {Frerejean}, \citenamefont {M\"{u}nninghoff},
  \citenamefont {Tinnemans}, \citenamefont {Hutchison}, \citenamefont {Genet},
  \citenamefont {Rowan}, \citenamefont {Rasing},\ and\ \citenamefont
  {Ebbesen}}]{chervy2016high-efficiency}%
  \BibitemOpen
  \bibfield  {author} {\bibinfo {author} {\bibfnamefont {T.}~\bibnamefont
  {Chervy}}, \bibinfo {author} {\bibfnamefont {J.}~\bibnamefont {Xu}}, \bibinfo
  {author} {\bibfnamefont {Y.}~\bibnamefont {Duan}}, \bibinfo {author}
  {\bibfnamefont {C.}~\bibnamefont {Wang}}, \bibinfo {author} {\bibfnamefont
  {L.}~\bibnamefont {Mager}}, \bibinfo {author} {\bibfnamefont
  {M.}~\bibnamefont {Frerejean}}, \bibinfo {author} {\bibfnamefont {J.~A.~W.}\
  \bibnamefont {M\"{u}nninghoff}}, \bibinfo {author} {\bibfnamefont
  {P.}~\bibnamefont {Tinnemans}}, \bibinfo {author} {\bibfnamefont {J.~A.}\
  \bibnamefont {Hutchison}}, \bibinfo {author} {\bibfnamefont {C.}~\bibnamefont
  {Genet}}, \bibinfo {author} {\bibfnamefont {A.~E.}\ \bibnamefont {Rowan}},
  \bibinfo {author} {\bibfnamefont {T.}~\bibnamefont {Rasing}},\ and\ \bibinfo
  {author} {\bibfnamefont {T.~W.}\ \bibnamefont {Ebbesen}},\ }\bibfield
  {title} {\bibinfo {title} {High-efficiency second-harmonic generation from
  hybrid light-matter states},\ }\href@noop {} {\bibfield  {journal} {\bibinfo
  {journal} {Nano Lett.}\ }\textbf {\bibinfo {volume} {16}},\ \bibinfo {pages}
  {7352} (\bibinfo {year} {2016})}\BibitemShut {NoStop}%
\bibitem [{\citenamefont {Kockum}\ \emph {et~al.}(2019)\citenamefont {Kockum},
  \citenamefont {Miranowicz}, \citenamefont {De~Liberato}, \citenamefont
  {Savasta},\ and\ \citenamefont {Nori}}]{kockum2019ultrastrong}%
  \BibitemOpen
  \bibfield  {author} {\bibinfo {author} {\bibfnamefont {A.~F.}\ \bibnamefont
  {Kockum}}, \bibinfo {author} {\bibfnamefont {A.}~\bibnamefont {Miranowicz}},
  \bibinfo {author} {\bibfnamefont {S.}~\bibnamefont {De~Liberato}}, \bibinfo
  {author} {\bibfnamefont {S.}~\bibnamefont {Savasta}},\ and\ \bibinfo {author}
  {\bibfnamefont {F.}~\bibnamefont {Nori}},\ }\bibfield  {title} {\bibinfo
  {title} {Ultrastrong coupling between light and matter},\ }\href@noop {}
  {\bibfield  {journal} {\bibinfo  {journal} {Nat. Rev. Phys.}\ }\textbf
  {\bibinfo {volume} {1}},\ \bibinfo {pages} {19} (\bibinfo {year}
  {2019})}\BibitemShut {NoStop}%
\bibitem [{\citenamefont {Liberko}\ \emph {et~al.}(2021)\citenamefont
  {Liberko}, \citenamefont {Busche}, \citenamefont {Seils}, \citenamefont
  {Bechtel}, \citenamefont {Rack}, \citenamefont {Masiello},\ and\
  \citenamefont {Camden}}]{liberko2021probing}%
  \BibitemOpen
  \bibfield  {author} {\bibinfo {author} {\bibfnamefont {J.~J.}\ \bibnamefont
  {Liberko}}, \bibinfo {author} {\bibfnamefont {J.~A.}\ \bibnamefont {Busche}},
  \bibinfo {author} {\bibfnamefont {R.}~\bibnamefont {Seils}}, \bibinfo
  {author} {\bibfnamefont {H.~A.}\ \bibnamefont {Bechtel}}, \bibinfo {author}
  {\bibfnamefont {P.~D.}\ \bibnamefont {Rack}}, \bibinfo {author}
  {\bibfnamefont {D.~J.}\ \bibnamefont {Masiello}},\ and\ \bibinfo {author}
  {\bibfnamefont {J.~P.}\ \bibnamefont {Camden}},\ }\bibfield  {title}
  {\bibinfo {title} {Probing nanoparticle substrate interactions with
  synchrotron infrared nanospectroscopy: {Coupling} gold nanorod
  {Fabry}-{P}\'{e}rot resonances with {$\mathrm{SiO}_2$} and {$h\mathrm{-BN}$}
  phonons},\ }\href@noop {} {\bibfield  {journal} {\bibinfo  {journal} {Phys.
  Rev. B}\ }\textbf {\bibinfo {volume} {104}},\ \bibinfo {pages} {035412}
  (\bibinfo {year} {2021})}\BibitemShut {NoStop}%
\bibitem [{\citenamefont {Pellegrini}\ \emph {et~al.}(2007)\citenamefont
  {Pellegrini}, \citenamefont {Mattei}, \citenamefont {Bello},\ and\
  \citenamefont {Mazzoldi}}]{pellegrini2007interacting}%
  \BibitemOpen
  \bibfield  {author} {\bibinfo {author} {\bibfnamefont {G.}~\bibnamefont
  {Pellegrini}}, \bibinfo {author} {\bibfnamefont {G.}~\bibnamefont {Mattei}},
  \bibinfo {author} {\bibfnamefont {V.}~\bibnamefont {Bello}},\ and\ \bibinfo
  {author} {\bibfnamefont {P.}~\bibnamefont {Mazzoldi}},\ }\bibfield  {title}
  {\bibinfo {title} {Interacting metal nanoparticles: Optical properties from
  nanoparticle dimers to core-satellite systems},\ }\href@noop {} {\bibfield
  {journal} {\bibinfo  {journal} {Mater. Sci. Eng. C}\ }\textbf {\bibinfo
  {volume} {27}},\ \bibinfo {pages} {1347} (\bibinfo {year}
  {2007})}\BibitemShut {NoStop}%
\bibitem [{\citenamefont {Olafsson}\ \emph {et~al.}(2020)\citenamefont
  {Olafsson}, \citenamefont {Busche}, \citenamefont {Araujo}, \citenamefont
  {Maiti}, \citenamefont {Idrobo}, \citenamefont {Gamelin}, \citenamefont
  {Masiello},\ and\ \citenamefont {Camden}}]{olafsson2020electron}%
  \BibitemOpen
  \bibfield  {author} {\bibinfo {author} {\bibfnamefont {A.}~\bibnamefont
  {Olafsson}}, \bibinfo {author} {\bibfnamefont {J.~A.}\ \bibnamefont
  {Busche}}, \bibinfo {author} {\bibfnamefont {J.~J.}\ \bibnamefont {Araujo}},
  \bibinfo {author} {\bibfnamefont {A.}~\bibnamefont {Maiti}}, \bibinfo
  {author} {\bibfnamefont {J.~C.}\ \bibnamefont {Idrobo}}, \bibinfo {author}
  {\bibfnamefont {D.~R.}\ \bibnamefont {Gamelin}}, \bibinfo {author}
  {\bibfnamefont {D.~J.}\ \bibnamefont {Masiello}},\ and\ \bibinfo {author}
  {\bibfnamefont {J.~P.}\ \bibnamefont {Camden}},\ }\bibfield  {title}
  {\bibinfo {title} {Electron beam infrared nano-ellipsometry of individual
  indium tin oxide nanocrystals},\ }\href@noop {} {\bibfield  {journal}
  {\bibinfo  {journal} {Nano Lett.}\ }\textbf {\bibinfo {volume} {20}},\
  \bibinfo {pages} {7987} (\bibinfo {year} {2020})}\BibitemShut {NoStop}%
\end{thebibliography}

\end{document}


\maketitle

\begin{figure}[H] 
\centering
\includegraphics[width = 0.9\textwidth]{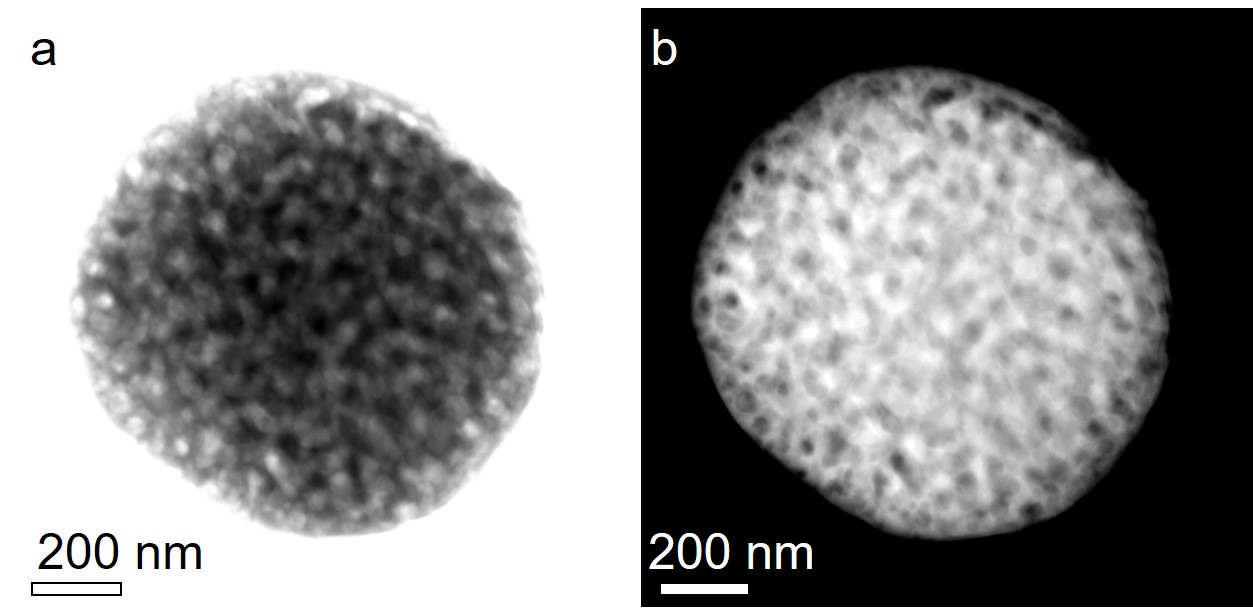}
\caption{
\textbf{Scanning transmission electron microscopy (STEM) images.} LiNbO$_3$ particle images in (a) bright-field and (b) dark-field modes.
\label{fig:figS1}
}
\end{figure}

\begin{figure}[H] 
\centering
\includegraphics[width = 0.5\textwidth]{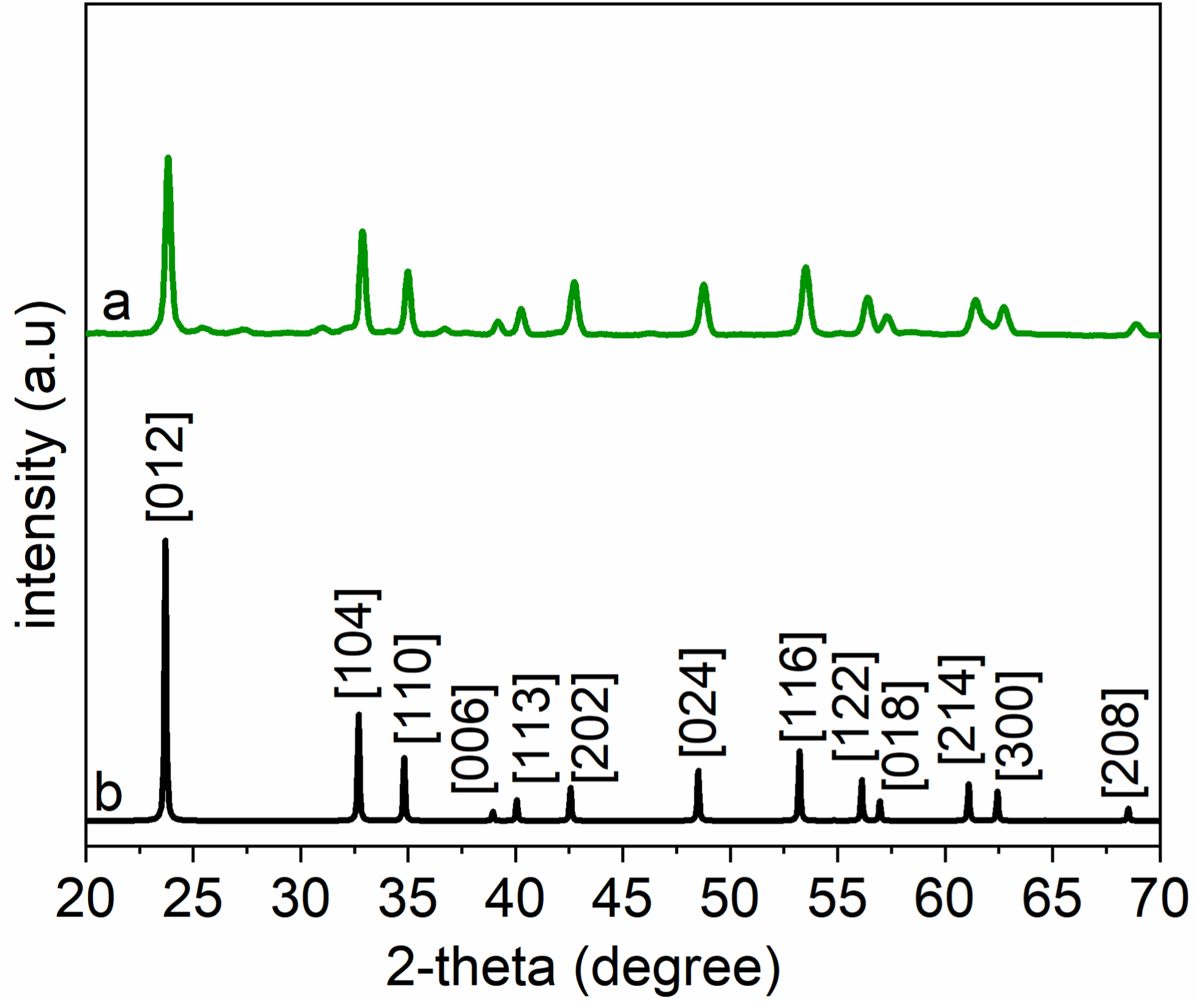}
\caption{
\textbf{Powder X-ray diffraction patterns.} (a) LiNbO$_3$ particles; and (b) a reported LiNbO$_3$ reference (JCPDS No. 020-0631).
\label{fig:figS2}
}
\end{figure}

\begin{figure}[H] 
\centering
\includegraphics[width = 0.5\textwidth]{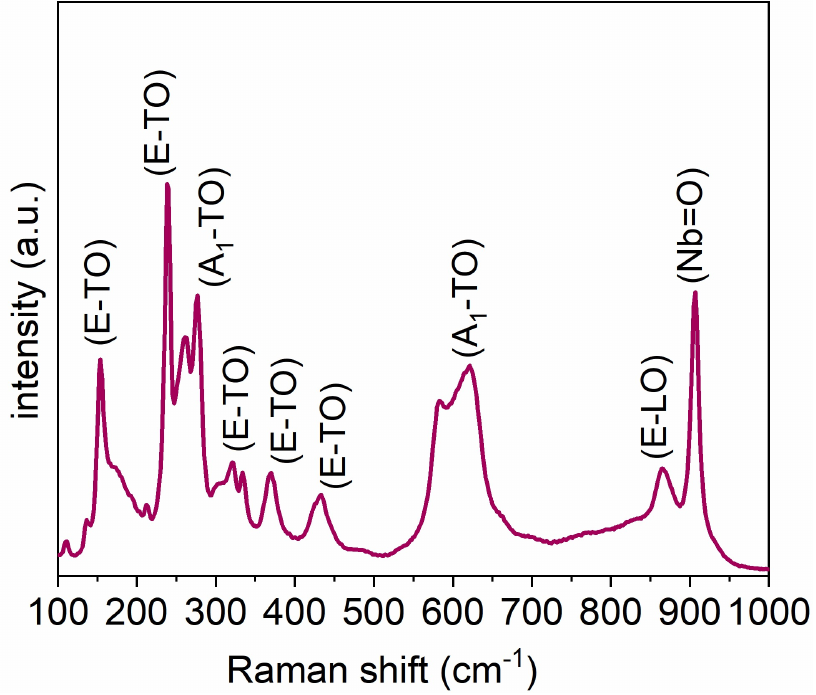}
\caption{
\textbf{Typical Raman spectrum for LiNbO$_{\bm{3}}$ particles.} The spectra indicated the formation of a pure rhombohedral phase in the products.
\label{fig:figS3}
}
\end{figure}

\begin{figure}[H] 
\centering
\includegraphics[width = 0.9\textwidth]{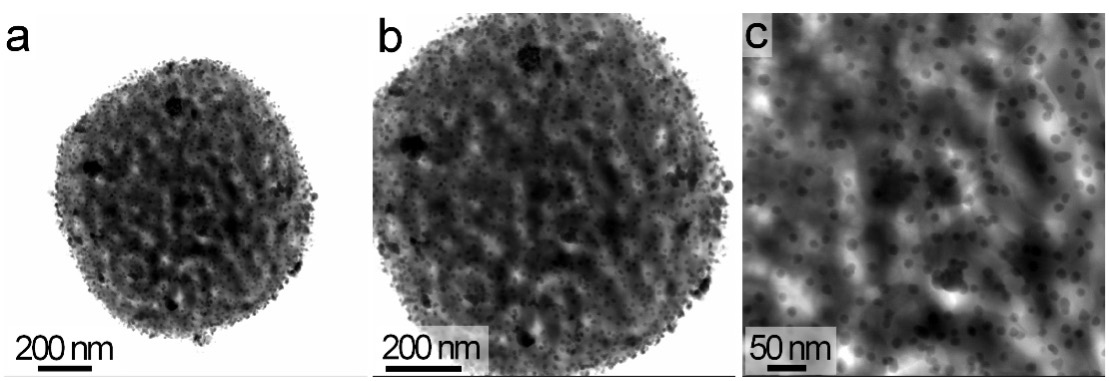}
\caption{
\textbf{STEM images of hybrid particles.} Assemblies of Au-LiNbO$_3$ hybrid particles as characterized by STEM operating in a bright-field mode.
\label{fig:figS4}
}
\end{figure}

\begin{figure}[H] 
\centering
\includegraphics[width = 0.6\textwidth]{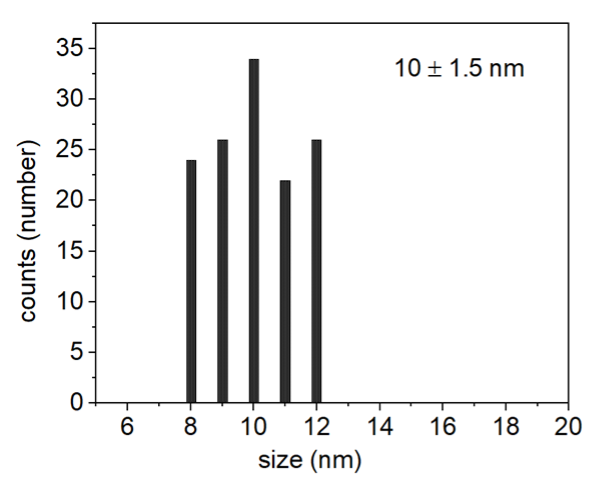}
\caption{
\textbf{Analysis of Au NP sizes.} Histogram depicting the size distribution of the diameters of the gold nanoparticles (NPs) present on the surfaces of the hybrid Au-LiNbO$_3$ nanostructures. This analysis included measurements obtained from 125 independent Au NPs. The variance of 1.5 nm is reported as one standard deviation from the calculated mean of 10 nm.
\label{fig:figS5}
}
\end{figure}

\begin{figure}[H] 
\centering
\includegraphics[width = 0.5\textwidth]{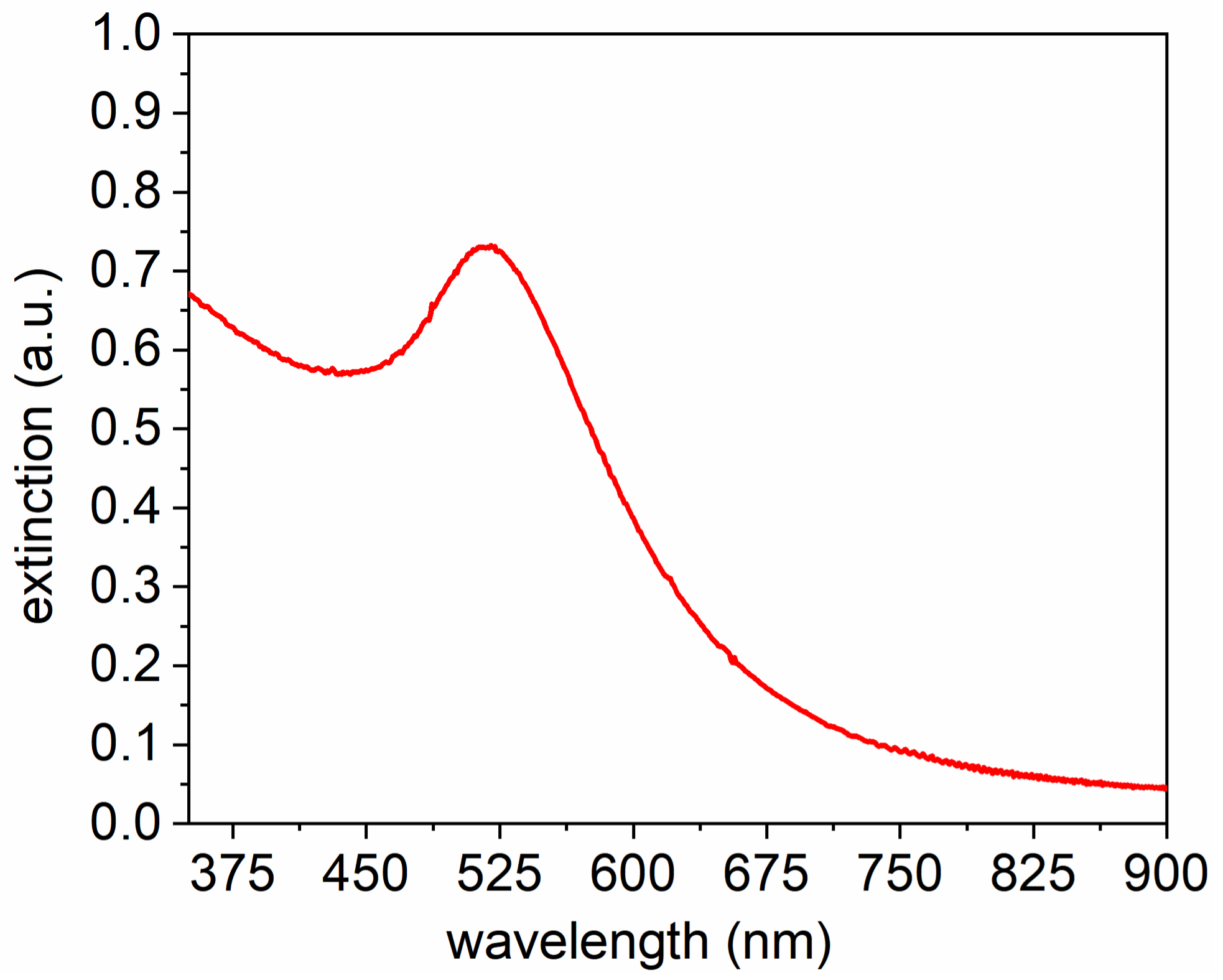}
\caption{
\textbf{Au NP absorbance.} Ultraviolet (UV)-visible absorbance spectrum of $\sim$10-nm diameter Au NPs suspended in an aqueous solution. This spectrum indicates the plasmonic band for the nanoparticles was centered at $\sim$520 nm, which is close to the 530 nm plasmonic band for hybrid Au-LiNbO$_3$ hybrid particles.
\label{fig:figS6}
}
\end{figure}

\begin{figure}[H] 
\centering
\includegraphics[width = 0.5\textwidth]{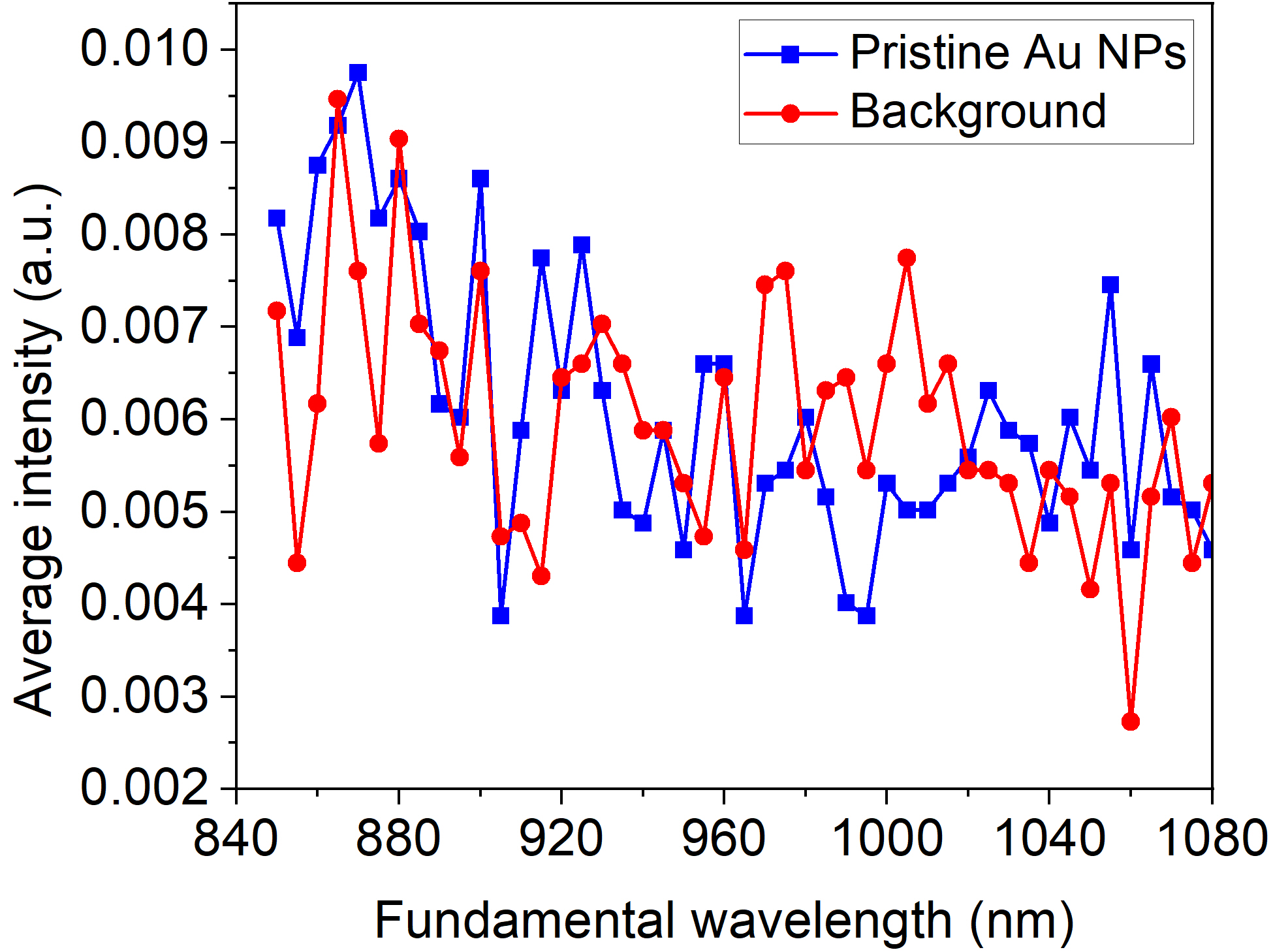}
\caption{
\textbf{Minimal SHG from Au NPs.} The SHG analyses for pristine $\sim$10-nm diameter gold nanoparticles recorded by sweeping the excitation laser wavelength from 850 to 1,080 nm with a step size of 5 nm. These results indicate a lack of SHG response in the centrosymmetric Au NPs.
\label{fig:figS7}
}
\end{figure}

\begin{figure}[H] 
\centering
\includegraphics[width = 0.9\textwidth]{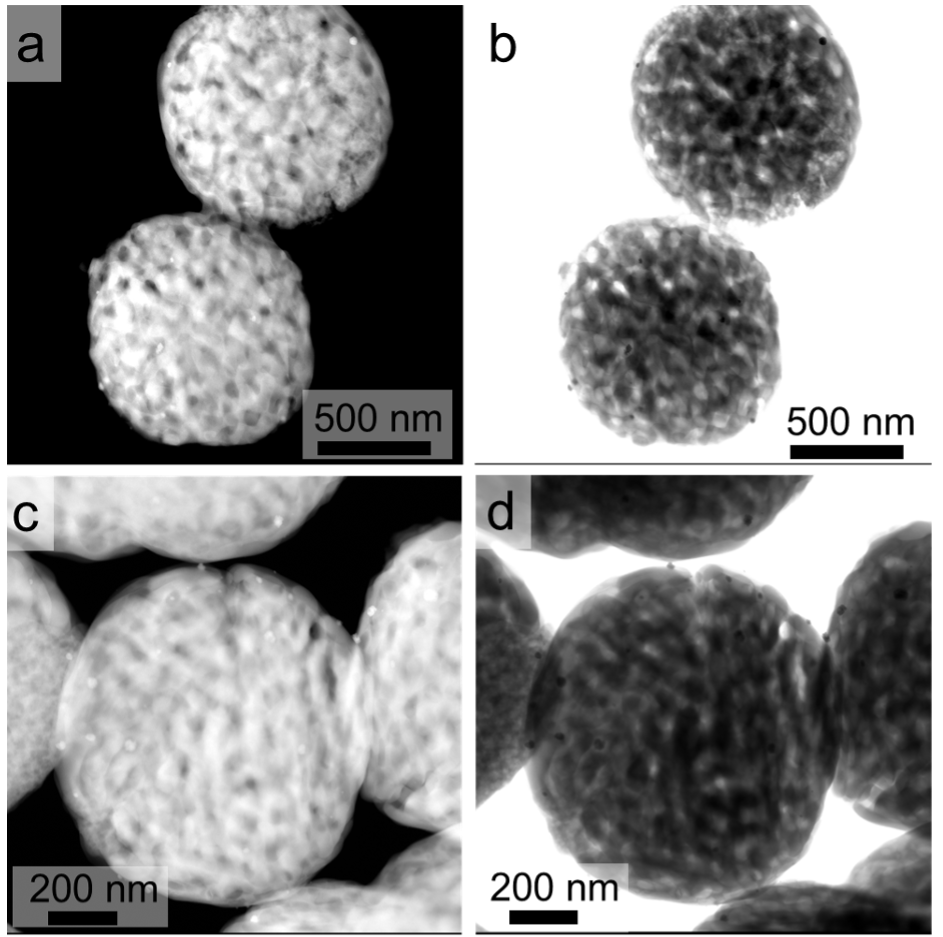}
\caption{
\textbf{Images of low NP-loading hybrid nanostructures.} Hybrid Au-LiNbO$_3$ particles prepared using 0.1 mL of a 5 mM HAuCl$_4$ solution to perform the \textit{in situ} synthesis of the Au NPs. The resulting products were characterized by scanning transmission electron microscopy (STEM) operating in: (a), (c) a high-angle annular dark-field (HAADF) mode; and (b), (d) a bright-field mode.
\label{fig:figS8}
}
\end{figure}

\begin{figure}[H] 
\centering
\includegraphics[width = 0.9\textwidth]{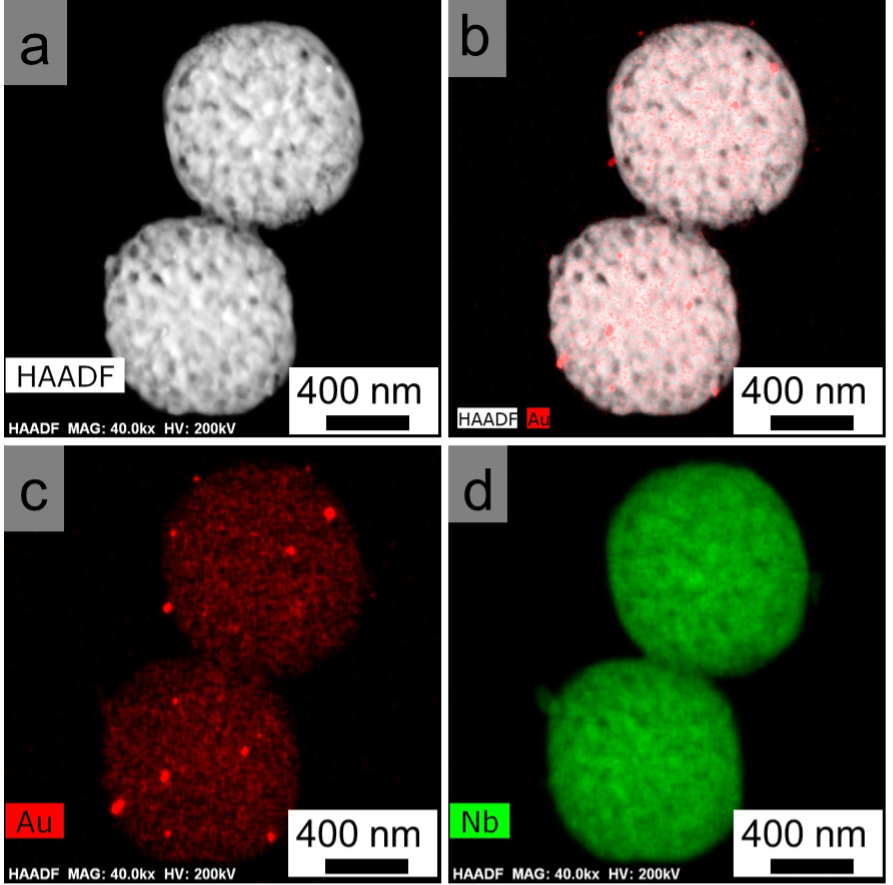}
\caption{
\textbf{Elemental analysis of low NP-loading hybrid nanostructures.} Energy dispersive X-ray spectroscopy (EDS) analysis of the hybrid Au-LiNbO$_3$ particles prepared using 0.1 mL of 5 mM HAuCl$_4$. These images show (a) a HAADF image obtained by STEM techniques, (b) an overlay of the Au NPs (as detected by EDS) on the HAADF image of the assemblies, (c) an EDS map of the Au NPs, and (d) an EDS map of Nb within these assemblies.
\label{fig:figS9}
}
\end{figure}

\begin{figure}[H] 
\centering
\includegraphics[width = 0.45\textwidth]{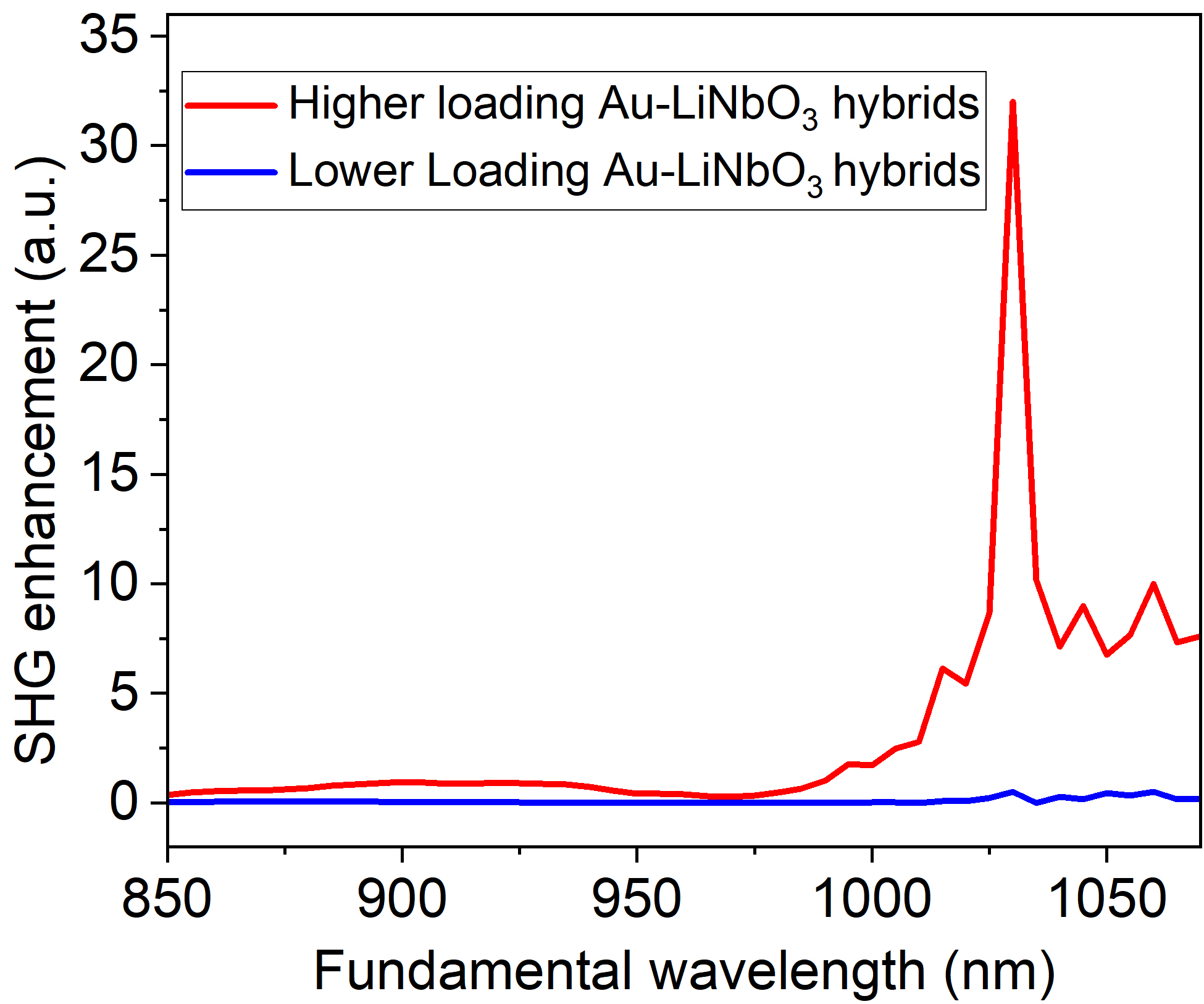}
\caption{
\textbf{SHG ehancement spectra comparison.} Evaluation of the enhancement in the SHG signal as a function of the fundamental wavelength (i.e., ranging from 850 to 1,070 nm) of the incident laser for the individual hybrid Au-LiNbO$_3$ particles prepared with a lower loading of Au NPs. After normalization of the SHG output of the individual Au-LiNbO$_3$ hybrids to the SHG output of the bare LiNbO$_3$, enhancement factors were plotted against the FWs of the measurement.
\label{fig:figS10}
}
\end{figure}

\begin{figure}[th!] 
\centering
\includegraphics[scale = 1.0]{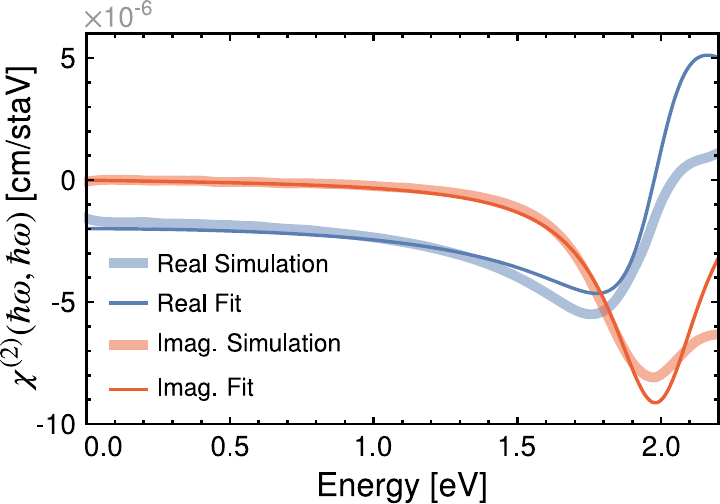}
\caption{
\textbf{LiNbO$_{\bm{3}}$ second-order susceptibility fit.} Comparison of the real and imaginary parts of the magnitude $\chi^{(2)}(\omega,\omega) \equiv \hat{\mathbf{n}}\cdot\bm{\chi}_2^{(2)}(\omega,\omega)\cdot\hat{\mathbf{n}}$ of the second-order susceptibility in the limit where the output frequency is twice the input frequency ($\hat{\mathbf{n}}$ is any real unit vector).
\label{fig:figS11}
}
\end{figure}

This PDF file includes:
\begin{itemize}
    \item Section \ref{sec:methodsSupp}: Materials and Methods provides brief details of the experimental and theoretical procedures used to generate the observables in the main text
    \item Section \ref{sec:moreTheory}: Additional Theoretical Details contains a long-form discussion of the construction of the near-field coupling and superradiance models
    \begin{itemize}
        \item Subsection \ref{sec:materialModels} details the material models used in this investigation
        \item Subsection \ref{sec:paramInf} provides derivations of the oscillator models of the bare Mie and plasmon resonances
        \item Subsection \ref{sec:scattering} provides all scattering calculations for the main text
        \item Subsection \ref{sec:nearField} shows the derivation of the coupled-oscillator model of the hybrid nanostructures from first principles
    \end{itemize}
    \item Section \ref{sec:moreFigures}: Additional Figures and Tables contains supplemental figures and tables for reference
    \begin{itemize}
        \item Figures \ref{fig:figS1} to \ref{fig:figS16}
        \item Tables \ref{tab:tabS1} to \ref{tab:tabS4}
    \end{itemize}
\end{itemize}

\section{Materials and Methods}\label{sec:methodsSupp}

\subsection{Experimental Details}

\subsubsection{Synthesis of Mesoporous Monodisperse Lithium Niobtate Particles}

All chemicals were used without further purification. Monodisperse lithium niobate particles were synthesized using our previously reported method. In brief, 0.8 mM of niobium n-butoxide [Nb(OBu)$_5$, 99\%, Alfa Aesar] was dissolved in 4 mL of ethanol. This solution was aged for 36 h in a desiccator in the presence of an open glass vial containing 25 mL of water to generate a humid atmosphere. The resulting gel-like precursor was mixed with 20 mL of an aqueous solution of 0.1 M lithium hydroxide monohydrate (LiOH$\cdot$H$_2$O, 99\%, Alfa Aesar) and sonicated for 20 min. A 10 mL aliquot of the resulting suspension was transferred to a 23 mL Teflon-lined autoclave (Model No. 4749, Parr Instruments Co., Moline, IL USA) and heated at 200 $^\circ$C for 48 h. After cooling to room temperature, white precipitates were isolated from the solution via a process of centrifugation (Model No. AccuSpin 400, Fisher Scientific) at 8,000 rpm for 15 min and decanting of the solution. These solids were washed by re-suspending them in 10 mL deionized water (18 M$\Omega\cdot$cm, produced using a Barnstead NANOpure DIamond water filtration system). This purification process was repeated for a total of three times. The purified product was dried at 70 $^\circ$C for 10 h to remove residual water prior to further analyses. The dried precipitates were calcined by heating from room temperature to 600 $^\circ$C at a rate of 5 $^\circ$C/min and held at 600 $^\circ$C for 45 min to induce complete crystallization.

\subsubsection{Synthesis of Gold-Lithium Niobate Hybrid Nanostructures}

An aqueous suspension of porous, monodisperse LiNbO$_3$ particles (0.5 mg/mL) was added and dispersed in 5 mL of water via sonication. Into this suspension, either 1 mL (for a higher loading of gold nanoparticles) or 0.1 mL (for a lower laoding of gold nanoparticles) of a 5 mM aqueous solution of gold(III) chloride trihydrate (HAuCl$_4\cdot3$H$_2$O, 99.9\%, Sigma Aldrich) was added, and the mixture stirred at 70 $^\circ$C for 3 h. This step was followed by the addition of 1 mL or 0.01 mL of 0.5\% (w/v) trisodium citrate dihydrate (C$_6$H$_5$Na$_3$O$_7\cdot2$H$_2$O, Sigma Aldrich, $\geq$99\%) aqueous solution into the reaction mixture. The reaction mixture was stirred further at 70 $^\circ$C for 1 h. The precipitates were washed two times with water. The as-synthesized hybrid nanostructures of LiNbO$_3$ with gold (Au) nanoparticles (NPs) were isolated from the unreacted excess gold salts and unbound Au NPs through centrifugation (Thermo Electron Corporation, IEC microlite microcentrifuge) at 2,000 rpm for 3 min, decanting of the supernatants, and re-dispersion of the isolated solids in DI water with the assistance of a vortexer for 3 min.

The morphology and dimensions of the LiNbO$_3$ particles were characterized using an FEI Osiris X-FEG 8 scanning/transmission electron microscope (TEM/STEM) operated at an accelerating voltage of 200 kV. Samples for TEM/STEM analyses were prepared by dispersing the purified products in ethanol followed by drop-casting 5 $\mu$L of each suspension onto separate TEM grids (300 mesh copper grids coated with Formvar/carbon) purchased from Cedarlane Labs. Each TEM grid was dried at $\sim$230 Torr for at least 20 min prior to analysis. Energy dispersive X-ray spectroscopy (EDS) analyses were performed using the FEI Osiris scanning/TEM, which was equipped with a Super-X EDS system with ChemiSTEM Technology integrating the signal from four spectrometers.

Purity, crystallinity, and phase of the LiNbO$_3$ particles were characterized using Raman spectroscopy and powder X-ray diffraction (XRD) techniques. Raman spectra were collected using a Renishaw inVia Raman microscope with a 50$\times$ SWD objective lens (Leica, 0.5 NA), and a 514 nm laser (argon-ion laser, Model No. Stellar-Pro 514/50) set to 100\% laser power with an exposure time of 30 s. The Raman spectrometer was calibrated by collecting the Raman spectrum of a polished silicon (Si) standard with a distinct peak centred at 520 cm$^{-1}$. The Raman spectra for the samples were acquired from 100 to 1,000 cm$^{-1}$ using a grating with 1,800 lines/mm. The XRD patterns of the samples were acquired with a Rigaku R-Axis Rapid diffractometer equipped with a 3 kW sealed tube copper source (K$\alpha$ radiation, $\lambda$ = 0.15418 nm) collimated to 0.5 mm. The samples were packed into a cylindrical recess drilled into glass microscope slides (Leica 1 mm Surgipath Snowcoat X-tra Micro Slides) for acquiring XRD patterns for the products.

The optical absorption spectra of the products were measured using an Agilent Technologies ultraviolet-visible spectrophotometer (Agilent 8453, Model No. G1103). For these measurements, the samples were suspended in water and held in 1 cm path length poly(methyl methacrylate) cuvettes (VWR$\mathrm{TM}$, Catalog No. 634-8537). The linear spectra of individual pristine LiNbO$_3$ and hybrid Au-LiNbO$_3$ particles were measured using a Zeiss M1m optical microscope operating in a dark-field spectroscopy setup. The light from a halogen lamp was focused onto the sample through a dark field objective (50$\times$, Zeiss Epiplan Neofluar). The scattered light was collected by the same objective and detected by an imaging spectrometer (Princeton Instrument Acton spectrometer equipped with a PIXIS 400 CCD detector cooled to ‒72°C). The linear scattering cross section spectra were obtained by normalizing the signal from a single particle to the signal coming directly from the lamp, after subtraction of the background in proximity to the particle.

The second harmonic generation (SHG) activity of the individual pristine LiNbO$_3$ and hybrid Au-LiNbO$_3$ particles was assessed using a Leica SP5 laser scanning confocal two photon microscope equipped with a Coherent Chameleon Vision II laser and a Zeiss LSM 510 MP confocal microscope. A dilute dispersion of nanostructures was drop-cast onto glass coverslips and brought into the focal point of the microscope. The SHG response was characterized by locating individual particles using a 63$\times$ oil immersion objective aperture and scanning the laser excitation from 850 nm to 1,070 nm as the fundamental wavelength.

\subsection{Theoretical Calculations}

\subsubsection{Bare Lithium Niobate Sphere Characterization}

Modeling the microsphere as a smooth, isotropic spherical particle, the nearly constant dielectric function of LiNbO$_3$ in the optical region of interest \cite{palik1998handbook} allows the microsphere's linear response near the SH to be described by the excitation of a set of 21 Mie resonances (SI, Section \ref{sec:suscFit}). We have established $\bm{\beta} = \{T',p',\ell',m'\}$ as a collective index to describe these modes such that the scattered electric field of each mode can be expanded as $\bm{\mathcal{E}}_\mathrm{sca}(\mathbf{r},\omega) = \sum_{\bm{\beta}}a_2^{-\ell - 2} \left[ \rho_{\bm{\beta}}(\omega) \mathbf{X}_{\bm{\beta}}(\mathbf{r},k_{\bm{\beta}}) + \rho_{\bm{\beta}}^*(-\omega) \mathbf{X}^*_{\bm{\beta}}(\mathbf{r},k_{\bm{\beta}}) \right]$ \cite{lalanne2018light}.

Using $\bm{\mathcal{E}}_\mathrm{sca}$ along with the magnetic fields of each mode, $\bm{\mathcal{B}}_\mathrm{sca}(\mathbf{r},\omega) = (c/i\omega)\nabla\times\bm{\mathcal{E}}_\mathrm{sca}(\mathbf{r},\omega)$, the scattered power averaged over a period of the SH frequency  can be calculated from Poynting's theorem as (Section \ref{sec:mieScatt}):
\begin{equation}\label{eq:P2}
P^{(1)}(2\omega_0) = \frac{c^3E_0^2}{4\pi\omega_0}\sum_{\bm{\beta}}\frac{1 - \delta_{p1}\delta_{m0}}{\omega_{\bm{\beta}}a_2^{2\ell + 4}}
|C_{\bm{\beta}}(\omega_0)|^2|\alpha_{\bm{\beta}}(2\omega_0)|^2.
\end{equation}
Here, $\alpha_{\bm{\beta}}(\omega) = (a_2^{\ell + 2} f_{\bm{\beta}}/ 2\omega_{\bm{\beta}})\exp(i\psi_{\bm{\beta}})/(\omega_{\bm{\beta}} - i\gamma_{\bm{\beta}}/2 - \omega)$ is the linear polarizability of the mode $\bm{\beta}$, with $C_{\bm{\beta}}(\omega_0)\sim E_0\chi_2^{(2)}(\omega_0,\omega_0)$ a unitless overlap coefficient. In general, $C_{\bm{\beta}}(\omega_0)$ quantifies the efficiency of photon upconversion between the fundamental resonances of the system at $\omega_0$ and a resonance $\bm{\beta}$ at the SH, and thus is large only if the fundamental resonances are driven strongly by the incident light. Also, $C_{\bm{\beta}}(\omega_0)$ is zero unless the fundamental and SH resonances satisfy specific symmetry requirements, which are detailed below. Finally, the explicit form of this overlap coefficient is complicated, but described in full in Section \ref{sec:oscModelConstr}, Eq. \eqref{eq:Cbeta}.

In general, the resonance frequencies of the Mie modes are independent of $p$ and $m$, although these indices do determine the strength of the response of a mode to a driving source of a given spatial symmetry. In detail, if we allow $\bm{\alpha} = \{T,p,\ell,m\}$ to be the collective index of one fundamental resonance of the LiNbO$_3$ particle and $\bm{\alpha}' = \{T'',p'',\ell'',m''\}$ to be the index of another, in the case where the sphere has a weak, isotropic, and frequency-dependent nonlinear susceptibility $\bm{\chi}_2^{(2)}(\omega',\omega - \omega') = \bm{1}_3\chi_2^{(2)}(\omega',\omega - \omega')$ only modes $\bm{\beta}$ for which $p + p' + p''$ is even and $m' \pm m \pm m'' = 0$ are driven by SHG. For example, with the microsphere driven by an $x$-polarized plane wave at the fundamental frequency such that $\bm{\alpha} = \{E,0,\ell,1\}$ or $\{M,1,\ell,1\}$, each of the 21 SH modes in the model has an index pair $(p',m') = (0,0)$, $(0,2)$, or $(1,2)$. In contrast, the dependence of a mode's response on $T'$ and $\ell'$ is determined by its spectral overlap with the source. In the energy window of the observed SHG enhancement, seven sets of modes can be significantly driven by the upconversion process: four sets of electric modes with $\ell' = 6$, 7, 9, and 10, and three sets of magnetic modes with $\ell' = 7$, 10, and 11. See Figure \ref{fig:figS13} for details.

\subsubsection{Characterization of the Lithium Niobate Radius and Dielectric Function}\label{sec:dielFit}

Section \ref{sec:paramInf} details the characterization of each of the LiNbO$_3$ Mie modes, each of which is assigned a resonance frequency $\omega_{\bm{\beta}}$, a damping rate $\gamma_{\bm{\beta}}$, and an oscillator strength $f_{\bm{\beta}}$ (equivalently, an effective mass $\mu_{\bm{\beta}} = e^2/a_2^3f_{\bm{\beta}}$) that determine its spectral position and response magnitude to external stimuli. This characterization is performed using a dielectric function $\epsilon_2 = 5.5 + 0.035i$ and a radius $a_2 = 500$ nm estimated from the experiment. The real part of $\epsilon_2$ is taken from estimates extracted from a set of nine single-particle scattering experiments conducted on bare LiNbO$_3$ spheres of radii 350 to 1000 nm. Figure \ref{fig:fig3}c shows results from one representative scattering experiment, with the remainder of the results given Figure \ref{fig:figS17} and in Table \ref{tab:tabS3}. Further, in conjunction with the choice of radius, the choice $\mathrm{Re}\{\epsilon_2\} = 5.5$ provides the best overlap between the spectral locations of the Mie resonances and the SHG enhancement peaks seen in Figure \ref{fig:fig4}c, as well as the best agreement between the relative peak heights.  We note that small ($\sim5\%$) increases (decreases) to $\mathrm{Re}\{\epsilon_2\}$ and $a_2$ lead to red (blue) shifts in the Mie resonance positions that can nullify each other, such appropriate choices with the ranges $a_2\pm20$ nm and $\mathrm{Re}\{\epsilon_2\}\pm0.1$ are likely to lead to similar results. The inset of Figure \ref{fig:fig4}c shows the range of Mie resonance energies available with $\pm4\%$ changes to $a_2$, and analogous results when varying $\mathrm{Re}\{\epsilon_2\}$ are shown in Figure \ref{fig:figS14}c.

The imaginary part of $\epsilon_2$, which is expected to be small \cite{palik1998handbook}, is chosen to agree well with the linewidths of the modes observed in the second harmonic data of Figure \ref{fig:fig4}c. The exact value has yet to be characterized via e.g. ellipsometry \cite{olafsson2020electron} and is difficult to estimate from the scattering data due to substrate effects (see Figure \ref{fig:figS13}a). The linewidths of the Mie resonances depend sensitively on the rate of internal material losses (see Figure \ref{fig:figS14}d) and the heights of the narrow Mie modes in both scattering and SHG enhancement spectra can vary noticeably with $\sim5\%$ changes to $\mathrm{Im}\{\epsilon_2\}$. However, small changes to $\mathrm{Im}\{\epsilon_2\}$ do not modify the peak positions within the SH enhancement spectrum such that we take $\mathrm{Im}\{\epsilon_2\}$ to be a fitting parameter that affects the overall magnitude of the theoretical SHG enhancements, but does not qualitatively affect the results. Further details are given in Section \ref{sec:materialModels}.

\subsubsection{Superradiant Nanoparticle Scattering}

Each NP is assumed to be spherical and isotropic with a Drude-Lorentz dielectric function $\epsilon_1(\omega)$ fit to the Au dielectric data of Ref. \citenum{olmon2012optical} (Section \ref{sec:materialModels}, Figure \ref{fig:figS16}, and Table \ref{tab:tabS4}) and an associated set of surface plasmon resonances within the spectral window of the observed SH emission. Figure \ref{fig:fig3}a demonstrates the excellent agreement between the resulting theoretical absorption cross section of the NP dipole plasmons using this model and extinction measurements performed on single hybrid nanostructures, highlighting the validity of the model and the uniformity of the NP responses in the experiment. We thus restrict our analysis to include only the dipole resonances of the NPs, which we label $\mathbf{d}_i(\omega)$, and allow each particle to lie at a position $\mathbf{r}_i$ on the surface of the LiNbO$_3$ microsphere.

The dynamics of the plasmon dipoles are inferred from numerical scattering calculations using both multiple-Mie scattering and boundary element method Maxwell solvers \cite{pellegrini2007interacting,hohenester2012mnpbem}. These simulations were performed on arrays of 10-nm diameter Au NPs in square grids with a 20 nm center-to-center spacing, as complete layers on spherical templates containing the $N\approx1000$ Au NPs present in the experiment are too large to compute directly. Moreover, grids reliably approximate the effects of interparticle coupling in the limit that the radii of the Au NPs and their separations are much smaller than the radius of the LiNbO$_3$ microsphere. Using these scaled down simulations, we extrapolated from smaller ensembles ($N\leq225$) the behaviors of larger collections of Au NPs. For comparative reasons, we used both regular grids and randomized grids of Au NPs to evaluate the effects of disorder on the amount of light scattered by each particle.

Our results (Figure \ref{fig:figS15}a) show that the maximum scattering cross section of each dipole grows roughly two orders of magnitude from $7.75\times10^{-3}$ nm$^2$ as $N$ varies from 1 to 1000, and that the exact magnitude of the growth depends on whether the dipole is oriented perpendicular (maximum of 0.355 nm$^2$) or parallel (1.18 nm$^2$) to the plane of the ensemble. Conversely, the absorption cross section maxima of each dipole converge quickly with ensemble size, as shown in Figure \ref{fig:figS15}a, moving from a single particle value of 13.5 nm$^2$ to perpendicular and the parallel values of 10.3 nm$^2$ and 15.7 nm$^2$, respectively, for ensembles of $N>50$ NPs. In addition, while randomness in the positions of the particles can modulate the per-particle scattering cross section maxima by $\pm20\%$ in comparison to a regular grid (Figure \ref{fig:figS14}b), an extreme value distribution fit to the recorded scattering cross section maxima of 100 simulations with randomized grids of $N = 16$ NPs shows that modifications beyond this range occur in $\sim$0.4\% of NPs. 

The regular-grid simulations are, thus, taken to be representative of the ensemble as a whole. Further, Figure \ref{fig:figS15}b shows that while the peak maxima of the scattering observables depend strongly on $N$, their lineshapes do not. Therefore, with the NPs' scattering strongly modified and their absorption changed relatively little with increasing $N$, we conclude that the dipole oscillator strengths are best estimated from the absorption data. Explicitly, each dipole is then treated quasistatically with a position- and orientation-dependent polarizability (see Section \ref{sec:ensembleAuInf}), while the ensemble-induced scattering enhancements are included phenomenologically through modification of the Rayleigh scattering rate of each dipole by a positive factor $A(N)$. When driven by the scattered fields of a bare LiNbO$_3$ sphere, the ensemble then radiates a time-averaged scattered power
\begin{equation}\label{eq:P1}
P_\mathrm{pl}(2\omega_0) \approx A(N)\frac{\gamma_1^\mathrm{rad}f_1a_1^3E_0^2}{2\gamma_1^2}\sum_{\bm{\beta}}\frac{K_{\bm{\beta}}}{a_2^{2\ell + 4}}\left|C_{\bm{\beta}}(\omega_0)\right|^2\left|\alpha_{\bm{\beta}}(2\omega_0)\right|^2.
\end{equation}
Here, $f_1$ is the orientation-averaged oscillator strength of the NP dipoles (Eq. [\ref{eq:f1Standard}]), and $K_{\bm{\beta}}$ (Eq. [\ref{eq:Kbeta}]) is an overlap coefficient that determines the strength of the ensemble's response to the LiNbO$_3$ fields.

See Section \ref{sec:scattering} for further details.

\subsubsection{SHG Enhancement via Near-Field Coupling}

The first and second-order terms in the solution of the Mie resonance equations of motion in Eq. \eqref{eq:EOMmain} are
\begin{equation}
\begin{split}
\rho_{\bm{\beta}}^{(1)}(\omega) &= \frac{f_{\bm{\beta}}a_2^{\ell + 2}(\Omega_{\bm{\beta}}^* + \omega)\mathrm{e}^{-i\psi_{\bm{\beta}}}}{2\omega_{\bm{\beta}}e\left(|\Omega_{\bm{\beta}}|^2 - \omega^2 - i\omega\gamma_{\bm{\beta}}\right)}F_{2\bm{\beta}}(\omega),\\
\rho_{\bm{\beta}}^{(2)}(\omega) &= \frac{f_{\bm{\beta}}a_2^{\ell + 2}(\Omega_{\bm{\beta}}^* + \omega)\mathrm{e}^{-i\psi_{\bm{\beta}}}}{2\omega_{\bm{\beta}}e^2\left(|\Omega_{\bm{\beta}}|^2 - \omega^2 - i\omega\gamma_{\bm{\beta}}\right)}\\
&\times\frac{ef_\nu a_1^3\eta_1(\omega)}{2\omega_1e^2\left(|\Omega_1|^2 - \omega^2 - i\omega\gamma_1\right)}\sum_\nu\sigma_{\bm{\beta}\nu}(\mathbf{r}_0)F_{1\nu}(\mathbf{r}_0,\omega),
\end{split}
\end{equation}
which detail the driving of each mode $\bm{\beta}$ through direct upconversion and through excitation of the NP by the upconversion process, respectively. Higher-order terms build in the consequences of multiple transfers of energy between the microsphere and the NP and become increasingly tedious to write explicitly, but can be analyzed compactly with
\begin{equation}
\begin{split}
\rho_{\bm{\beta}}^{(n > 2)}(\omega) &= \sum_{\nu}\sum_{\bm{\beta}'\neq\bm{\beta}}\frac{f_{\bm{\beta}}f_\nu a_2^{\ell + 2} a_1^3}{4e^4\omega_{\bm{\beta}}\omega_1}\frac{\eta_1(\omega)\eta_{\bm{\beta}}(\omega)}{\left(|\Omega_{\bm{\beta}}|^2 - \omega^2 - i\omega\gamma_{\bm{\beta}}\right)\left(|\Omega_1|^2 - \omega^2 - i\omega\gamma_1\right)}\\
&\times \frac{\sigma_{\bm{\beta}\nu}(\mathbf{r}_0)}{a_2^{\ell' - 1}}\left[\sigma_{\bm{\beta}'\nu}(\mathbf{r}_0)\rho_{\bm{\beta}'}^{(n - 2)}(\omega) +\sigma_{\bm{\beta}'\nu}^*(\mathbf{r}_0)\rho_{\bm{\beta}'}^{(n - 2)*}(-\omega)\right],
\end{split}
\end{equation}
from which it is clear that, with $|\rho_{\bm{\beta}}^{(2)}(\omega)| \ll |\rho_{\bm{\beta}}^{(1)}(\omega)|$, terms beyond the third order provide only small corrections.

To account for all $N$ NPs, we label each NP with the index $i \in[1,N]$ such that $\mathbf{r}_0\to\mathbf{r}_i$ and $d_\nu(\omega) \to d_{i\nu}(\omega)$, but allow the NPs to have identical radii and their plasmons to have identical resonance frequencies, linewidths, phase offsets, and oscillator strengths. Therefore, the coupling term in Eq. \eqref{eq:EOMmain} is replaced by the sum $\sum_{i\nu}\sigma_{\bm{\beta}\nu}(\mathbf{r}_i)d_{i\nu}(\omega)$ and similar sums in each successive term of the perturbation expansion must be adjusted accordingly. In this work, these sums are carried out by placing three dipoles $d_{i\nu}(\omega)$ at $N = \text{1,000}$ points arranged on a sphere of radius $a_1 + a_2$ in a Fibonacci lattice \cite{gonzalez2010measurement}. Calculations with randomized points display only small variations from the results of the regular Fibonacci array in good agreement with the random-ensemble scattering results of Figure \ref{fig:fig3}b and are thus not shown.

\section{Additional Theoretical Details}\label{sec:moreTheory}

\subsection{Construction of the Material Models from Data}\label{sec:materialModels}

In order to maximize the agreement between our theoretical models and the SH enhancement data, we construct a model dielectric function for Au and a dielectric function and second-order susceptibility and LiNbO$_3$ using our own single particle scattering data, as well as from ellipsometry data and atomistic numerical simulations from Refs. \citenum{olmon2012optical} and \citenum{riefer2012linear}, respectively. The Au details are given in Section \ref{sec:dielFit} and the LiNbO$_3$ details are given in Section \ref{sec:suscFit}.

\subsubsection{Fitting of the Au Lorentz-Drude Dielectric Model}\label{sec:dielFitAu}

The oscillator parameters of the dipole plasmons within the Au NPs modeled in this investigation are calculated using a simple Drude-Lorentz model of the dielectric function of gold. This dielectric function is given by
\begin{equation}\label{eq:eps1}
\epsilon_1(\omega) = 1 - \frac{\omega_{p1}^2}{\omega^2 + \mathrm{i}\omega\Gamma_1} + \sum_{i=2}^3\frac{\omega_{pi}^2}{\Lambda_i^2 - \omega^2 - \mathrm{i}\omega\Gamma_i},
\end{equation}
and its form and parameters are inferred from ellipsometry data published in Ref. \citenum{olmon2012optical}. In particular, as shown in Figure \ref{fig:figS16}a, a Drude-model ($\omega_{p2}$, $\omega_{p3}\to0$) fit via nonlinear least squares methods to the gold dielectric function provides a good approximation to the dielectric data for photon energies $\hbar\omega < 2.0$ eV, but this model cannot capture the effects of interband transitions that become increasingly important at higher energies. Indeed, a Drude model is provided by Ref. \citenum{olmon2012optical}, but it is not used in this analysis as it produces inaccurate predictions of the dipole plasmon energies of Au NPs, which exist well above the 2.0 eV threshold.

To produce a more accurate Au NP model in the region between 2.0 and 2.8 eV, the addition of two Lorentz oscillators to the Drude model dielectric function was found to be sufficient. The NP dipole plasmons have resonance energies near 2.5 eV, well below the upper bound of the single-oscillator Drude-Lorentz model's accuracy near 2.9 eV (see Figure \ref{fig:figS16}a). The parameters for the Drude-Lorentz dielectric model were inferred by first fitting a simplified Drude model
\begin{equation}
\epsilon_1(\omega)\approx - \frac{\omega_{p1}^2}{\omega^2 + \Gamma_1^2} + \mathrm{i}\frac{\omega_{p1}^2\Gamma_1}{\omega^3 + \omega\Gamma_1}
\end{equation}
to the real and imaginary parts of single-crystal dielectric data of Ref. \citenum{olmon2012optical} at low energies $\omega\ll\Lambda_2$. These fits produced estimates $\hbar\omega_{p1} = 7.20\pm0.9$ eV and $\hbar\Gamma_1 = 71.1\pm13$ meV, respectively, which were then used as initial guesses for fits using the more robust Drude-Lorentz model. Focusing on the energy window between 0.8 and 2.8 eV, the Drude-Lorentz fits were made using the regularized function $\mathrm{Re}\{\epsilon_1(\omega)\} + \eta(\omega)\mathrm{Im}\{\epsilon_1(\omega)\}$ to simultaneously fit the model to both components of the complex dielectric data while avoiding bias toward the data's much larger real part in the energy window between 0.8 and 2.4 eV. 

Fits using simplex, differential evolution, simulated annealing, and random search fitting methods \cite{Mathematica} produced stable results using weighting functions $\eta(\omega) = \lambda_1\exp(-\lambda_2\omega)$ with $\lambda_1 = 40\pm5$ and $\lambda_2 = (6.5\pm1)\times10^{-16}$ $\mathrm{s}/\mathrm{rad}$. The central values for either factor are used in this work. Taking the average of the results of the four fitting routines, the model parameters used in the following discussion are given in Table \ref{tab:tabS4}.

\subsubsection{Fitting of the Lithium Niobate Dielectric Model and Nonlinear Susceptibility}\label{sec:suscFit}

The radii of the samples examined in Figure \ref{fig:fig4}c have radii of $a_2 \approx 500$ nm, and we use this value in the following analysis. To estimate $\epsilon_2$, nine scattering spectra were collected using bare LiNbO$_3$ particles with radii between $350$ nm and $1000$ nm, with the radii estimated from microscopy analyses with a precision of $\pm50$ nm. Mie theory scattering spectra were compared to the experimental data, but peak splitting and broadening generated by the substrate in the experimental data restricts the utility of the ideal substrate-free model.

More precisely, in agreement with Ref. \citenum{savo2020broadband}, the presence of a substrate in the scattering spectra generates peak broadening that can in most cases be captured to acceptable approximation by a Mie model with phenomenologically added internal damping. However, as is also demonstrated in Ref. \citenum{savo2020broadband}, the substrate can also induce splitting, shifting, and even peak amplification that are not uniform and are not reproducible by simple Mie theory. These effects can be seen in Figure \ref{fig:figS17}, where single peaks in a Mie theory model are spectrally aligned with multiply-peaked features in the data, and more obviously in Figure \ref{fig:figS13}. The latter compares Mie theory to simulations that include a substrate, showing that a phenomenologically damped dielectric model reproduces some substrate effects (peak suppression, broadening) but not others (enhancements, shifts of narrow peaks). 

Thus, estimates of the real part of $\epsilon_2$ were performed by aligning the visible broad peaks of the theory and experiment in the spectral window between 720 nm and 380 nm, and the absolute peak widths and relative heights were ignored. Narrow features in the data are also ignored as they are most likely to be shifted by the substrate in a manner not captured by Mie theory, as evidenced by the narrow feature near 2.5 eV in Figure \ref{fig:figS13}a. Results using these comparison guidelines are shown in Table \ref{tab:tabS2}.

Within the range of possible values of $\mathrm{Re}\{\epsilon_2\}$ shown therein, the value 5.5 provides the best agreement between the spectral positions and relative magnitudes of the SH peaks in the theory and experiment, assuming $a_2 = 500$ nm. As is mentioned in the main text and demonstrated in Figures \ref{fig:fig4}c (inset) and \ref{fig:figS14}c, small tweaks to $\epsilon_2$ and $a_2$ will provide similar Mie spectra such that the particular values chosen within a narrow range ($\pm\sim\!5\%$) are irrelevant to the conclusions of this investigation. 

The imaginary part of $\epsilon_2$ is taken to be $0.035$. This value is likely larger than the true value extracted from earlier ellipsometry experiments \cite{palik1998handbook}, as the linewidths of the Mie resonances of the LiNbO$_3$ sphere are increased by the presence of a substrate. Because the magnitude of this increase is difficult to infer from first-principles models, we instead choose $\mathrm{Im}\{\epsilon_2\}$ to agree well with the linewidths of the modes observed in the second harmonic data of Figure \ref{fig:fig4}c,b. Within a range of $\pm\sim\!10\%$ of the chosen value, $\mathrm{Im}\{\epsilon_2\}$ simply scales the magnitude of the SH enhancements, such that, like the choices of $\mathrm{Re}\{\epsilon_2\}$ and $a_2$, the precise value of $\mathrm{Im}\{\epsilon_2\}$ does not affect our conclusions.

The second-order susceptibility of the LiNbO$_3$, $\bm{\chi}_2^{(2)}(\omega',\omega - \omega')$ was fit to the simulated data from Ref. \citenum{riefer2012linear} using an anharmonic oscillator model \cite{shen2003principles} in the approximation that the LiNbO$_3$ has an isotropic, homogeneous response within the microsphere boundaries. Explicitly, the susceptibility is given by:
\begin{equation}
\begin{split}
\bm{\chi}_2^{(2)}&(\omega',\omega - \omega') = \\
&\bm{1}\chi^{(2)}_\infty - \bm{1}\frac{s}{(\Lambda^2 - \omega^2 - \mathrm{i}\omega\Gamma)(\Lambda^2 - \omega'^2 - \mathrm{i}\omega'\Gamma)(\Lambda^2 - [\omega - \omega']^2 - \mathrm{i}[\omega - \omega']\Gamma)},
\end{split}
\end{equation}
wherein $\Lambda$ and $\Gamma$ are the natural frequency and linewidth of the Lorentz oscillator, respectively. These features are used to characterize the response of the LiNbO$_3$ carriers. $\chi^{(2)}_\infty$ is a constant offset that approximates the shift of the real part of $\bm{\chi}_2^{(2)}$ at low energies by higher-energy transitions. Further, $s$ is the characteristic anharmonic oscillator strength with units $\mathrm{cm}\cdot \mathrm{s}^6/\mathrm{statV}$ and is taken to be very small such that $sE_0\ll\Lambda^6$ with $E_0$ the characteristic strength of the laser field. Finally, $\bm{1}_3$ is the $3\times3$ identity matrix.

To fit the data, we evaluated the second-order susceptibility in the limit that $\omega = 2\omega'$, i.e. with the assumption that the driving laser at $\omega' = \omega_0$ is very narrow and the observation frequency is always at $2\omega_0$. With this restriction, the expression for the susceptibility simplifies to $\bm{\chi}_2^{(2)}(\omega_0,\omega_0) = \bm{1}\chi_\infty - \bm{1}s/(\Lambda^2 - 4\omega_0^2 - 2\mathrm{i}\omega_0\Gamma)(\Lambda^2 - \omega_0^2 - \mathrm{i}\omega_0\Gamma)^2$ and can be easily visualized as is shown in Figure \ref{fig:figS11}.

As was done with the dielectric function of Au in Section \ref{sec:dielFit}, the model was fit to the real part, imaginary part, and absolute value of the data using simplex, differential evolution, simulated annealing, and random search nonlinear least squares methods. An average of the results of each of the four methods was collected for the fit to each function of the data, and the parameter average that produced the best fit to the data in the experimentally relevant energy range was selected. In this case, the fits to the absolute value of the data were superior in the region between $\sim$1.0 eV -- 1.5 eV analyzed in Figure \ref{fig:fig4}, returning parameter values of $\hbar\Lambda = 4.00$ eV, $\hbar\Gamma = 748$ meV, $s = 4.08\times10^{-3}$ $\mathrm{cm}\cdot\mathrm{s^6}$/statV, and $\chi_\infty^{(2)} = -9.97\times10^{-7}$ cm/statV.

The estimation of the dielectric function of the LiNbO$_3$ is described in the main text. We note here that the corresponding linear susceptibility model of LiNbO$_3$ to the second-order function described above is a Lorentz-model dielectric $\chi^{(1)}(\omega) = \chi_\infty^{(1)} + f/(\Lambda^2 - \omega^2 - \mathrm{i}\omega\Gamma)$. However, in the case where $\Lambda$ is sufficiently detuned from the fundamental and second harmonic frequencies, $\chi^{(1)}(\omega)$ is well-approximated as a dielectric. In our case, $\hbar\Lambda - \hbar\Gamma > 2\omega_0$ such that the constant-dielectric approximation is appropriate and in agreement with Ref. \citenum{palik1998handbook}.

\subsection{Estimation of the Mode Oscillator Parameters}\label{sec:paramInf}

In this section, we use the dielectric models we have constructed to infer the parameters for oscillators models of the plasmon and Mie resonances of the coupled Au-LiNbO$_3$ nanostructures. Bare Au and LiNbO$_3$ particles are discussed in Sections \ref{sec:bareAuInf} and \ref{sec:bareLiNbO3Inf}, while the effects of weak coupling between members of the NP ensemble are discussed in Section \ref{sec:ensembleAuInf}.

\subsubsection{Bare Au NP Dipoles}\label{sec:bareAuInf}

The relation of the oscillator parameters of the Mie and plasmon resonances to the dielectric parameters of their respective particles is done via the particles' response functions. These are shown explicitly in Eqs. \eqref{eq:gResp}--\eqref{eq:ABresp} (see Section \ref{sec:oscModelConstr}), and provide a direct link between the Au and LiNbO$_3$ dielectric functions $\epsilon_1(\omega)$ and $\epsilon_2$, respectively, the resonance freqeuncies $\omega$, damping rates $\gamma$, masses $\mu$, and phase offsets $\psi$ of each mode, and the physical observables of the particles.

\begin{figure}[th!] 
\centering
\includegraphics[scale = 1.0]{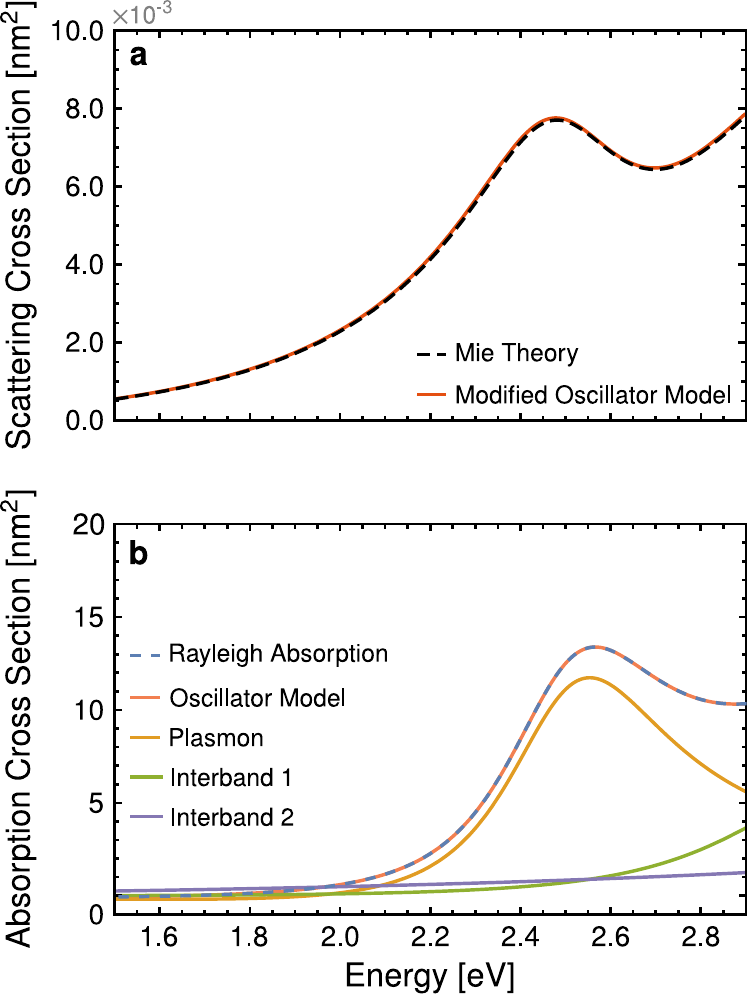}
\caption{
\textbf{Model Au NP plasmon spectra.} Comparison of the plasmon oscillator model with the (a) scattering and (b) absorption cross sections of the dipole modes of a 5-nm radius Au sphere, respectively. The corresponding Mie and Rayleigh observables are calculated directly from the Au dielectric function as shown in Ref. \citenum{bohren2004absorption}. Panel (a) highlights the more dominant role played by high-energy resonances in Au on the blue side of the plasmon spectrum, while panel (b) shows the separate and important contributions to the total absorption cross section from the plasmon and interband oscillators in the energy range of the observed SHG.
\label{fig:figS12}
}
\end{figure}

\begin{figure}[th!] 
\centering
\includegraphics[width = 0.45\textwidth]{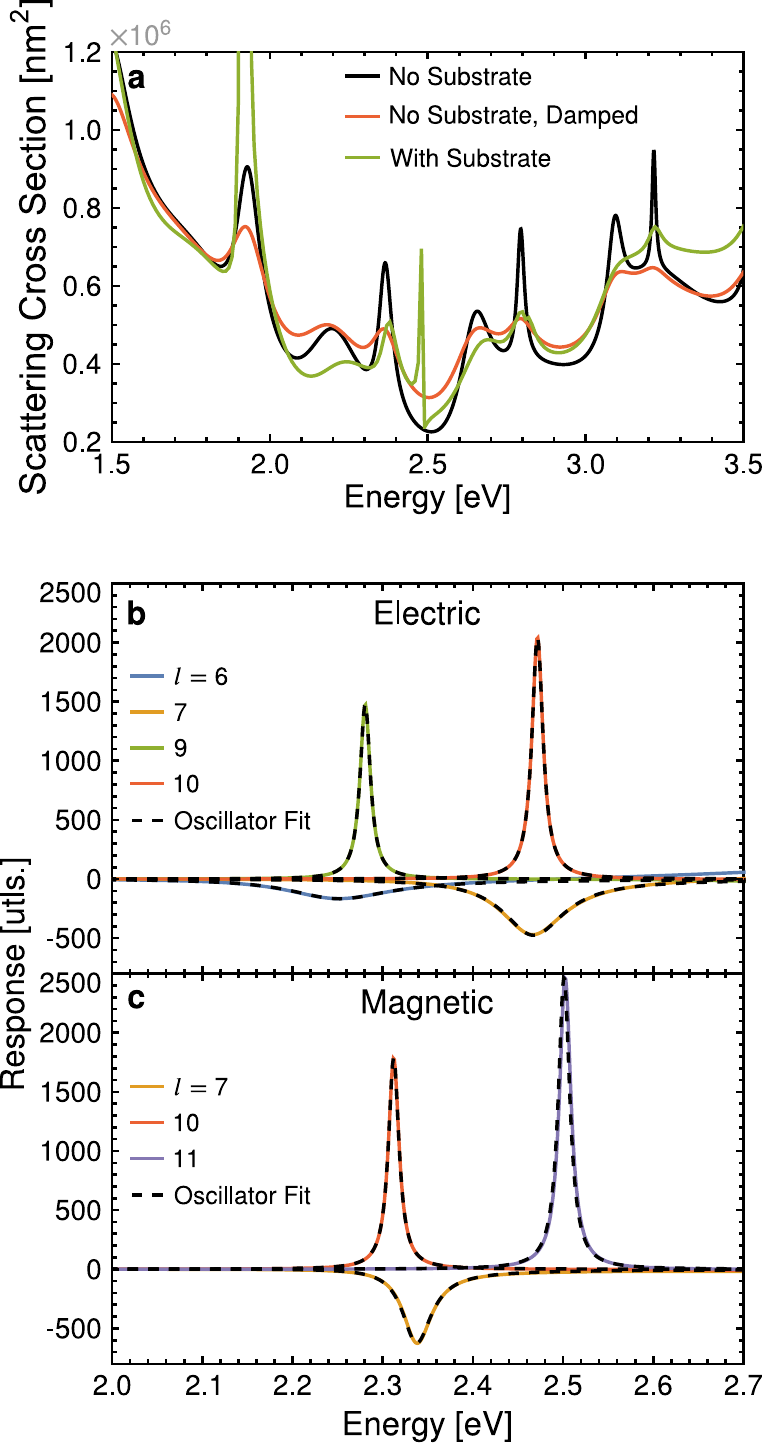}
\caption{
\textbf{Analysis of LiNbO$_{\bm{3}}$ mode structure.} (a) Comparison of the scattering cross section of a 250 nm radius sphere with dielectric function $\epsilon = 5.0$ with spectra generated by two strategies for accounting for substrate effects. In red, the modification of Mie theory with a complex dielectric function $\epsilon = 5.0 + 0.2\mathrm{i}$, and in green, the simulated spectrum of the unmodified sphere on a semi-infinite substrate of dielectric 2.25 \cite{marcos2016self-consistent}. (b), (c) Numerical fit (dashed lines) of the imaginary parts of the oscillator model response functions of the (b) electric- and (c) magnetic-type LiNbO$_3$ Mie modes to the exact functions (solid lines) $(\sqrt{\epsilon_2})^\ell A_{p\ell m}^<(\omega) - 1$ and $(\sqrt{\epsilon_2})^\ell B_{p\ell m}^<(\omega) - 1$, respectively (see Eq. [\ref{eq:ABresp}]).
\label{fig:figS13}
}
\end{figure}

\begin{figure}[th!] 
\centering
\includegraphics[scale = 1.0]{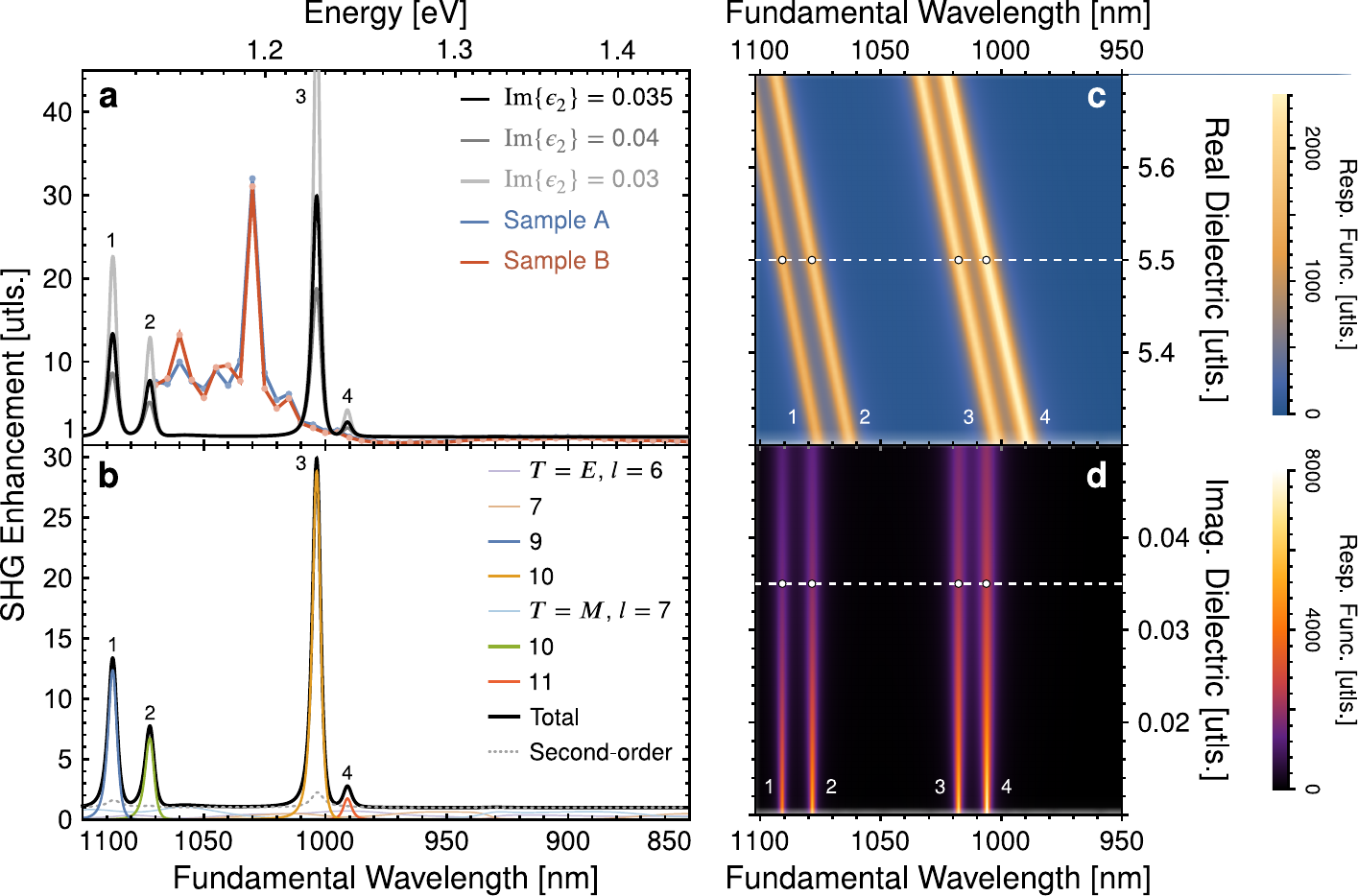}
\caption{
\textbf{Additional details of theoretical SHG enhancement spectra.} (a) Reproduction of the theoretical and experimental SHG enhancement spectra from Figure \ref{fig:fig4}c. The theory (black)is shown without phenomenological shifts, along with a demonstration of the strong changes in the enhancement magnitude with small increases (dark gray) or decreases (light gray) to the imaginary part of $\epsilon_2$. The four theoretical peaks are numbered for convenience and are plotted along with experimental SHG data (Sample A [blue] and Sample B [red]) for comparison. (b) Reconstruction of the theoretical SHG enhancement spectrum (black) from its constituent parts, highlighting the contributions of the modes of varying values of $T$ and $\ell$ (colors). For simplicity, the contributions of modes of common $T$ and $\ell$ are added together. The total SHG signal calculated to second order (gray, dashed) is shown as well, highlighting the importance of the nanoscopic energy-transfer described by the third-order contributions. (c) Demonstration of the strong dependence of the resonance positions of four main peaks of the theoretical spectrum on the real part of $\epsilon_2$ with $\mathrm{Im}\{\epsilon_2\} = 0.035$. The plotted curve (shown in color) at each value of $\mathrm{Re}\{\epsilon_2\}$ is the imaginary part of the sum of the mode response functions (see Eq. [\ref{eq:ABresp}]) for the four peaks. The white dashed line indicates the value $\mathrm{Re}\{\epsilon_2\} = 5.5$ used in the main text, while the white dots lite at the locations of the numbered peaks in (a) and (b). (d) Showcase of the rapid decrease of $\gamma_{\bm{\beta}}$ with $\mathrm{Im}\{\epsilon_2\}$. The color plot is a collection of curves as in (c) but with the imaginary part of the LiNbO$_3$ dielectric allowed to vary and $\mathrm{Re}\{\epsilon_2\} = 5.5$. The dashed white line indicates the value $\mathrm{Im}\{\epsilon_2\} = 0.035$ used in the main text.
\label{fig:figS14}
}
\end{figure}

\begin{figure}[th!] 
\centering
\includegraphics[width = 0.9\textwidth]{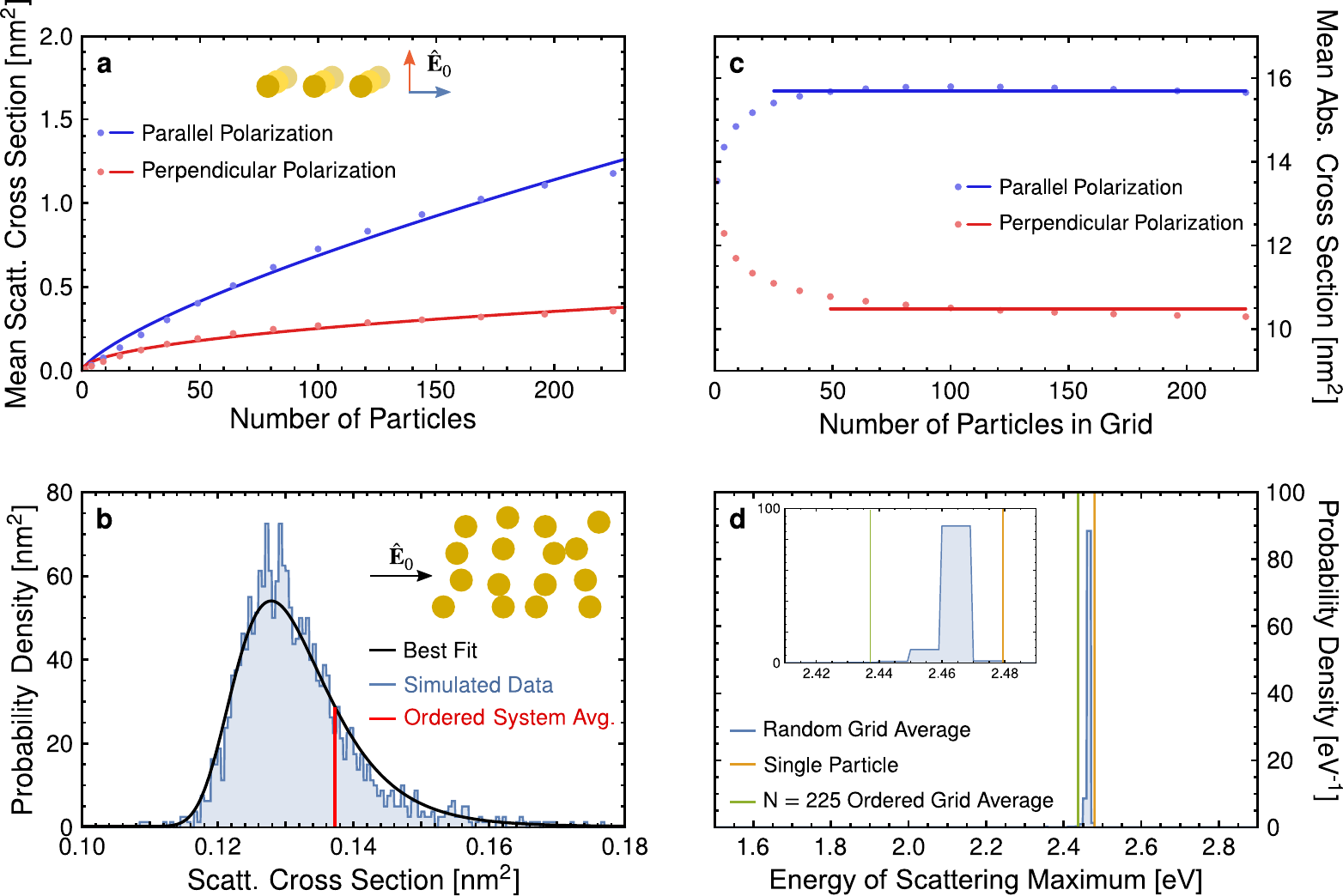}
\caption{
\textbf{Model Au NP ensemble behaviors.} (a) Simulated average scattering cross sections of Au nanospheres organized in square grids of $N\leq$225 particles with light polarized parallel to the grid plane (blue points) and perpendicular (red points). The spheres have radii of $a_1 = 5$ nm and a minimum interparticle spacing of $10$ nm. Fits to the data (lines) using a simple exponential function are used to extrapolate approximate upper bounds for the values of these cross sections when $N=\text{1,000}$. (b) Distribution of scattering cross sections from 100 simulations of disordered grids of $N=16$ Au NPs. The grids were arranged as in (a) but with each NP randomly shifted between -$a_1$ and $a_1$ in the in-plane directions and up to $\pm2a_1$ along the out-of-plane axis such that no two particles overlap. The observed probability density (blue) is fit to an extreme value distribution (black) to estimate the proportion of the particles with cross sections lower and higher than the average from the ordered grid of 16 particles (red). (c) Demonstration that the absorption cross sections of the Au NPs approach stable values as the number of particles in the ensemble increases, both for parallel (blue) and perpendicular (red) incident light. Points show ensemble averages of the per-particle cross section maxima for ensembles of size $N = 1$ to 225, while the solid lines show the extracted averages of 15.7 nm$^2$ for parallel light and 10.5 nm$^2$ for perpendicular. The averages are drawn over the range of data points from which they are calculated. (d) Probability density of the spectral position of the scattering maxima of the NPs from 100 simulations of $N=16$ NP ensembles. The energy of the scattering maximum of a single particle (see Figure \ref{fig:figS12}a) and the average of the energies of the scattering maxima from a simulation of an $N=225$ NP ensemble are shown for reference. Inset: A zoomed-in picture of the probability density, with identical quantities and units plotted on the axes as the main figure.
\label{fig:figS15}
}
\end{figure}

\begin{figure}[th!] 
\centering
\includegraphics[width = 0.95\textwidth]{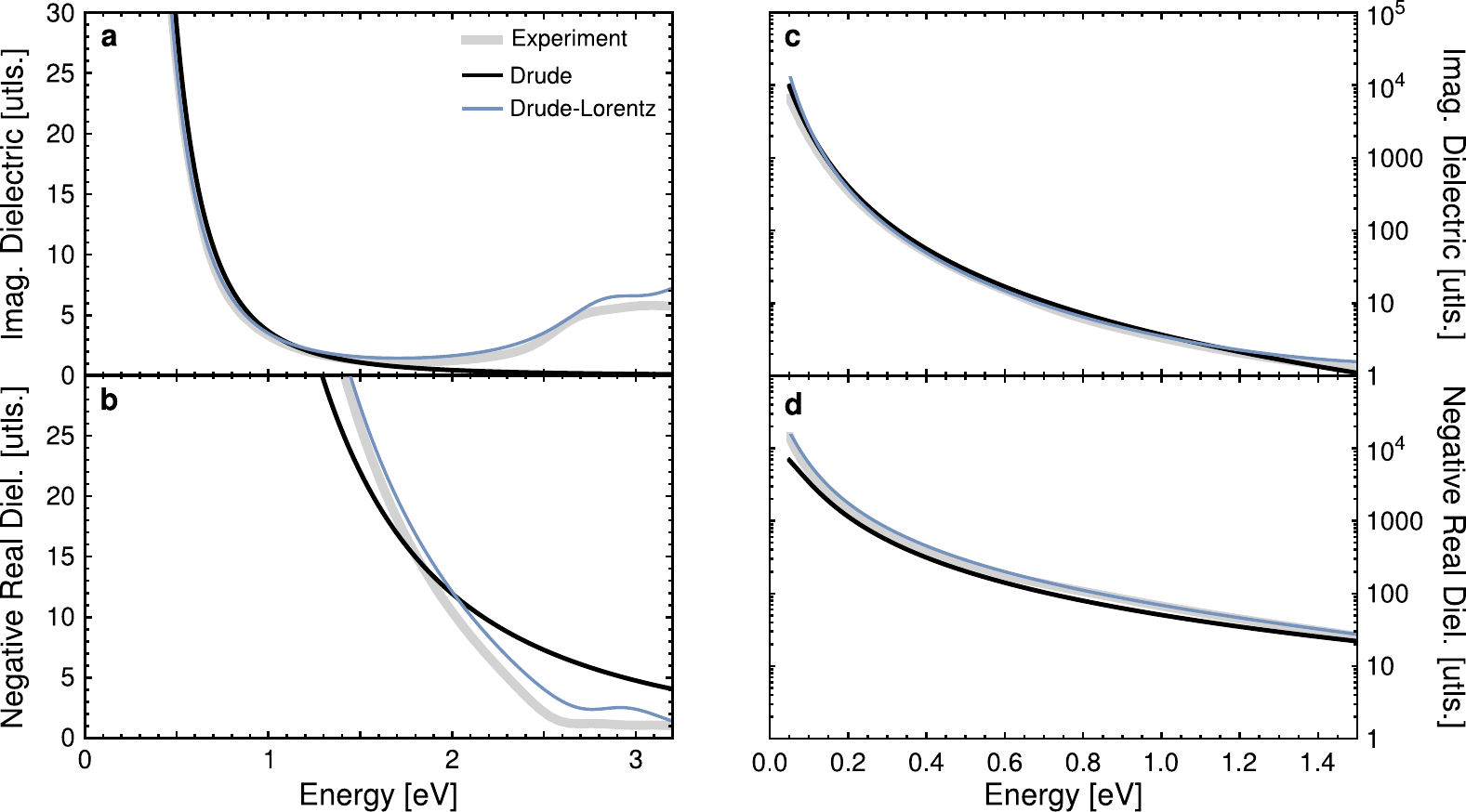}
\caption{
\textbf{Au dielectric function fit.} Demonstration of the agreement between Drude and Drude-Lorentz dielectric models (black, blue, red) of $\epsilon_1(\omega)$ and dielectric data from Ref. \citenum{olmon2012optical}, both in the near-IR to near-UV range (a,b) and in the IR range (c,d).
\label{fig:figS16}
}
\end{figure}

\begin{figure}[th!] 
\centering
\includegraphics[width = 0.45\textwidth]{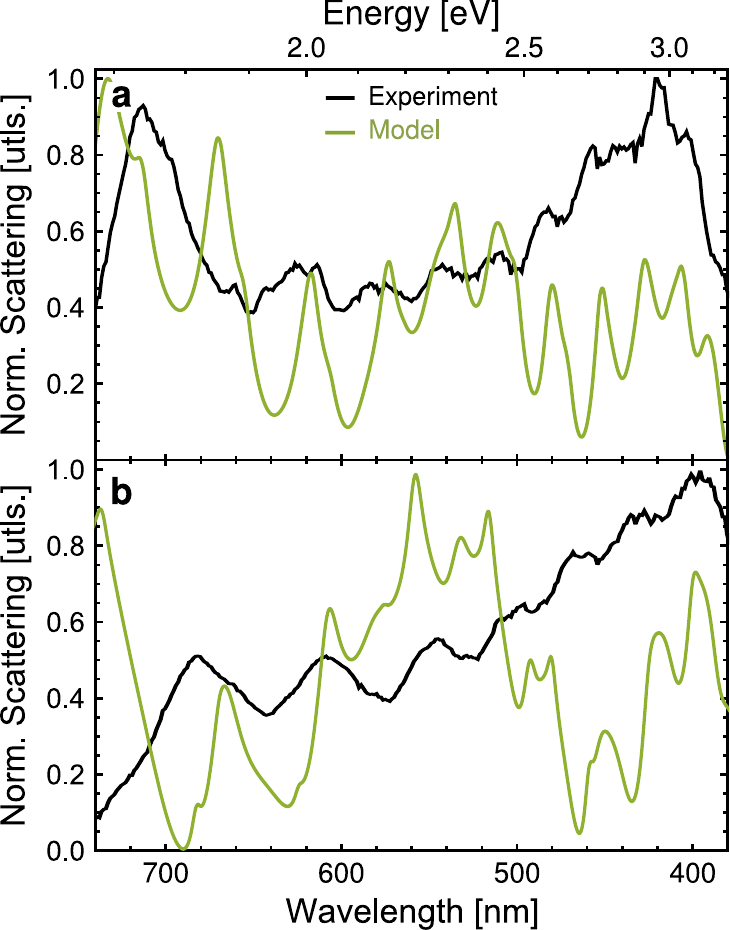}
\caption{
\textbf{Additional Mie scattering data.} Comparison of Mie theory (green) to individual bare LiNbO$_3$ scattering data (black) for samples of relatively high (a) and low (b) dielectric function. The experimental and theoretical data are normalized to range from 0 to 1 within this selected frequency window. The theoretical curves are produced using dielectric values of 7.0 + 0.075i (a) and 5.0 + 0.05i (b) such that, as in Figure \ref{fig:fig3}c, the number and spectral position of the sharper visible peaks in either spectrum are in agreement.
\label{fig:figS17}
}
\end{figure}

For example, with $\epsilon_1(\omega)$ characterized in Section \ref{sec:dielFit}, the response function of Eq. \eqref{eq:gResp2} can be used used to construct the absorption cross section,
\begin{equation}\label{eq:absCrossSect}
\begin{split}
\sigma_1^\mathrm{abs}(\omega) &= \frac{4\pi\omega}{c}\mathrm{Im}\left\{\frac{e^2}{2\omega_1\mu_1}\left(\frac{\mathrm{e}^{\mathrm{i}\psi_1}}{\Omega_1 - \omega} + \frac{\mathrm{e}^{-\mathrm{i}\psi_1}}{\Omega_1^* + \omega}\right) + \sum_{i = 1}^2\frac{e^2}{2\omega_{L_i}\mu_{L_i}}\left(\frac{\mathrm{e}^{\mathrm{i}\psi_{L_i}}}{\Omega_{L_i} - \omega} + \frac{\mathrm{e}^{-\mathrm{i}\psi_{L_i}}}{\Omega_{L_i}^* + \omega}\right)\right\},
\end{split}
\end{equation}
of each of the Au NP dipole modes, in which the plasmon oscillator parameters (subscript 1) and interband resonance parameters (subscript $L_i$) can be exactly expressed as functions of the dielectric parameters of $\epsilon_1(\omega)$. These functions are impossible to write explicitly, as the eigenfrequencies $\Omega_{1,L_i} = \omega_{1,L_i} - \mathrm{i}\gamma_{1,L_i}/2$ are solutions to a sextic polynomial. Nevertheless, the oscillator parameters are simple to infer numerically by fitting Eq. \eqref{eq:absCrossSect} to the standard form $\sigma_1^\mathrm{abs}(\omega) = (4\pi\omega/c)a_1^3\mathrm{Im}\{[\epsilon_1(\omega) - 1]/[\epsilon_1(\omega) + 2]\}$ as shown in Figure \ref{fig:figS12}b with the NP radius $a_1$ set to 5 nm. 

It is well-known that plasmon resonance energies are redshifted when the metal nanoparticle supporting them comes in close contact with a dielectric substrate with a large refractive index \cite{cherqui2018multipolar}. As the LiNbO$_3$ spheres in this investigation are much larger than the Au NPs and have a constant dielectric function larger than 5 (see Section \ref{sec:bareLiNbO3Inf}), some plasmon redshifting is expected. We incorporate the redshifts by lowering the Au NP dipole by 0.23 eV to align the NP absorption maximum with the observed maximum of hybrid-particle extinction measurements in Figure \ref{fig:fig3}a. The oscillator parameters used in this work are given in Table \ref{tab:tabS1}.

However, to model the effects of radiation on the plasmon motion, one must modify the results extracted from Eq. \eqref{eq:absCrossSect}. More concretely, in the formal quasistatic approximation under which the absorption cross section is derived, only the rate of Au carrier losses to heat will appear in the cross section expression and any back-action of radiation on the dipole plasmon will be ignored. This back-action generally results in broadening of the plasmon lineshape, such that one can phenomenologically expand $\gamma_1 \to \gamma_1^\mathrm{rad} + \gamma_1^\mathrm{NR}$ to provide the plasmon damping rate with both a radiative (rad) and nonradiative (NR) component. 

While approximate, this simple formulation of scattering loss works well in the limit $\gamma_1^\mathrm{rad}\ll\gamma_1^\mathrm{NR}$. The validity of the approximation can be demonstrated by comparing the Rayleigh scattering cross section
\begin{equation}\label{eq:scaCrossSect}
\begin{split}
\sigma_1^\mathrm{sca}(\omega) &= \frac{8\pi a_1^6 \omega^4}{3c^4}\left|\frac{e^2}{2\omega_1\mu_1}\left(\frac{\mathrm{e}^{\mathrm{i}\psi_1}}{\omega_1 - \mathrm{i}(\gamma_1^\mathrm{NR} + \gamma_1^\mathrm{rad})/2 - \omega} + \frac{\mathrm{e}^{-\mathrm{i}\psi_1}}{\omega_1 + \mathrm{i}(\gamma_1^\mathrm{NR} + \gamma_1^\mathrm{rad})/2 + \omega}\right)\right.\\[0.5em]
&\left.+ \sum_{i = 1}^2\frac{e^2}{2\omega_{L_i}\mu_{L_i}}\left(\frac{\mathrm{e}^{\mathrm{i}\psi_{L_i}}}{\omega_{L_i} - \mathrm{i}(\gamma_{L_i}^\mathrm{NR} + \gamma_{L_i}^\mathrm{rad})/2 - \omega} + \frac{\mathrm{e}^{-\mathrm{i}\psi_{L_i}}}{\omega_{L_i} + \mathrm{i}(\gamma_{L_i}^\mathrm{NR} + \gamma_{L_i}^\mathrm{rad})/2 + \omega}\right) \right|^2,
\end{split}
\end{equation}
in which the interband damping rates have been similarly transformed as $\gamma_{L_i}\to\gamma_{L_i}^\mathrm{NR} + \gamma_{L_i}^\mathrm{rad}$ to account for radiation broadening and red shifting, to the exact Mie theory scattering cross section of a spherical Au NP with a 5-nm radius. Numerical fits of the modified oscillator model to the Mie scattering lineshape are generally unstable and suggest radiation-induced changes to both $\gamma_1$ and $\gamma_{L_i}$ are roughly $1\%$ or smaller, so we instead approximate the radiative damping rates of the dipole plasmons and interband resonances using their Larmor formulae. Assuming $\gamma_1^\mathrm{NR}\approx\gamma_1$, we have $\gamma_1^\mathrm{rad} = 2e^2\omega_1^2/3c^3\mu_1 = 1.01\times10^{-4}\gamma_1^\mathrm{NR}$. The radiation from the Lorentz oscillator resonances is similarly minimal, with $\gamma_{L_1}^\mathrm{rad} = 7.15\times10^{-5}\gamma_{L_1}^\mathrm{NR}$, $\gamma_{L_1}^\mathrm{rad} = 3.67\times10^{-2}\gamma_{L_2}^\mathrm{NR}$, and $\gamma_{L_i}^\mathrm{NR}\approx\gamma_{L_i}$. Figure \ref{fig:figS12}a shows the excellent agreement between the Larmor-modified oscillator model and Mie scattering calculations.

\subsubsection{Ensemble-Modified Au NP Dipoles}\label{sec:ensembleAuInf}

To describe in a concise way the scattering observables of the dipole plasmons in the NPs surrounding the LiNbO$_3$ sphere, it is first necessary to quantify the alterations to the scattering behaviors of each NP caused by its neighbors. In other words, due to inter-NP interactions, we \textit{cannot} take the masses $\mu_i$ (or, equivalently, the oscillator strengths $f_i = e^2/a_1^3\mu_i$) of the $N$ NPs of the ensemble to be their bare values $\mu_1$ (or $e^2/a_1^3\mu_1$).

We can begin this process by letting the polarizability of each NP be a tensor dependent on its position, such that differences between the ensemble perpendicular and parallel dipole oscillations of each NP can be captured. Explicitly, we let $\alpha_1(\omega)\to\bm{\alpha}_i(\omega) = \alpha_\perp(\omega)\hat{\mathbf{r}}_i\hat{\mathbf{r}}_i + \alpha_\parallel(\omega)(\hat{\bm{\theta}}_i\hat{\bm{\theta}}_i + \hat{\bm{\phi}}_i\hat{\bm{\phi}}_i)$, wherein the unit vectors $\hat{\mathbf{r}}_i = \hat{\mathbf{r}}(\theta_i,\phi_i)$, $\hat{\bm{\theta}}_i = \hat{\bm{\theta}}(\theta_i,\phi_i)$, and $\hat{\bm{\phi}}_i = \hat{\bm{\phi}}(\theta_i,\phi_i)$ are simply the spherical unit vectors evaluated at the angular position of the $i^\mathrm{th}$ dipole. The tensor elements $\alpha_{\perp,\parallel}(\omega) = f_{\perp,\parallel}(a_1^3/2\omega_1)(\exp{\mathrm{i}\psi_{\perp,\parallel}}/[\Omega_1 - \omega] + \exp{-\mathrm{i}\psi_{\perp,\parallel}}/[\Omega_1^* + \omega])$ are the polarizabilities of the dipole components at $\mathbf{r}_i$ that are oriented perpendicular and parallel to the shell of Au NPs, respectively, and have magnitudes characterized by the oscillator strengths $f_{\perp,\parallel}$ and phases determined by the angles $\psi_{\perp,\parallel}$. 

We can define the terms of $\bm{\alpha}_i(\omega)$ with poles at $\pm\Omega_1$ as 
\begin{equation}
\bm{\alpha}_i^{(+)}(\omega) = \left(\frac{f_\perp a_1^3}{2\omega_1}\frac{\mathrm{e}^{\mathrm{i}\psi_\perp}}{\Omega_1 - \omega} \right)\hat{\mathbf{r}}_i\hat{\mathbf{r}}_i+ \left(\frac{f_\parallel a_1^3}{2\omega_1}\frac{\mathrm{e}^{\mathrm{i}\psi_\parallel}}{\Omega_1 - \omega}\right)(\hat{\bm{\theta}}_i\hat{\bm{\theta}}_i + \hat{\bm{\phi}}_i\hat{\bm{\phi}_i})
\end{equation}
\sloppy and $\bm{\alpha}_i^{(-)}(\omega) = \bm{\alpha}_i^{(+)*}(-\omega)$ and further define the components of either as $\alpha_{\perp,\parallel}^{(\pm)}(\omega) = a_1^3f_{\perp,\parallel}\mathrm{e}^{\mathrm{i}\psi_{\perp,\parallel}}/2\omega_1(\Omega_1 \mp \omega)$, respectively. Therefore, upon excitation by an impinging field $\mathbf{E}(\mathbf{r}_i,\omega)$, the dipole set up in the $i^\mathrm{th}$ NP is $\mathbf{d}_i(\omega,\hat{\bm{\epsilon}}_i) = \bm{\alpha}_i(\omega)\cdot\mathbf{E}(\mathbf{r}_i,\omega)\approx E(\mathbf{r}_i,\omega)\Theta(\omega)\bm{\alpha}_i^{(+)}(\omega)\cdot \hat{\bm{\epsilon}}_i + E(\mathbf{r}_i,\omega)\Theta(-\omega)\bm{\alpha}_i^{(-)}(\omega)\cdot\hat{\bm{\epsilon}}_i^*$. Here, $E(\mathbf{r}_i,\omega)$ is the (real) magnitude of the electric field at $\mathbf{r}_i$, $\hat{\bm{\epsilon}}_i$ is the complex polarization unit vector that describes the phases and orientations of the field components, and $\Theta(\omega)$ is the Heaviside function. Letting $\mathbf{d}^{(\pm)}(\omega,\hat{\bm{\epsilon}}_i)$ be the dipole terms valid at positive and negative frequencies, one finds the magnitude of $\mathbf{d}^{(+)}$ is
\begin{equation}\label{eq:dipoleMag}
\norm{\mathbf{d}_i^{(+)}(\omega,\hat{\bm{\epsilon}}_i)} \approx \frac{f_i(\hat{\bm{\epsilon}}_i)a_1^3}{2\omega_1}\frac{1}{|\Omega_1 - \omega|}\Theta(\omega)E(\mathbf{r}_i,\omega).
\end{equation}
We can, therefore, define $f_i(\hat{\bm{\epsilon}}_i) = \sqrt{f_\perp^2|\hat{\mathbf{r}}_i\cdot\hat{\bm{\epsilon}}_i|^2 + f_\parallel^2(|\hat{\bm{\theta}}_i\cdot\hat{\bm{\epsilon}}_i|^2 + |\hat{\bm{\phi}}_i\cdot\hat{\bm{\epsilon}}_i|^2)}$ as the orientation-dependent dipole plasmon oscillator strength.

With $\hat{\bm{\epsilon}}_i$ oriented strictly perpendicular to the ensemble, one can see that $f_i(\hat{\mathbf{r}}_i) = f_\perp$. Similarly, $f_i(\hat{\bm{\theta}}_i) = f_\parallel$. Figure \ref{fig:figS15}c shows the simulated absorption cross section maxima of NPs in ensembles excited with these two choices of $\hat{\bm{\epsilon}}_i$, where it can be seen that for $N \gtrsim 50$ the absorption cross section of each particle is independent of $N$. Analytically, the absorption cross sections of ensemble perpendicular and parallel dipoles are simply $\sigma^\mathrm{abs}_{\perp,\parallel}(\omega) = (4\pi\omega/c)R_\mathrm{abs}\mathrm{Im}\{\alpha_{\perp,\parallel}(\omega)\}$, where $R_\mathrm{abs} = 1.16$ phenomenologically accounts for the absorption contribution of the Lorentz oscillators in the region $\omega < 2.8$ eV. With $\cos\psi_{\perp,\parallel}\approx\cos\psi_1\approx1$, the cross section maxima are then $\sigma^\mathrm{abs}_{\perp,\parallel}(\omega_1)\approx 4\pi R_\mathrm{abs} f_{\perp,\parallel}a_1^3/c\gamma_1$ such that the oscillator strengths implied by Figure \ref{fig:figS14}c are $f_\parallel = 2.37\times10^{30}$ s$^{-2}$ and $f_\perp/f_\parallel = 0.668$.

Finally, to define the orientation-averaged plasmon oscillator strength $f_1$ used in the main text, we let an electric field $\mathbf{E}_0(\mathbf{r},\omega) = E_0\pi[\exp(2\mathrm{i}k_0x)\delta(\omega - 2\omega_0) + \exp(-2\mathrm{i}k_0x)\delta(\omega + 2\omega_0)]\hat{\mathbf{z}}$ excite the NP ensemble, producing $f_i(\hat{\mathbf{z}}) = \sqrt{f_\perp^2\cos^2\theta_i + f_\parallel^2\sin^2\theta_i}$. We then take a formal average 
\begin{equation}\label{eq:formalAvg}
\begin{split}
\langle f_i(\hat{\mathbf{z}})\rangle_i = \frac{\int_0^{2\pi}\int_0^\pi f_i(\hat{\mathbf{z}})(t)\sin\theta_i\;\mathrm{d}\theta_i\mathrm{d}\phi_i}{\int_0^{2\pi}\int_0^\pi\sin\theta_i\;\mathrm{d}\theta_i\mathrm{d}\phi_i}
\end{split}
\end{equation}
over all of the possible $\hat{\mathbf{r}}_i$ and define $f_1 = \langle f_i(\hat{\mathbf{z}}) \rangle_i$, giving
\begin{equation}\label{eq:f1Standard}
f_1 = \frac{1}{2}\left(f_\perp + \frac{f_\parallel^2\cos^{-1}\left\{\frac{f_\perp}{f_\parallel}\right\}}{\sqrt{f_\parallel^2 - f_\perp^2}}\right) = 2.13\times10^{30} \;\,\mathrm{s}^{-2}.
\end{equation}

\subsubsection{Bare Lithium Niobate Mie Resonances}\label{sec:bareLiNbO3Inf}

Inference of the oscillator parameters of the Mie resonances $\bm{\beta}$ of the LiNbO$_3$ microsphere is more straightforward, as their damping rates $\gamma_{\bm{\beta}}$ contain only losses to radiation. Thus, only a single observable is needed to extract the full set of parameters from each mode. We use the response functions shown in Eq. \eqref{eq:ABresp} to fit $\gamma_{\bm{\beta}}$ as well as the resonance frequencies $\omega_{\bm{\beta}}$, effective masses $\mu_{\bm{\beta}}$ (equivalently, the oscillator strenghts $f_{\bm{\beta}} = e^2/a_2^3\mu_{\bm{\beta}}$), and phase offsets $\psi_{\bm{\beta}}$. Explicitly, we let
\begin{equation}
\begin{split}
(\sqrt{\epsilon_2})^\ell A_{p\ell m}^<(\omega) - 1 &\approx -\frac{e^2}{a_2^3}\frac{1}{2\omega_{M\ell}\mu_{M\ell}}\left(\frac{\mathrm{e}^{\mathrm{i}\psi_{M\ell}}}{\Omega_{M\ell} - \omega} + \frac{\mathrm{e}^{-\mathrm{i}\psi_{M\ell}}}{\Omega_{M\ell}^* + \omega}\right), \\
(\sqrt{\epsilon_2})^\ell B_{p\ell m}^<(\omega) - 1 &\approx -\frac{e^2}{a_2^3}\frac{1}{2\omega_{E\ell}\mu_{E\ell}}\left(\frac{\mathrm{e}^{\mathrm{i}\psi_{E\ell}}}{\Omega_{E\ell} - \omega} + \frac{\mathrm{e}^{-\mathrm{i}\psi_{E\ell}}}{\Omega_{E\ell}^* + \omega}\right).
\end{split}
\end{equation}
Note that, in contrast with Eq. \eqref{eq:ABresp}, the above expressions only consider a single resonance for a given $\bm{\beta}$. See Section \ref{sec:nearField} for details. The results of the fits are shown graphically in Figure \ref{fig:figS15}b,c and are tabulated in Table \ref{tab:tabS2}.

\subsection{Derivation of the Scattering Observables}\label{sec:scattering}

We provide here a more thorough derivation of the superradiant scattering enhancement ratio. First, in Section \ref{sec:mieScatt}, we describe the scattered fields and power of the modes of a dielectric sphere with a diameter on the order of a wavelength of the scattered light. Second, in sections Section \ref{sec:dipScattInd} we develop the scattering observables of the ensemble of NPs surrounding a dielectric sphere. In Section \ref{sec:pRatio}, we complete the calculation of the ratio of scattered powers of a NP-dressed and bare LiNbO$_3$ dielectric sphere.

\subsubsection{Mie Modes of the Lithium Niobate Sphere Driven by a Plane Wave}\label{sec:mieScatt}

The time averaged radiated power from each of the LiNbO$_3$ SH modes can be calculated from Poynting's theorem. Beginning with
\begin{equation}\label{eq:Esca2}
\bm{\mathcal{E}}_\mathrm{sca}(\mathbf{r},\omega) = \sum_{\bm{\beta}}\frac{1}{a_2^{\ell + 2}}\left(\rho_{\bm{\beta}}(\omega)\mathbf{X}_{\bm{\beta}}(\mathbf{r},k_{\bm{\beta}}) + \rho_{\bm{\beta}}^*(-\omega)\mathbf{X}^*_{\bm{\beta}}(\mathbf{r},k_{\bm{\beta}})\right),
\end{equation}
wherein $(2)$ signifies that the fields arise from a second-order scattering process, $\rho_{\bm{\beta}}(\omega)$ are the magnitudes of the moments of the modes $\bm{\beta}$, and $\mathbf{X}_{\bm{\beta}}(\mathbf{r},\omega)$ are the regularized vector spherical harmonics
\begin{equation}\label{eq:Xharmonics}
\begin{split}
\mathbf{M}_{p\ell m}(\mathbf{r},k) &= \sqrt{(2 - \delta_{m0})\frac{2\ell + 1}{\ell(\ell + 1)}\frac{(\ell - m)!}{(\ell + m)!}}\left[\frac{(-1)^{p + 1}m}{\sin\theta}h_\ell^{(1)}(kr)P_{\ell m}(\cos\theta)S_{p + 1}(m\phi)\hat{\bm{\theta}}\right.\\
&\left.- h_\ell^{(1)}(kr)\frac{\partial P_{\ell m}(\cos\theta)}{\partial\theta}S_p(m\phi)\hat{\bm{\phi}}\right],\\[0.5em]
\mathbf{N}_{p\ell m}(\mathbf{r},k) &= \sqrt{(2 - \delta_{m0})\frac{2\ell + 1}{\ell(\ell + 1)}\frac{(\ell - m)!}{(\ell + m)!}}\left[\frac{\ell(\ell + 1)}{kr}h_\ell^{(1)}(kr)P_{\ell m}(\cos\theta)S_p(m\phi)\hat{\mathbf{r}}\right.\\
&\hspace{-0.1\textwidth}\left.+ \frac{1}{kr}\frac{\partial\{rh_\ell^{(1)}(kr)\}}{\partial r}\left(\frac{\partial P_{\ell m}(\cos\theta)}{\partial\theta}S_p(m\phi)\hat{\bm{\theta}} + \frac{(-1)^{p+1}m}{\sin\theta}P_{\ell m}(\cos\theta)S_{p+1}(m\phi)\hat{\bm{\phi}}\right)\right]
\end{split}
\end{equation}
for $T = M$ and $E$, respectively, the calculation of the Poynting vector is straightforward.  Here, $h_\ell^{(1)}(x)$ are the spherical Hankel functions of the first kind, $P_{\ell m}(x)$ are the associated Legendre polynomials, $S_p(x) = \cos(x)\delta_{p\;\mathrm{even}} + \sin(x)\delta_{p\;\mathrm{odd}}$, and $k = \omega/c$. Further, with
\begin{equation}\label{eq:Bsca2}
\bm{\mathcal{B}}_\mathrm{sca}(\mathbf{r},\omega) = \sum_{\bm{\beta}}\frac{c}{\mathrm{i}\omega a_2^{\ell + 2}}\left[\rho_{\bm{\beta}}(\omega)\nabla\times\mathbf{X}_{\bm{\beta}}(\mathbf{r},k_{\bm{\beta}}) + \rho_{\bm{\beta}}^*(-\omega)\nabla\times\mathbf{X}^*_{\bm{\beta}}(\mathbf{r},k_{\bm{\beta}})\right]
\end{equation}
from Faraday's law,
\begin{equation}\label{eq:XcurlXint}
\begin{split}
\lim_{r\to\infty}\int_0^{2\pi}\int_0^\pi&\mathbf{X}_{\bm{\beta}}(\mathbf{r},k_{\bm{\beta}})\times\nabla\times\mathbf{X}_{\bm{\beta}'}^*(\mathbf{r},k_{\bm{\beta}'})\cdot\hat{\mathbf{r}}r^2\sin\theta\;\mathrm{d}\theta\,\mathrm{d}\phi\\
&= 4\pi(1 - \delta_{p1}\delta_{m0})\frac{(-1)^{\ell + 1}\mathrm{i}^{2\ell + 1}}{k_{\bm{\beta}}}\delta_{\bm{\beta}\bm{\beta}'},
\end{split}
\end{equation}
and
\begin{equation}\label{eq:sinc}
\frac{\omega_0}{\pi}\int_{-\pi/2\omega_0}^{\pi/2\omega_0}\mathrm{e}^{\mathrm{i}(\omega - \omega')t}\;\mathrm{d}t = \frac{2\omega_0}{\pi}\frac{\sin\!\left(\frac{\pi}{2\omega_0}[\omega - \omega']\right)}{\omega - \omega'},
\end{equation}
the time-averaged scattered power from the dielectric sphere can be rapidly simplified from
\begin{equation}
\begin{split}
&\bar{P}_2(2\omega_0) \equiv \langle P_{\bm{\beta}}(t) \rangle_{2\pi/2\omega_0}\\
&= \frac{2\omega_0}{2\pi}\frac{c}{4\pi}\int_{-\pi/2\omega_0}^{\pi/2\omega_0}\oint\bm{\mathcal{E}}_\mathrm{sca}(\mathbf{r},t)\times\bm{\mathcal{B}}_\mathrm{sca}(\mathbf{r},t)\cdot\mathrm{d}\mathbf{a}\,\mathrm{d}t\\
&= \lim_{r\to\infty}\frac{\omega_0c}{2\pi^2}\int_0^{2\pi}\int_0^\pi\iint\frac{\sin\left(\frac{\pi}{2\omega_0}[\omega - \omega']\right)}{\omega - \omega'}\bm{\mathcal{E}}_\mathrm{sca}(\mathbf{r},\omega)\times\bm{\mathcal{B}}_\mathrm{sca}^*(\mathbf{r},\omega')r^2\sin\theta\cdot\hat{\mathbf{r}}\;\frac{\mathrm{d}\omega\,\mathrm{d}\omega'}{4\pi^2}\,\mathrm{d}\theta\,\mathrm{d}\phi,
\end{split}
\end{equation}
wherein the surface integral is taken to be across a sphere of radius $r\to\infty$, $\langle\cdot\rangle_\tau$ is the time-average operator over a period $\tau$, and the Fourier transforms of the fields $\bm{\mathcal{F}}_\mathrm{sca}(\mathbf{r},t) = \int\bm{\mathcal{F}}_\mathrm{sca}(\mathbf{r},\omega)\exp(-\mathrm{i}\omega t)\;\mathrm{d}\omega/2\pi$ have been used. Letting the moments of the sphere be driven by the second-harmonic upconversion of an incoming plane wave of frequency $\omega$ and electric field strength $E_0$, we can say $\rho_{\bm{\beta}}(\omega) = C_{\bm{\beta}}(\omega)\alpha_{\bm{\beta}}(\omega)E_0\pi[\delta(\omega - 2\omega_0) + \delta(\omega + 2\omega_0)]$ with $\alpha_{\bm{\beta}}(\omega) = (a_2^{\ell + 2} f_{\bm{\beta}}/ 2\omega_{\bm{\beta}})\exp(i\psi_{\bm{\beta}})/(\omega_{\bm{\beta}} - i\gamma_{\bm{\beta}}/2 - \omega)$ the linear polarizability of the $\bm{\beta}^\mathrm{th}$ mode and $C_{\bm{\beta}}(\omega)$ an overlap coefficient defined in Eq. \eqref{eq:Cbeta}. The final result is
\begin{equation}
\bar{P}_2(2\omega_0) = \sum_{\bm{\beta}}(1 - \delta_{p1}\delta_{m0})\frac{c^2E_0^2}{4\pi\omega_0k_{\bm{\beta}}a_2^{2\ell + 4}}|C_{\bm{\beta}}(2\omega_0)|^2|\alpha_{\bm{\beta}}(2\omega_0)|^2.
\end{equation}

Importantly, the radiated power from the Mie modes contains no cross-terms due to the orthogonality condition imposed by Eq. \eqref{eq:XcurlXint}. Further, due to the good agreement between an oscillator model of each mode and its electromagnetic response (see Section \ref{sec:paramInf}), the scattered power can also be modeled mechanically. To do so, it is important to first define a generalized coordinate $q_{\bm{\beta}}(\omega)$ to represent the displacement magnitude of the moments $\rho_{\bm{\beta}}(\omega)$ that obeys the reality condition $q_{\bm{\beta}}(-\omega) = q_{\bm{\beta}}^*(\omega)$. We will choose the definition $q_{\bm{\beta}}(\omega) = [\rho_{\bm{\beta}}(\omega) + \rho_{\bm{\beta}}^*(-\omega)]/2ea_2^{\ell - 1}$ such that $q_{\bm{\beta}}(t) = \mathrm{Re}\{\rho_{\bm{\beta}}(t)\}/ea_2^{\ell - 1}$ is a real quantity. 

With $q_{\bm{\beta}}(t)$ defined, we model losses to radiation as a weak, frequency-independent linear damping process. In this case,
\begin{equation}
\begin{split}
P_2(t) &= -\sum_{\bm{\beta}}\dot{q}_{\bm{\beta}}(t)F_{\bm{\beta}}^\mathrm{rad}(t),\\
&= \sum_{\bm{\beta}}A_{\bm{\beta}}\mu_{\bm{\beta}}\gamma_{\bm{\beta}}\dot{q}_{\bm{\beta}}^2(t),
\end{split}
\end{equation}
where $A_{\bm{\beta}}$ is a unitless proportionality constant that relates the mechanical quantities to the electromagnetics. This leads directly to
\begin{equation}
\bar{P}_2(2\omega_0) = \sum_{\bm{\beta}}A_{\bm{\beta}}\gamma_{\bm{\beta}}\mu_{\bm{\beta}}\frac{\omega_0^2 E_0^2}{2e^2a_2^{2\ell - 2}}|C_{\bm{\beta}}(2\omega_0)|^2|\alpha_{\bm{\beta}}(2\omega_0)|^2
\end{equation}
via Eq. \eqref{eq:sinc} and the identity $q_{\bm{\beta}}(t) = \int(-\mathrm{i}\omega)\exp(-\mathrm{i}\omega t)q_{\bm{\beta}}(\omega)\;\mathrm{d}\omega/2\pi$ such that $A_{\bm{\beta}} = (1 - \delta_{p1}\delta_{m0})(e^2c^3/2\pi\omega_0^3a_2^6\mu_{\bm{\beta}}\omega_{\bm{\beta}}\gamma_{\bm{\beta}}) = (1 - \delta_{p1}\delta_{m0})(f_{\bm{\beta}}c^3/2\pi\omega_0^3a_2^3\omega_{\bm{\beta}}\gamma_{\bm{\beta}})$. Finally, to arrive at Eq. \eqref{eq:P2} and the notation of the main text we substitute the explicit form of $A_{\bm{\beta}}$ and make the simplification $\bar{P}_2(2\omega_0)\to P^{(1)}(2\omega_0)$ to the notation.

\subsubsection{Plasmon Dipoles Driven by the Scattered Lithium Niobate Electric Field}\label{sec:dipScattInd}

The radiated power by the ensemble of Au NPs can be straightforwardly derived using a mechanical model parameterized using the values outlined in Sections \ref{sec:bareAuInf} and \ref{sec:ensembleAuInf}. More explicitly, the time-averaged scattered power by the $i^\mathrm{th}$ dipole plasmon of the Au NP ensemble when the collection is driven by an external field can be calculated in a straightforward manner using a well-known mechanical model of dipole radiation
\begin{equation}
P_i(t) = -\dot{\mathbf{x}}_i(t)\cdot A_i(N,\hat{\bm{\epsilon}}_i)\mathbf{F}_i^\mathrm{rad}(t),
\end{equation}
wherein $\mathbf{x}_i(t) = \mathbf{d}_i(t)/e$ is the coordinate characterizing the magnitude of the dipole plasmon located at $\mathbf{r}_i$ and oriented along $\hat{\mathbf{x}}_i$ and $A_i(N,\hat{\bm{\epsilon}}_i)$ is a phenomenological enhancement factor that builds in the ensemble-induced scattering enhancements seen in Figure \ref{fig:fig4}c. Similarly to the orientation-dependent oscillator strengths, we take $A_i(N,\hat{\bm{\epsilon}}_i) = \sqrt{A_\perp^2(N)|\hat{\mathbf{r}}_i\cdot\hat{\bm{\epsilon}}_i|^2 + A_\parallel^2(N)(|\hat{\bm{\theta}}_i\cdot\hat{\bm{\epsilon}}_i|^2 + |\hat{\bm{\phi}}_i\cdot\hat{\bm{\epsilon}}_i|^2)}$ where $A_{\perp,\parallel}(N)$ are the enhancements experience by ensemble-perpendicular and parallel dipoles, respectively, in an ensemble of $N$ NPs.

With the radiation back-force $\mathbf{F}_i^\mathrm{rad}(t)$ treated as a damping force, one can let $\mathbf{F}_i^\mathrm{rad}(t) = -\mu_i(\hat{\bm{\epsilon}}_i)\gamma_1^\mathrm{rad}\dot{\mathbf{x}}_i(t)$, with $\mu_i(\hat{\bm{\epsilon}}_i) = e^2/a_1^3f_i(\hat{\bm{\epsilon}}_i)$ the mass of the $i^\mathrm{th}$ plasmon, such that an average over a period $\tau = 2\pi/2\omega_0$ of the second harmonic frequency gives:
\begin{equation}\label{eq:timeAvgPi}
\begin{split}
\bar{P}_i(2\omega_0) &\equiv \langle P_i(t)\rangle_{2\pi/2\omega_0}\\
&= A_i(N,\hat{\bm{\epsilon}}_i)\frac{\omega_0}{\pi}\int_{-\pi/2\omega_0}^{\pi/2\omega_0}\mu_i(\hat{\bm{\epsilon}}_i)\gamma_1^\mathrm{rad}\dot{x}_i^2(t)\;\mathrm{d}t.
\end{split}
\end{equation}
\sloppy Inserting the identity $\dot{x}_i(t) = \int(-\mathrm{i}\omega) x_i(\omega)\exp(\mathrm{i}\omega t)\;\mathrm{d}\omega/2\pi$ twice and letting $\mathbf{x}_i(\omega) = \bm{\alpha}_i(\omega)\cdot\mathbf{E}_\mathrm{sca}^{(2)}(\mathbf{r},\omega)/e$, one finds
\begin{equation}\label{eq:timeAvgPi2}
\begin{split}
\bar{P}_i(2\omega_0) &= A_i(N,\hat{\bm{\epsilon}}_i)\frac{\omega_0\gamma_1^\mathrm{rad}}{4\pi^3 a_1^3f_i(\hat{\bm{\epsilon}}_i)}\iint\int_{-\pi/2\omega_0}^{\pi/2\omega_0}\omega\omega'\left[\bm{\alpha_i}(\omega)\cdot\mathbf{E}_\mathrm{sca}^{(2)}(\mathbf{r}_i,\omega)\right]\cdot\left[\bm{\alpha_i}^*(\omega')\cdot\mathbf{E}_\mathrm{sca}^{(2)*}(\mathbf{r}_i,\omega')\right]\\
&\times\mathrm{e}^{-\mathrm{i}(\omega - \omega')t}\;\mathrm{d}t\,\mathrm{d}\omega\,\mathrm{d}\omega'.
\end{split}
\end{equation}
Application of the identity of Eq. \eqref{eq:sinc} and neglect of terms proportional to $\rho_{\bm{\beta}}(-2\omega_0)\ll\rho_{\bm{\beta}}(2\omega_0)$ provides
\begin{equation}
\begin{split}
\bar{P}_i(2\omega_0) &\approx A_i(N,\hat{\bm{\epsilon}}_i)\frac{2\gamma_1^\mathrm{rad}}{ a_1^3f_i(\hat{\bm{\epsilon}}_i)}\sum_{\bm{\beta}\bm{\beta}'}E_0^2\omega_0^2a_2^{-\ell-\ell'-4}C_{\bm{\beta}}(2\omega_0)C_{\bm{\beta}'}^*(2\omega_0)\alpha_{\bm{\beta}}(2\omega_0)\alpha_{\bm{\beta}'}^*(2\omega_0)\\
&\times\left[\bm{\alpha}_i(2\omega_0)\cdot\mathbf{X}_{\bm{\beta}}(\mathbf{r}_i,k_{\bm{\beta}})\right]\cdot\left[\bm{\alpha}_i(2\omega_0)\cdot\mathbf{X}_{\bm{\beta}'}(\mathbf{r}_i,k_{\bm{\beta}'})\right]^*.
\end{split}
\end{equation}
Finally, we average over the power scattered from each dipole $i$ to define the power scattered from the typical dipole. Numerically, one can show that terms proportional to $\langle A_i(N,\hat{\bm{\epsilon}}_i)\left[\bm{\alpha}_i(2\omega_0)\cdot\mathbf{X}_{\bm{\beta}}(\mathbf{r}_i,k_{\bm{\beta}})\right]\cdot\left[\bm{\alpha}_i(2\omega_0)\cdot\mathbf{X}_{\bm{\beta}'}(\mathbf{r}_i,k_{\bm{\beta}'})\right]^*/f_i(\hat{\bm{\epsilon}}_i)\rangle_i$ are at least two orders of magnitude smaller when $\bm{\beta}\neq\bm{\beta}'$ than when $\bm{\beta} = \bm{\beta}'$. Thus, we can safely neglect the cross terms of the sum and define $\bar{P}_1(\omega) = \langle\bar{P}_i(\omega)\rangle_i$ such that
\begin{equation}\label{eq:timeAvgPiFinal}
\bar{P}_1(2\omega_0) \approx \frac{2\gamma_1^\mathrm{rad}}{a_1^3}E_0^2\omega_0^2\sum_{\bm{\beta}}\frac{|C_{\bm{\beta}}(2\omega_0)|^2}{a_2^{2\ell + 4}}|\alpha_{\bm{\beta}}(2\omega_0)|^2\left\langle A_i(N,\hat{\bm{\epsilon}}_i)\frac{\norm{\bm{\alpha}_i(2\omega_0)\cdot\mathbf{X}_{\bm{\beta}}(\mathbf{r}_i,k_{\bm{\beta}})}^2}{f_i(\hat{\bm{\epsilon}}_i)}\right\rangle_i.
\end{equation}
Rectification of Eq. \eqref{eq:timeAvgPiFinal} with Eq. \eqref{eq:P1} and the notation of the main text can be achieved by substituting appropriately the dimensionless constant
\begin{equation}\label{eq:Kbeta}
\begin{split}
&K_{\bm{\beta}} = \frac{\omega_1^2\gamma_1^2}{A(N)f_1a_1^6}\left\langle \frac{A_i(N,\hat{\bm{\epsilon}}_i)\norm{\bm{\alpha}_i(2\omega_0)\cdot\mathbf{X}_{\bm{\beta}}(\mathbf{r}_i,k_{\bm{\beta}})}^2}{f_i(\hat{\bm{\epsilon}}_i)}\right\rangle_i\\
&\approx \left\langle \frac{A_i(N,\hat{\bm{\epsilon}}_i)}{A(N)}\frac{f_\perp^2|\mathbf{X}_{\bm{\beta}}(\mathbf{r}_i,k_{\bm{\beta}})\cdot\hat{\mathbf{r}}_i|^2 + f_\parallel^2\left(|\mathbf{X}_{\bm{\beta}}(\mathbf{r}_i,k_{\bm{\beta}})\cdot\hat{\bm{\theta}}_i|^2 + |\mathbf{X}_{\bm{\beta}}(\mathbf{r}_i,k_{\bm{\beta}})\cdot\hat{\bm{\phi}}_i|^2\right)}{f_1f_i(\hat{\bm{\epsilon}}_i)}\right\rangle_i
\end{split}
\end{equation}
and simplifying the notation such that $\bar{P}_1(2\omega_0)\to P_\mathrm{pl}(2\omega_0)$. Here, we have taken
\begin{equation}
A(N) = \frac{1}{2}\left(A_\perp(N) + \frac{A_\parallel^2(N)\cos^{-1}\left\{\frac{A_\perp(N)}{A_\parallel(N)}\right\}}{\sqrt{A_\parallel^2(N) - A_\perp^2(N)}}\right)
\end{equation}
similar to the definition of the angle-averaged oscillator strength $f_1$. 

Thus, all that is left to characterize in $\bar{P}_1(2\omega_0)$ are the phenomenological constants $A_{\perp,\parallel}(N)$. This can be achieved by replacing $\bm{\mathcal{E}}_\mathrm{sca}(\mathbf{r}_i,\omega)$ with a field $\mathbf{E}_0(\mathbf{r}_i,\omega) = E_0\pi[\delta(\omega - 2\omega_0) + \delta(\omega + 2\omega_0)]\hat{\mathbf{e}}_{\perp,\parallel}$ with $\hat{\mathbf{e}}_{\perp,\parallel} = \hat{\mathbf{r}}_i$ and $\hat{\bm{\theta}}_i$, respectively, in Eq. \eqref{eq:timeAvgPi2}. The results can then be compared to the simulations detailed in Figure \ref{fig:fig4}c, which provide the simulated scattering cross sections of a grid of spheres upon which a plane wave polarized parallel or perpendicular to the ensemble plane is impinged, such that only ensemble-parallel or perpendicular dipoles are excited. The use of $\mathbf{E}_0(\mathbf{r}_i,\omega)$ in the theory faithfully approximates the simulated arrangement in the limit where the spacing between NPs is much smaller than the impinging light wavelength and the LiNbO$_3$ sphere radius $a_2$.

Explicitly, the power scattered by the $i^\mathrm{th}$ dipole is 
\begin{equation}\label{eq:timeAvgPiPW}
\bar{P}_i^\mathrm{PW}(2\omega_0,\hat{\mathbf{e}}_{\perp,\parallel}) = A_{\perp,\parallel}(N)\frac{2\gamma_1^\mathrm{rad}}{a_1^3f_{\perp,\parallel}}E_0^2\omega_0^2\norm{\bm{\alpha}_i(2\omega_0)\cdot\hat{\mathbf{e}}_{\perp,\parallel}}^2
\end{equation}
and, after an angular average,
\begin{equation}\label{eq:timeAvgP1PW}
\bar{P}_1^\mathrm{PW}(2\omega_0,\hat{\mathbf{e}}_{\perp,\parallel}) \approx A(N)f_{\perp,\parallel} \frac{a_1^3\gamma_1^\mathrm{rad}}{2\omega_1^2}\frac{E_0^2\omega_0^2}{|\Omega_1 - 2\omega_0|^2}
\end{equation}
where the superscript PW signifies that the driving source in a plane wave and $\bar{P}_1^\mathrm{PW}(2\omega_0) = \langle\bar{P}_i^\mathrm{PW}(2\omega_0)\rangle_i$. The associated scattering cross section to $\bar{P}_1^\mathrm{PW}(2\omega_0)$ that can be directly compared to the simulation is given by
\begin{equation}\label{eq:scattCrossPWcomponents}
\sigma^\mathrm{sca}_{\perp,\parallel}(N,2\omega_0) = R_\mathrm{sca}A_{\perp,\parallel}(N)f_{\perp,\parallel} \frac{4\pi a_1^3\gamma_1^\mathrm{rad}}{c\omega_1^2}\frac{\omega_0^2}{|\Omega_1 - 2\omega_0|^2}
\end{equation}
wherein $R_\mathrm{sca} = 6.00$ is a scaling constant that builds in the contribution of the Lorentz resonances of Au at low energies. With all other constants already well-characterized, we find the best fits to the simulated scattering data are $A_\perp(N) = 2.71N^{0.489}$ and $A_\parallel(N) = 2.43N^{0.730}$ such that $A(1000) = 301$. Finally, we find that due to the similar average magnitudes over $\mathbf{r}_i$ of the harmonics $\mathbf{X}_{\bm{\beta}}(\mathbf{r}_i,k_{\bm{\beta}})$, $K_{\bm{\beta}}$ varies little between modes, with $\mathrm{max}_{\bm{\beta}}\{K_{\bm{\beta}}\}= K_{M,0,10,2} = 10.2\times10^{-3}$ and $\mathrm{min}_{\bm{\beta}}\{K_{\bm{\beta}}\}= K_{E,0,10,0} = 5.04\times10^{-3}$.

\subsubsection{Ratio of Time-Averaged Scattered Powers}\label{sec:pRatio}

One can see from Eqs. \eqref{eq:P2} and \eqref{eq:P1} that the superradiant scattering enhancement ratio $N\bar{P}_1(2\omega_0)/\bar{P}_2(2\omega_0)$ is greatly simplified by defining the spectral profile the associated spectral profile 
\begin{equation}\label{eq:kappa}
\kappa(\omega) = \frac{\sum_{\bm{\beta}}K_{\bm{\beta}}|C_{\bm{\beta}}(\omega)|^2|\alpha_{\bm{\beta}}(\omega)|^2/a_2^{2\ell}}{\sum_{\bm{\beta}} f_2|C_{\bm{\beta}}(\omega)|^2|\alpha_{\bm{\beta}}(\omega)|^2/\gamma_2\omega_{\bm{\beta}}a_2^{2\ell}}
\end{equation}
wherein $f_2 = \langle f_{\bm{\beta}}\rangle_{\bm{\beta}}$ and $\gamma_2 = \langle\gamma_{\bm{\beta}}\rangle_{\bm{\beta}}$ are the average Mie oscillator strength and damping rate, respectively. As $K_{\bm{\beta}}$ is small for all of the relevant modes of the system, so too is $\kappa(\omega)$ for $\omega$ in the optical region, with a value between $1.67\times10^{-3}$ and $1.78\times10^{-3}$.

\subsection{Derivation of the Near-Field Enhanced Scattering}\label{sec:nearField}

In this section, the derivation of the coupled-oscillator model of the hybrid Au-LiNbO$_3$ nanostructures is developed from first principles. Section \ref{sec:first} details the calculations of the relevant field quantities and Section \ref{sec:oscModelConstr} translates these field quantities into an osicllator model.

\subsubsection{Nonlinear Optics First Principles}\label{sec:first}

The oscillator model of the modes of the LiNbO$_3$ microsphere is generated from the solutions to the coupled nonlinear wave equation for the total electric field
\begin{equation}\label{eq:nonLinWave0}
\nabla\times\nabla\times\mathbf{E}(\mathbf{r},\omega) - \frac{\omega^2}{c^2}\mathbf{E}(\mathbf{r},\omega) - \frac{\omega^2}{c^2}\left[\mathbf{P}^{(1)}(\mathbf{r},\omega) + \mathbf{P}^{(2)}(\mathbf{r},\omega)\right] = \frac{4\pi\mathrm{i}\omega}{c^2}\mathbf{J}_0(\mathbf{r},\omega),
\end{equation}
wherein $\mathbf{J}_0(\mathbf{r},\omega)$ is the current density of the laser. The polarization fields $\mathbf{P}^{(1)}(\mathbf{r},\omega) = \chi_2^{(1)}(\mathbf{r})\mathbf{E}(\mathbf{r},\omega)$ and 
\begin{equation}
\mathbf{P}^{(2)}(\mathbf{r},\omega) = \int_{-\infty}^\infty\mathbf{E}(\mathbf{r},\omega - \omega')\cdot\bm{\chi}_2^{(2)}(\mathbf{r};\omega',\omega - \omega')\cdot\mathbf{E}(\mathbf{r},\omega')\;\frac{\mathrm{d}\omega'}{2\pi}
\end{equation}
build in the first- and second-order material response of the LiNbO$_3$ sphere, respectively. The sphere's electric susceptibilities are allowed to vary in space such that $\chi_2^{(1)}(\mathbf{r}) = \chi_2^{(1)}\Theta(r \leq a_2)$ and $\bm{\chi}_2^{(2)}(\mathbf{r};\omega',\omega - \omega') = \bm{1}_3\chi_2^{(2)}(\omega',\omega - \omega')\Theta(r \leq a_2)$ with $\bm{1}_3$ the $3\times3$ identity matrix. The sphere's dielectric function is then given by $\epsilon_2(\mathbf{r}) = 1 + 4\pi\chi_2^{(1)}(\mathbf{r})$.

From here, the electric field can be perturbatively expanded with the assumption that $|\bm{\chi}_2^{(2)}(\mathbf{r};\omega',\omega - \omega')|$ is a small quantity. Explicitly, we let $\mathbf{E}(\mathbf{r},\omega) = \mathbf{E}_0(\mathbf{r},\omega) + \sum_{n = 2}^\infty\mathbf{E}_0^{(n)}(\mathbf{r},\omega) + \sum_{n = 1}^\infty\mathbf{E}_\mathrm{sca}^{(n)}(\mathbf{r},\omega)$, wherein the first term is the laser's electric field and the terms in the sums contribute $n^\mathrm{th}$-order corrections to the vacuum-like and scattered fields $\mathbf{E}_\mathrm{vac}(\mathbf{r},\omega) = \sum_{n = 2}^\infty\mathbf{E}_0^{(n)}(\mathbf{r},\omega)$ and $\mathbf{E}_\mathrm{sca}(\mathbf{r},\omega) = \sum_{n=1}^\infty\mathbf{E}_\mathrm{sca}^{(n)}(\mathbf{r},\omega)$, respectively, that are set up by the polarized sphere. Subtracting the laser field equation $\nabla\times\nabla\times\mathbf{E}_0(\mathbf{r},\omega) - (\omega^2/c^2)\mathbf{E}_0(\mathbf{r},\omega) = (4\pi\mathrm{i}\omega/c^2)\mathbf{J}_0(\mathbf{r},\omega)$ from Eq. \eqref{eq:nonLinWave0}, one finds
\begin{equation}\label{eq:nonLinWave}
\begin{split}
\left(\{\nabla\times\nabla\times\} - \epsilon_2(\mathbf{r})\frac{\omega^2}{c^2}\right)\mathbf{E}_\mathrm{sca}^{(1)}(\mathbf{r},\omega) &= \frac{4\pi\mathrm{i}\omega}{c^2}\mathbf{J}^{(1)}(\mathbf{r},\omega),\\
\left(\{\nabla\times\nabla\times\} - \epsilon_2(\mathbf{r})\frac{\omega^2}{c^2}\right)\left\{\mathbf{E}_\mathrm{sca}^{(2)}(\mathbf{r},\omega) + \mathbf{E}_0^{(2)}(\mathbf{r},\omega)\right\} &= \frac{4\pi\mathrm{i}\omega}{c^2}\mathbf{J}^{(2)}(\mathbf{r},\omega).
\end{split}
\end{equation} 

The currents in right-hand-sides of Eq. \eqref{eq:nonLinWave} are bound currents that are only nonzero where the first- and second-order susceptibilities of the LiNbO$_3$ sphere are nonzero, respectively, such that
\begin{equation}\label{eq:currents}
\begin{split}
\mathbf{J}^{(1)}(\mathbf{r},\omega) &= -\mathrm{i}\omega\chi_2^{(1)}(\mathbf{r})\mathbf{E}_0(\mathbf{r},\omega),\\
\mathbf{J}^{(2)}(\mathbf{r},\omega) &= -\mathrm{i}\omega\int_{-\infty}^\infty\left(\mathbf{E}_0(\mathbf{r},\omega') + \mathbf{E}_\mathrm{sca}^{(1)}(\mathbf{r},\omega')\right)\\
&\cdot\bm{\chi}_2^{(2)}(\mathbf{r};\omega',\omega - \omega')\cdot\left(\mathbf{E}_0(\mathbf{r},\omega - \omega') + \mathbf{E}_\mathrm{sca}^{(1)}(\mathbf{r},\omega - \omega')\right)\;\frac{\mathrm{d}\omega'}{2\pi}.
\end{split}
\end{equation}
The former current can straightforwardly be seen to be large where $\omega = \pm\omega_0$ if the laser light is a monochromatic plane wave with an electric field
\begin{equation}
\begin{split}
\mathbf{E}_0(\mathbf{r},\omega) &= E_0\left[\pi\delta(\omega - \omega_0)\mathrm{e}^{\mathrm{i}\omega_0 z/c} + \pi\delta(\omega + \omega_0)\mathrm{e}^{-\mathrm{i}\omega_0 z/c}\right]\hat{\mathbf{x}},
\end{split}
\end{equation}
while for the same incoming field the leading factor of $\omega$ in the latter restricts the integral to be zero unless $\omega' = \pm\omega_0$ and $\omega = 2\omega'$. The currents drive the total fields through the Green's function solution to either wave equation,
\begin{equation}\label{eq:waveEqSoln}
\mathbf{E}_\mathrm{sca}^{(n)}(\mathbf{r},\omega) + \mathbf{E}_0^{(n)}(\mathbf{r},\omega) = \frac{4\pi\mathrm{i}\omega}{c}\int\mathbf{G}_\mathrm{LNO}(\mathbf{r},\mathbf{r}';\omega)\cdot\frac{\mathbf{J}^{(n)}(\mathbf{r}',\omega)}{c}\;\mathrm{d}^3\mathbf{r}',
\end{equation}
where the spherical dyadic Green's function $\mathbf{G}_\mathrm{LNO}(\mathbf{r},\mathbf{r}';\omega)$ is given in the literature \cite{tai1994dyadic}. The Green's function, as we show below, is separable as $\mathbf{G}_\mathrm{LNO}(\mathbf{r},\mathbf{r}';\omega) = \mathbf{G}_\mathrm{sca}(\mathbf{r},\mathbf{r}';\omega) + \mathbf{G}_0(\mathbf{r},\mathbf{r};\omega)$ into a ``scattering'' part $\mathbf{G}_\mathrm{sca}(\mathbf{r},\mathbf{r}';\omega)$ and a ``vacuum-like'' part $\mathbf{G}_0(\mathbf{r},\mathbf{r};\omega)$ such that
\begin{equation}
\begin{split}
\mathbf{E}_\mathrm{sca}^{(n)}(\mathbf{r},\omega)= \frac{4\pi\mathrm{i}\omega}{c}\int\mathbf{G}_\mathrm{sca}(\mathbf{r},\mathbf{r}';\omega)\cdot\frac{\mathbf{J}^{(n)}(\mathbf{r}',\omega)}{c}\;\mathrm{d}^3\mathbf{r}',\\
\mathbf{E}_0^{(n)}(\mathbf{r},\omega) = \frac{4\pi\mathrm{i}\omega}{c}\int\mathbf{G}_0(\mathbf{r},\mathbf{r}';\omega)\cdot\frac{\mathbf{J}^{(n)}(\mathbf{r}',\omega)}{c}\;\mathrm{d}^3\mathbf{r}'.
\end{split}
\end{equation}

A similar solution can be found for the scattered electric field of the Au NPs. The incident field and second-order current $\mathbf{J}^{(2)}(\mathbf{r},\omega)$ can drive the response of each NP in the same manner that they drive the scattered fields of the LiNbO$_3$ sphere, such that we can build a model that includes only these two sources and temporarily neglects interactions between the NPs and microsphere. Ignoring the effects of radiation, the scattered potential of the NP is
\begin{equation}\label{eq:gaussNP}
-\nabla\cdot\epsilon_1(\mathbf{r},\omega)\nabla\left\{\Phi_\mathrm{NP}(\mathbf{r},\omega) + \Phi_0^{(2)}(\mathbf{r},\omega)\right\} = 4\pi\chi_1(\mathbf{r},\omega)\left[\nabla\cdot\nabla\Phi_0(\mathbf{r},\omega) + \frac{4\pi\mathrm{i}}{\omega}\nabla\cdot\mathbf{J}^{(2)}(\mathbf{r},\omega)\right]
\end{equation}
for $\chi_1(\mathbf{r},\omega) = \chi_1(\omega)\Theta(\mathbf{r}\in V_1)$ and $\mathbf{r}$ not on the boundary of the NP's volume $V_1$. Assuming the laser wavelength $\lambda_0 = 2\pi c/\omega_0$ is much longer than the extent of the NP, the incident potential can be written as $\Phi_0(\mathbf{r},\omega)\approx \mathbf{r}\cdot\lim_{k\to0}\mathbf{E}_0(\mathbf{r},\omega) = xE_0\pi[\delta(\omega - \omega_0) + \delta(\omega + \omega_0)]$, where $k = \omega/c$. Similarly, $-\nabla\Phi_0^{(2)}(\mathbf{r},\omega) = \lim_{k\to0}\mathbf{E}_0^{(2)}(\mathbf{r},\omega)$.

Assuming that the solutions to Eq. \eqref{eq:gaussNP} are near zero at $\pm\omega_0$, i.e. the NP does not respond strongly near the fundamental frequency of the laser, the term proportional to $\Phi_0(\mathbf{r},\omega)$ on the RHS can be dropped. Thus, with the dielectric function $\epsilon_1(\mathbf{r},\omega) = \epsilon_1(\omega)\Theta(r\leq a_1) + 1\Theta(r\geq a_1)$ defined such that the NP lies at the origin, one finds:
\begin{equation}\label{eq:PoissonEqSoln}
\Phi_\mathrm{NP}(\mathrm{r},\omega) + \Phi_0^{(2)}(\mathbf{r},\omega) \approx \int G_\mathrm{NP}(\mathbf{r},\mathbf{r}';\omega)\chi_1(\mathbf{r},\omega)\frac{4\pi\mathrm{i}}{\omega}\nabla\cdot\mathbf{J}^{(2)}(\mathbf{r},\omega)\;\mathrm{d}^3\mathbf{r}',
\end{equation}
wherein the (scalar) Green's function $G_\mathrm{NP}(\mathbf{r},\mathbf{r}',\omega)$ is the standard solution to the Poisson equation in spherical coordinates \cite{olafsson2020electron}. Much like the dyadic Green's function for the LiNbO$_3$ fields, $G_\mathrm{NP}(\mathbf{r},\mathbf{r}',\omega) = G_\mathrm{sca}(\mathbf{r},\mathbf{r}',\omega) + G_\mathrm{0}(\mathbf{r},\mathbf{r}',\omega)$ is separable into a scattering part and a vacuum-like or free-space part. The former encodes the resonances of the NP, while the latter serves simply to satisfy the principle of superposition and is otherwise unimportant.

\subsubsection{Construction of the Oscillator Model}\label{sec:oscModelConstr}

The above definitions are clarified by the forms of both $\mathbf{G}_\mathrm{LNO}$ and $G_\mathrm{NP}$, which are cumbersome but manageable. The former, in the case where both the source charges and the observer are outside the sphere's surface, is given by
\begin{equation}
\begin{split}
G_\mathrm{NP}(\mathbf{r},\mathbf{r}';\omega)&\Theta(r > a_1)\Theta(r' > a_1) = \sum_{p = 0}^1\sum_{\ell = 1}^\infty\sum_{m = 0}^\ell(2 - \delta_{m0})\frac{(\ell - m)!}{(\ell + m)!}\frac{\ell[1 - \epsilon_1(\omega)]}{\ell\epsilon_1(\omega) + (\ell + 1)}\\
&\times \frac{a_1^{2\ell + 1}}{r^{\ell + 1}r'^{\ell + 1}}P_{\ell m}(\cos\theta)P_{\ell m}(\cos\theta')S_p(m\phi)S_p(m\phi') + \frac{1}{|\mathbf{r} - \mathbf{r}'|}
\end{split}
\end{equation}
for a sphere centered at the origin. Each of the mode functions $(a_1^{\ell + 1}/r^{\ell + 1})P_{\ell m}(\cos\theta)S_p(m\phi)$ describes the spatial variation of the observables of an electric multipole mode with a characteristic response function \cite{cherqui2016characterizing}
\begin{equation}\label{eq:gResp}
g_{p\ell m}(\omega) = \frac{\ell[1 - \epsilon_1(\omega)]}{\ell\epsilon_1(\omega) + (\ell + 1)}
\end{equation}
that describes its oscillations in time. Here, $\ell$ and $m$ give the order and degree of each mode's corresponding spherical harmonic and $p$ the reflection symmetry of each mode across the $x$-axis.

The Green's function of the LiNbO$_3$ sphere can be similarly expanded using a set of so-called quasinormal geometric resonances of both magnetic and electric multipole symmetry \cite{lalanne2018light} with response functions that depend on the same spherical harmonic symmetry parameters $p,\ell,m$. For an observer inside the sphere, this Green's function is:
\begin{equation}
\begin{split}
\mathbf{G}_\mathrm{LNO}&(\mathbf{r},\mathbf{r}';\omega)\Theta(r < a_2) = \frac{\mathrm{i}\omega}{4\pi c}\sum_{p\ell m}\left[C_{p\ell m}^>(\omega)\bm{\mathcal{M}}_{p\ell m}(\mathbf{r},\sqrt{\epsilon_2}k)\mathbf{M}_{p\ell m}(\mathbf{r}',k)\right.\\
&\left.+D_{p\ell m}^>(\omega)\bm{\mathcal{N}}_{p\ell m}(\mathbf{r},\sqrt{\epsilon_2}k)\mathbf{N}_{p\ell m}(\mathbf{r}',k)\right]\Theta(r < a_2)\Theta(r' > a_2)\\
&+\sqrt{\epsilon_2}\frac{\mathrm{i}\omega}{4\pi c}\sum_{p\ell m}\left[C_{p\ell m}^<(\omega)\bm{\mathcal{M}}_{p\ell m}(\mathbf{r},\sqrt{\epsilon_2}k)\bm{\mathcal{M}}_{p\ell m}(\mathbf{r}',\sqrt{\epsilon_2}k) \right.\\
&+\left.D_{p\ell m}^<(\omega)\bm{\mathcal{N}}_{p\ell m}(\mathbf{r},\sqrt{\epsilon_2}k)\bm{\mathcal{N}}_{p\ell m}(\mathbf{r}',\sqrt{\epsilon_2}k)\right]\Theta(r < a_2)\Theta(r' < a_2)\\
&+ \mathbf{G}_0(\mathbf{r},\mathbf{r}';\sqrt{\epsilon_2}\omega)\Theta(r < a_2)\Theta(r' < a_2),
\end{split}
\end{equation}
where again the sphere is centered at the origin. The mode functions $\bm{\mathcal{M}}_{p\ell m}$ and $\bm{\mathcal{N}}_{p\ell m}$ are identical to the regularized vector spherical harmonics given in \eqref{eq:Xharmonics} but with the spherical Hankel functions replaced with spherical Bessel functions, $h_{\ell}^{(1)}(x)\to j_{\ell}(x)$. The Green's function for an observer outside the sphere is, similarly:
\begin{equation}
\begin{split}
\mathbf{G}_\mathrm{LNO}&(\mathbf{r},\mathbf{r}';\omega)\Theta(r > a_2) = \frac{\mathrm{i}\omega}{4\pi c}\sum_{p\ell m}\left[A_{p\ell m}^>(\omega)\mathbf{M}_{p\ell m}(\mathbf{r},k)\mathbf{M}_{p\ell m}(\mathbf{r}',k)\right.\\
&\left.+B_{p\ell m}^>(\omega)\mathbf{N}_{p\ell m}(\mathbf{r},k)\mathbf{N}_{p\ell m}(\mathbf{r}',k)\right]\Theta(r > a_2)\Theta(r' > a_2)\\
&+\sqrt{\epsilon_2}\frac{\mathrm{i}\omega}{4\pi c}\sum_{p\ell m}\left[A_{p\ell m}^<(\omega)\mathbf{M}_{p\ell m}(\mathbf{r},k)\bm{\mathcal{M}}_{p\ell m}(\mathbf{r}',\sqrt{\epsilon_2}k) \right.\\
&+\left.B_{p\ell m}^<(\omega)\mathbf{N}_{p\ell m}(\mathbf{r},k)\bm{\mathcal{N}}_{p\ell m}(\mathbf{r}',\sqrt{\epsilon_2}k)\right]\Theta(r > a_2)\Theta(r' < a_2)\\
&+ \mathbf{G}_0(\mathbf{r},\mathbf{r}';\omega)\Theta(r > a_2)\Theta(r' > a_2),
\end{split}
\end{equation}
with the forms of the unwieldy response functions $A_{p\ell m}(\omega)$, $B_{p\ell m}(\omega)$, $C_{p\ell m}(\omega)$, and $D_{p\ell m}(\omega)$ described in Ref. \citenum{tai1994dyadic}. Further, $\mathbf{G}_0(\mathbf{r},\mathbf{r}';\omega)$ is the Green's function of free space.

To compare the responses of the NP and microsphere, we can see that $B_{p\ell m}^<(\omega)$ obeys the simple relation $\lim_{k\to0}(\sqrt{\epsilon_2})^\ell B_{p\ell m}(\omega) - 1 = \ell(1 - \epsilon_2)/(\ell\epsilon_2 + \ell + 1)$. This gives us a foundation from which to make quantitative comparisons between the oscillator parameters assigned to each response in either particle. In detail, using the Drude-Lorentz dielectric model of Au from Section \ref{sec:dielFit}, we can see that the dipole response function $g_1(\omega)$ describes the oscillations of three coupled modes, one which has primarily free-electron character (the dipole plasmon) two others which are mostly comprised of the material's interband resonances. Explicitly,
\begin{equation}\label{eq:gResp2}
g_{p1m}(\omega) = -\frac{e^2}{a_1^3}\left[\frac{1}{2\omega_1 \mu_1}\left(\frac{\mathrm{e}^{\mathrm{i}\psi_1}}{\Omega_1 - \omega} + \frac{\mathrm{e}^{-\mathrm{i}\psi_1}}{\Omega_1^* + \omega}\right) + \sum_{i=1}^2\frac{1}{2\omega_{L_i} \mu_{L_i}}\left(\frac{\mathrm{e}^{\mathrm{i}\psi_{L_i}}}{\Omega_{L_i} - \omega} + \frac{\mathrm{e}^{-\mathrm{i}\psi_{L_i}}}{\Omega_{L_i}^* + \omega}\right)\right],
\end{equation}
wherein the mass $\mu$, resonance frequency $\omega$, complex eigenvalue $\Omega$, and phase offset $\psi$ of the plasmon are labeled with subscript 1, and the oscillator parameters of the fictitious oscillator are labeled with subscripts $L_i$. Explicit values of these parameters are given in Section \ref{sec:paramInf}.

As is also shown in Section \ref{sec:paramInf}, there are many modes of the LiNbO$_3$ sphere that have significant response magnitudes in the range of laser energies in which SHG is observed ($\sim$2.3--2.4 eV, see Figure \ref{fig:fig4}c). To simplify their description, we use an expansion of the LiNbO$_3$ mode response functions:
\begin{equation}\label{eq:ABresp}
\begin{split}
(\sqrt{\epsilon_2})^\ell A_{p\ell m}^<(\omega) - 1 &\approx -\frac{e^2}{a_2^3}\sum_j\frac{1}{2\omega_{M\ell j}\mu_{M\ell j}}\left(\frac{\mathrm{e}^{\mathrm{i}\psi_{M\ell j}}}{\Omega_{M\ell j} - \omega} + \frac{\mathrm{e}^{-\mathrm{i}\psi_{M\ell j}}}{\Omega_{M\ell j}^* + \omega}\right), \\
(\sqrt{\epsilon_2})^\ell B_{p\ell m}^<(\omega) - 1 &\approx -\frac{e^2}{a_2^3}\sum_j\frac{1}{2\omega_{E\ell j}\mu_{E\ell j}}\left(\frac{\mathrm{e}^{\mathrm{i}\psi_{E\ell j}}}{\Omega_{E\ell j} - \omega} + \frac{\mathrm{e}^{-\mathrm{i}\psi_{E\ell j}}}{\Omega_{E\ell j}^* + \omega}\right),
\end{split}
\end{equation}
where the labels $M,E$ denote magnetic- and electric-type oscillator parameters and the indices $j$ label the different resonances of common angular symmetry $p,\ell,m$ but different node structure along the radial coordinate. Because only a single mode of the set of modes $j$ lies in the energetic range of interest, we will drop the sum over $j$ and the corresponding labels in the following discussion.

With explicit value of the relevant LiNbO$_3$ oscillator parameters given in Section \ref{sec:paramInf}, the formal definitions of the oscillator moments $d_\nu(\omega)$ of the dipole plasmons oriented along the three cardinal axes $\nu = r,\theta,\phi$ as well the LiNbO$_3$ Mie multipoles $\bm{\beta}$ can be defined analytically. In particular, with a vector spherical harmonic expansion of the laser field\cite{bohren2004absorption}
\begin{equation}
\mathbf{E}_0(\mathbf{r},\omega) = E_0\left[\pi\delta(\omega - \omega_0) + \pi\delta(\omega + \omega_0)\right]\sum_\ell\mathrm{i}^\ell\frac{2\ell + 1}{\ell(\ell + 1)}\left[\bm{\mathcal{M}}_{1\ell 1}(\mathbf{r},k) - \mathrm{i}\bm{\mathcal{N}}_{0\ell 1}(\mathbf{r},k)\right],
\end{equation}
the second-order fields of a \textit{bare} LiNbO$_3$ microsphere centered at the origin can be written as
\begin{equation}
\begin{split}
\mathbf{E}_\mathrm{sca}^{(2)}(\mathbf{r},\omega) &= \sum_{\bm{\beta}}\frac{1}{a_2^{\ell + 2}}\left[\rho_{\bm{\beta}}(\omega)\mathbf{X}_{\bm{\beta}}(\mathbf{r},k_{\bm{\beta}}) + \rho_{\bm{\beta}}^*(-\omega)\mathbf{X}_{\bm{\beta}}^*(\mathbf{r},k_{\bm{\beta}})\right],\\
\mathbf{E}_0^{(2)}(\mathbf{r},\omega) &= -\mathrm{i}\frac{\omega_0^9}{4c^9}\pi\delta(\omega - 2\omega_0)\epsilon_2(\epsilon_2 - 1)^2E_0^2\chi_2^{(2)}(\omega_0,\omega_0)\\
&\times\sum_{\bm{\gamma}\bm{\alpha}\bm{\alpha}'}w_{\bm{\alpha}}(\omega_0)w_{\bm{\alpha}'}(\omega_0)I_{\bm{\gamma}\bm{\alpha}\bm{\alpha}'}(k_{\bm{\gamma}},\sqrt{\epsilon_2}k_0,\sqrt{\epsilon_2}k_0;0,a_2)\bm{\mathcal{X}}_{\bm{\gamma}}(\mathbf{r},\omega) + \mathrm{c.c.r.}
\end{split}
\end{equation}
Here, as defined in the main text, $\rho_{\bm{\beta}}(\omega)$ are the multipole moment magnitudes of the Mie resonances of the sphere. Their explicit form is complicated and is detailed below. The index $\bm{\gamma} = \{T''',p''',\ell''',m'''\}$ is another collective Mie index like $\bm{\alpha}$ and $\bm{\beta}$. It labels modes that contribute to the field near $2\omega_0$ but, unlike $\bm{\beta}$, is not restricted to counting only modes with strongly resonant behavior. In the calculations of this work, it is taken to label all Mie modes with $\ell > 1$ up to a cutoff $\ell = 12$ after which the contribution of successive terms in the sum is negligible. Further, the term $\mathrm{c.c.r.}$ is simply the frequency reversed ($\omega\to-\omega$) complex conjugate of the first term of the second equality above such that $\mathbf{E}_0^{(2)}(\mathbf{r},\omega)$ satisfies the Fourier reality condition $\mathbf{E}_0^{(2)*}(\mathbf{r},\omega) = \mathbf{E}_0^{(2)}(\mathbf{r},-\omega)$.

The functions $w_{\bm{\alpha}}(\omega)$ inside the definition of the second-order vacuum-like field are the weights 
\begin{equation}
w_{\bm{\alpha}}(\omega_0) = 
\begin{cases}
\mathrm{i}^\ell\sqrt{\dfrac{2\ell + 1}{2\ell(\ell + 1)}\dfrac{(\ell + 1)!}{(\ell - 1)!}}C_{p\ell m}^<(\omega)R_\ell(k_0,\sqrt{\epsilon_2}k_0;0,a_2), & T = M;\\[1.0em]
\!\begin{aligned}
-\mathrm{i}^{\ell + 1}\sqrt{\frac{2\ell + 1}{2\ell(\ell + 1)}\frac{(\ell + 1)!}{(\ell - 1)!}}&D_{p\ell m}^<(\omega)\left[\frac{\ell + 1}{2\ell + 1}R_{\ell - 1}(k_0,\sqrt{\epsilon_2}k_0;0,a_2)\right.\\
&+\left. \frac{\ell}{2\ell + 1}R_{\ell + 1}(k_0,\sqrt{\epsilon_2}k_0;0,a_2) \right]
\end{aligned}, & T = E;
\end{cases}
\end{equation}
wherein 
\begin{equation}
\begin{split}
R_\ell(k,k';a,b) &= \int_a^b r^2j_\ell(kr)j_\ell(k'r)\;\mathrm{d}r\\
&= \frac{r^2}{k^2 - k'^2}\left[k'^2j_\ell(kr)j_{\ell - 1}(k'r) - k^2j_{\ell - 1}(kr)j_\ell(k'r)\right]_a^b
\end{split}
\end{equation}
is an overlap integral over the radial components $j_\ell(kr)$ of the fundamental mode functions. Further, the functions $I_{\bm{\alpha}_1\bm{\alpha}_2\bm{\alpha}_3}$ are overlap triple integrals
\begin{equation}
\begin{split}
I_{\bm{\alpha}_1\bm{\alpha}_2\bm{\alpha}_3}(k_1,k_2,k_3;a,b) &= \int_0^{2\pi}\int_0^\pi\int_a^b \bm{\mathcal{X}}_{\bm{\alpha}_1}(\mathbf{r}',k_1)\cdot\left[\bm{\mathcal{X}}_{\bm{\alpha}_2}(\mathbf{r}',k_2)\cdot\bm{1}_3\cdot\bm{\mathcal{X}}_{\bm{\alpha}_3}(\mathbf{r}',k_3)\right]\\
&\times r'^2\sin\theta'\;\mathrm{d}r'\,\mathrm{d}\theta'\,\mathrm{d}\phi',
\end{split}
\end{equation}
wherein $\bm{\mathcal{X}}_{Tp\ell m}(\mathbf{r},k) = \bm{\mathcal{M}}_{p\ell m}(\mathbf{r},k)\delta_{T,M} + \bm{\mathcal{N}}_{p\ell m}(\mathbf{r},k)\delta_{T,E}$.

Using the definition of the dipole polarizabilities $\alpha_\nu(\omega)$ given in Section \ref{sec:ensembleAuInf}, we can define the motion of each component of a single Au dipole and each Mie resonance in the absence of their mutual coupling as
\begin{equation}\label{eq:momentsUncoupled}
\begin{split}
d_\nu(\omega) &= \frac{e^2}{2\omega_1\mu_\nu}\left(\frac{\mathrm{e}^{\mathrm{i}\psi_1}}{\Omega_1 - \omega} + \frac{\mathrm{e}^{-\mathrm{i}\psi_1}}{\Omega_1^* + \omega}\right)\hat{\mathbf{e}}_\nu\cdot\mathbf{E}_0^{(2)}(\mathbf{r}_0,\omega),\\
\rho_{\bm{\beta}}(\omega) &=  \frac{e^2a_2^{\ell-2}}{2\omega_{\bm{\beta}}\mu_{\bm{\beta}}}\frac{\mathrm{e}^{\mathrm{i}\psi_{\bm{\beta}}}}{\Omega_{\bm{\beta}} - \omega}E_0\pi\delta(\omega - 2\omega_0)C_{\bm{\beta}}(\omega_0).
\end{split}
\end{equation}
In accordance with the main text, the dipole is assumed to exist at $\mathbf{r}_0 = (a_1 + a_2)\hat{\mathbf{r}}(\theta_0,\phi_0)$ and to be driven by the second-order vacuum-like field. We ignore any modifications to $\mathbf{E}_0^{(2)}(\mathbf{r},\omega)$ that are produced by $d_\nu(\omega)$. The Mie resonances of the LiNbO$_3$ are driven through the upconversion process, as detailed by the constants
\begin{equation}\label{eq:Cbeta}
\begin{split} 
C_{\bm{\beta}}(\omega_0) &= \mathrm{i}\frac{\omega_0^9}{4c^9}\frac{\epsilon_2^\frac{3}{2}(\epsilon_2 - 1)^2}{(\sqrt{\epsilon_2})^\ell}E_0\chi_2^{(2)}(\omega_0,\omega_0)\\
&\times\sum_{\bm{\alpha}\bm{\alpha}'}w_{\bm{\alpha}}(\omega_0)w_{\bm{\alpha'}}(\omega_0)I_{\bm{\beta}\bm{\alpha}\bm{\alpha'}}(\sqrt{\epsilon_2}k_{\bm{\beta}},\sqrt{\epsilon_2}k_0,\sqrt{\epsilon_2}k_0;0,a_2).
\end{split}
\end{equation}

Finally, to reproduce the equations of motion of Eq. \eqref{eq:EOMmain}, we must define the external forces acting on the particle moments and introduce the coupling forces to Eq. \eqref{eq:momentsUncoupled}. The former can be quickly written as $F_{1\nu}(\omega) = e\hat{\mathbf{e}}_\nu\cdot\mathbf{E}_0^{(2)}(\mathbf{r}_0,\omega)$ and $F_{2\bm{\beta}}(\omega) = eE_0\pi\delta(\omega - 2\omega_0)C_{\bm{\beta}}(\omega)$. The latter arise from the interaction energy $U(t) = -\mathbf{d}(t)\cdot\mathbf{E}_\mathrm{sca}^{(2)}(\mathbf{r}_0,t)$, where $\mathbf{d}(t) = \sum_\nu d_\nu(t)\hat{\mathbf{e}}_\nu$. With the identities
\begin{equation}
\begin{split}
-a_2^{\ell - 1}\frac{\partial U(t)}{\partial \rho_{\bm{\beta}}(t)} & = \sigma_{\bm{\beta}\nu}d_\nu(t),\\
-a_2^{\ell - 1}\frac{\partial U(t)}{\partial d_{\nu}(t)} &= \sigma_{\bm{\beta}\nu}\rho_{\bm{\beta}}(t),
\end{split}
\end{equation}
one is immediately delivered the equations of motion in the main text with $\sigma_{\bm{\beta}\nu} = (e^2/a_2^3)\hat{\mathbf{e}}_\nu\cdot\mathbf{X}_{\bm{\beta}}(\mathbf{r}_0,k_{\bm{\beta}})$. A detailed description of the solutions to the equations of motion are given in Section \ref{sec:methodsSupp}, and an analysis of the contribution of the various modes of the LiNbO$_3$ sphere to enhancement signal is shown in Figure \ref{fig:figS14}.

In closing, we note that, as the solutions to the equations of motion discussed in the main text and highlighted in Section \ref{sec:methodsSupp} involve a second perturbation expansion of the equations of motion of each Mie resonance, two sets of superscripts arise that correspond to two separate expansions. In order to clarify the notation, we can see that no terms in the solution to the wave equation are kept beyond second order, such that we can replace superscripted variable names with script names. Further, we can drop the superscripts of the first-order terms altogether, such that, elsewhere in the SI and main text, we let $\mathbf{E}_\mathrm{sca}^{(2)}\to\bm{\mathcal{E}}_\mathrm{sca}$, $\mathbf{E}_0^{(2)}\to\bm{\mathcal{E}}_0$, and $\mathbf{E}_\mathrm{sca}^{(1)}\to\mathbf{E}_\mathrm{sca}$.

\pagebreak
\section{Additional Figures and Tables}\label{sec:moreFigures}

\pagebreak

\pagebreak

\pagebreak

\pagebreak

\pagebreak



\pagebreak

\pagebreak

\pagebreak

\pagebreak


\pagebreak

\pagebreak

\pagebreak

\pagebreak

\pagebreak

\pagebreak

\pagebreak

\pagebreak
\begin{figure}[th!] 
\centering
\includegraphics[width = 0.80\textwidth]{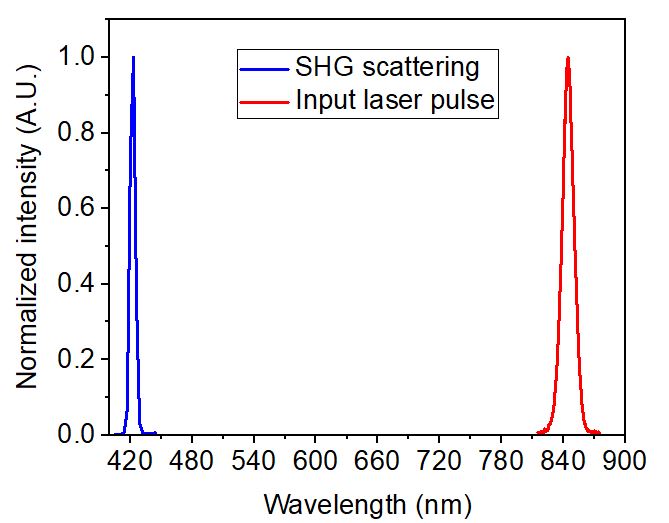}
\caption{
\textbf{Input and SHG lineshapes.} Normalized lineshapes of the input laser (red) and SHG (blue) power spectra. 
\label{fig:figS18}
}
\end{figure}

\pagebreak
\begin{table}[H]
\centering
\begin{tabular}{l | l }
Parameter & Value\\
\hline\hline
$\hbar\omega_{1}$ (bare) & 2.49 eV\\
$\hbar\omega_{1}$ (LiNbO$_3$-adjacent) & 2.26 eV\\
$\hbar\omega_{L_1}$ & 3.07 eV\\
$\hbar\omega_{L_2}$ & 6.78 eV\\
\hline
$\hbar\gamma_{1}$ & 450 meV\\
$\hbar\gamma_{L_1}$ & 688 meV\\
$\hbar\gamma_{L_2}$ & 529 meV\\
\hline
$\psi_{1}$ & -433 mrad\\
$\psi_{L_1}$ & -260 mrad\\
$\psi_{L_2}$ & 26.5 mrad\\
\hline
$\mu_{1}$ & $1.20\times10^{-30}$ g\\
$\mu_{L_1}$ & $1.66\times10^{-30}$ g\\
$\mu_{L_2}$ & $2.04\times10^{-32}$ g\\
\end{tabular}
\caption{\textbf{Au NP oscillator parameters.} Table of oscillator parameters for the model of Au used in this work.}
\label{tab:tabS1}
\end{table}
\clearpage

\begin{table}[H]
\centering
\begin{tabular}{ c | c | c | c | c }
Mode indices & $\hbar\omega_{\bm{\beta}}$ [eV] & $\hbar\gamma_{\bm{\beta}}$ [eV] & $\mu_{\bm{\beta}}\times10^{38}$ [g] & $\psi_{\bm{\beta}}$ [rad]\\
\hline\hline
$T = E$, $\ell = 6$ & 2.24 & 0.137 & 1.56 & -0.139\\
$\ell = 7$ & 2.46 & 0.0898 & 0.824 & -0.0539\\
$\ell = 9$ & 2.28 & 0.0140 & 1.70 & 3.14\\
$\ell = 10$ & 2.47 & 0.0149 & 1.06 & 3.14\\
\hline
$T = M$, $\ell = 7$ & 2.34 & 0.0394 & 1.39 & -0.127\\
$\ell = 10$ & 2.31 & 0.0144 & 1.35 & 3.14\\
$\ell = 11$ & 2.50 & 0.0155 & 0.835 & 3.14
\end{tabular}
\caption{\textbf{LiNbO$_{\bm{3}}$ microsphere oscillator parameters.} Oscillator model parameters of the modeled Mie resonances of the LiNbO$_3$ microsphere.}
\label{tab:tabS2}
\end{table}
\clearpage

\begin{table}[H]
\centering
\begin{tabular}{c | c | c}
Sample Number & Radius [nm] & Real Dielectric Constant\\
\hline
1 & 350 & 6.3\\
2 & 350 & 6.3\\
3 & 400 & 6.0\\
4 & 550 & 5.7\\
5 & 600 & 5.8\\
6 & 600 & 7.0\\
7 & 600 & 5.0\\
8 & 1000 & 6.3\\
9 & 1000 & 5.5
\end{tabular}
\caption{\textbf{Estimated LiNbO$_{\bm{3}}$ microsphere parameters.} Estimates of the dielectric constant of LiNbO$_3$ extracted from Mie theory reproductions of the scattering spectra of nine bare mesoporous LiNbO$_3$ microspheres in the spectra region 1.5--3.5 eV. The radii of the particles were measured by microscopy techniques with a precision of $\pm$50 nm. The real parts of the dielectric functions are estimated to the nearest 0.1.}
\label{tab:tabS3}
\end{table}
\clearpage

\begin{table}[H]
\centering
\begin{tabular}{c | c }
Parameter & Value\\
\hline\hline
$\hbar\omega_{p1}$ & 8.94 eV\\
$\hbar\omega_{p2}$ & 2.58 eV\\
$\hbar\omega_{p3}$ & 6.31 eV\\
\hline
$\hbar\Gamma_{1}$ & 42.8 meV\\
$\hbar\Gamma_{2}$ & 660 meV\\
$\hbar\Gamma_{3}$ & 964 meV\\
\hline
$\hbar\Lambda_2$ & 2.84 eV\\
$\hbar\Lambda_3$ & 3.71 eV
\end{tabular}
\caption{\textbf{Au Drude-Lorentz model parameters.} Table of Drude-Lorentz dielectric function values for the model of Au used in this work. See Figure \ref{fig:figS16} for further details.}
\label{tab:tabS4}
\end{table}
\clearpage



